\documentclass[11pt]{article}
\usepackage{amsthm}
\usepackage{graphicx}
\usepackage{amsmath}
\usepackage{amssymb}
\usepackage{amsfonts}
\usepackage[english]{babel}
\usepackage[]{graphics}
\usepackage{slashed} 

\begin{document}

\begin{flushright}

\baselineskip=12pt

\hfill{ }\\

\end{flushright}

\begin{center}

\baselineskip=24pt

{\Large\bf First numerical approach to a Grosse-Wulkenhaar  model }

\vskip 0.5 cm

\vspace{1cm}

{ \Large Bernardino Spisso} \\ \normalsize
\emph{ Mathematisches Institut der Westf\"alischen Wilhelms-Universit\"at} \\
\emph{ Einsteinstraße 62, D-48149 M\"unster, Germany}\\
\emph{e-mail:nispisso@tin.it}
\end{center}
\vskip 1 cm
\begin{abstract}
A numerical investigation of a
non-commutative field theory defined via the spectral action
principle is conducted. The
construction of this triple relies on an 8-dimensional Clifford
algebra. Following to the standard procedure of non-commutative
geometry, the spectral action is computed for the product of the triple
$(\mathcal{A}_4,\mathcal{H}_4,\mathcal{D}_4)$ with a matrix-valued
spectral triple. Using Monte Carlo simulation we study various quantities such as
the energy density, the specific heat density and some order parameters varying
   the matrix size and the independent parameters of the model.
\end{abstract}

\section{Introduction }

The main object of this work is a particular non-commutative field theory  which is  derived using the spectral 
action principle and then treated numerically. Non-commutativity can be found in many fields of physics 
like quantum field theories, string theory, condensed matter physics. The first application of  non-commutativity 
into physics is  dated from the middle of the last century inspired by the ideas of quantum mechanics, where  starting from classical 
mechanics,  the commutative algebra of functions on the phase space is replaced by a non-commutative operator algebra on a Hilbert space.
The duality between ordinary spaces $M$ and proper commutative algebras is expressed by the Gel'fand-Naimark theorem  which states the
fact that the algebra of all continuous functions on $M$ is the only possible
type of commutative $C^*$-algebra. Additionally, given a commutative $C^*$-algebra $C$, it is
possible to reconstruct a Hausdorff topological space $M$ in order to obtain that $C$ is the algebra
of continuous functions on $M$. The study of commutative
$C^*$-algebras is equivalent to the study of topological Hausdorff spaces.
The previous duality has inspired the identification, in non-commutative geometry, of  
some algebraical objects as a category of non-commutative topological spaces.
Alain Connes \cite{Connes}, one of the founders of non-commutative geometry, has proposed a candidate for the objects of such  category,
the spectral triples $(\mathcal{A},\mathcal{H}, \mathcal{D})$ \cite{Connes-1}, composed by an algebra $\mathcal{A}$, an Hilbert
space $\mathcal{H}$ on which $\mathcal{A}$ is represented and an  selfadjoint operator $\mathcal{D}$.
In fact, every compact oriented Riemannian manifold can be used to define a spectral triple, 
this kind of manifold $M$ characterizes a  Dirac operator on self-adjoint Clifford module bundles
over $M$. Connes, after a conjecture in 1996 \cite{Connes3}
and some considerable attempts of Rennie and Varilly \cite{Rennie-vari},
 proved the so called reconstruction theorem \cite{ConnesRec} for commutative
spectral triples satisfying various axioms, showing that 
exists a compact oriented smooth manifold $X$ such
that $A = C^\infty(X)$ is the algebra of smooth functions on $X$ and
every compact oriented smooth manifold emerges in this way.
Pushed by the aim of reformulating the standard model of particles in a non-commutative  way \cite{Connes2,Connes3},  
Connes has introduced the almost-commutative spectral triple extending the axioms of the reconstruction theorem to a non-commutative algebra.
A first attempt to formulate a field theory for a truly  non-commutative algebra was obtained replacing in the usual field theory action the point-wise multiplication of the fields with a non-commutative one, namely a $\star$-product.
The fields now belongs to  $\mathbb{R}^4_\Theta$,  a vector space defined by an enough regular class functions on $\mathbb{R}^4_\Theta$ equipped with the Moyal product:
\begin{equation*}
(f \star g)(x) = \int \int d^4 y \frac{d^4 k}{(2\pi)^4} f(x+\frac{1}{2}\Theta \cdot k) g(x+y) e^{i\langle k,y\rangle} 
\end{equation*}
Where $\Theta$ is a skew-symmetric matrix.
A very important question  about non-commutative quantum field theory \cite{ncft3}, is whether or
not the quantum theory is well-defined or in other words if it is renormalizable or not.
At first sight, the non-locality of the non-commutate action induced by the $\star$-product in
the position space can induce us to fear some problems for the renormalization. In fact it was discovered  \cite{ncft3}   that 
after computing the Feynman rules for such theory and deriving the loop amplitude we find that the non-local interaction terms in the action induce an oscillatory factors (involving loop
momenta) in the Feynman
integrals. Studying the structure of Feynman diagrams for the  action,  Filk \cite{one-loop-nc3} has showed  
that are present two types of loop diagram the planar and non-planar diagrams. The planar diagrams do not have this oscillatory factors coming from the non-local interaction terms, 
and therefore the corresponding integrals are the same as in usual quantum field theory.
On the other hand, all non-planar diagrams have the oscillatory factors involving loop momenta.
Due to this terms  the renormalization of quantum field theories on the non-commutative $\mathbb{R}^n$ is not achieved 
and these models show a phenomenon called UV/IR-mixing \cite{UV-mix}. Chepelev and Roiban \cite{UV-mix1} analyses UV/IR-mixing to all orders, 
the conclusion of the power-counting theorem is that field theories on non-commutative $\mathbb{R}^n$
are not renormalizable if the divergence of their commutative counterparts are higher than logarithmic.

A great step towards the non-commutative field theory was made when H.Grosse and R.Wulkenhaar \cite{phi4-non}, found a non-commutative $\varphi^4$-theory renormalizable action which develops additional marginal coupling, corresponding to an harmonic oscillator potential for the real-valued free field $\varphi$ on $\mathbb{R}^4_\Theta$ :
\begin{equation}
S[\varphi]=\int d^4x \left(\frac{1}{2} \varphi \star (-\Delta+\Omega^2 \tilde{x}^2 + \mu^2)\star\varphi + \frac{\lambda}{4}\varphi \star \varphi \star \varphi \star \varphi\right)(x) \nonumber
\end{equation}
Where $x = 2\Theta^{-1} \cdot x, \ \lambda \in \mathbb{R} \ \Omega \in [0,1]$, and $\mu$ is a real parameter.
Using the Moyal matrix base, which turns the  $\star$-product into a standard (infinite) matrix product, H.Grosse and R.Wulkenhaar  were able to prove the perturbative renormalizability of the theory \cite{matrix-renorm}.
Afterward, R.Wulkenhaar et al. \cite{nc-renorm} found an alternative simpler normalization proof using  multi-scale analysis in matrix base, showing the equivalence of various renormalization schemes. A last, but useful, renormalization proof was formulated  using Symanzik type hyperbolic polynomials \cite{nc-renorm2}.

The non-commutative model treated in this work is a sort of extension, via spectral action principle, of the scalar W-G model, in which we are interested to formulate a  Yang-Mills theory in renormalizable way on Moyal space. We can expect that usual Yang-Mills theory on Moyal space without modifications of the action by something similar to an oscillator potential, to be not renormalizable \cite{UV-mix}. 
Additionally, the Moyal space with usual Dirac operator is a spectral triple, the corresponding spectral action was computed in \cite{nc-S}, with the result that it is the usual not renormalizable action on Moyal plane. 
In \cite{8-dim} H.Grosse and R.Wulkenhaar, in order to obtain a gauge theory with an oscillator potential via the spectral action principle, used a Dirac operator constructed  using the statement $ \mathcal{D}^2= H$  where the four dimensional Laplacian is substituted  by the four dimensional oscillator Hamiltonian  $H = -\Delta+\Omega^2 ||x||^2$.  
The idea  behind is that the spectral dimension is defined through the Dirac operator so the spectral dimension  defined
by such Dirac operator is related to the harmonic oscillator phase space dimension. It turns out that to write down an Dirac operator, so that its square equals the 4D harmonic oscillator Hamiltonian, is an easy task using eight dimension Clifford algebra. In addition, can be shown that using this Dirac operator on 4D-Moyal space, is possible define  an eight-dimensional spectral triple. After defined the Dirac operator with the desired spectrum it is considered the total spectral triple as the tensor product
of the "oscillating" spectral triple $(\mathcal{A}_4, \mathcal{H}_4, \mathcal{D}_4 )$ with an almost-commutative triple
and then is perform the previous described procedure of non-commutative geometry to compute
the spectral action. We notice that matrix algebra introduces an extension of the standard  potential in the commutative case, in fact the  scalar field $\phi$ and the $X_\mu$ fields are present together in a potential of the form\footnote{Einstein notation on repeated indices is used.} $(\alpha X_\mu \star X^\mu  + \beta \bar{\varphi} \star \varphi - 1)^2$, with $\alpha, \beta \in \mathbb{R}^+$ and $X_\mu(x) = (\Theta^{-1} )_{\mu}^\nu x_\nu + A_\mu(x)$ is a covariant coordinate.

The high  non-triviality of the vacuum makes very difficult to explicit the vacuum configuration of the system in \cite{Goursac1} A. de Goursac, J.C. Wallet, and R. Wulkenhaar, using the matrix base formalism, have found an expressions from vacuum solutions deriving them from the relevant solutions of the equations of motion.
Although, the  complexity  of the vacuum configuration makes the perturbative approach very complicated, in order to conduct some investigations will be considered a non-perturbative scheme  using a discretized matrix model of the action in which the fields become matrices, the star product become the matrix multiplication and the integral turns in a matrix trace.

Now comes in to play the numerical treatment, the standard method is to approximate the space by discrete points, for example using a lattice approximation  and then calculate the observables over that set of points \cite{lattice}. 
Since an approximation in the position space is not suitable due to the oscillator factor of the Moyal product, instead  the lattice approximation, will be used the matrix Moyal base, which was already used in the first renormalization proof of $\varphi^4$-model restricted to finite matrices.  Hence, will be performed a Monte Carlo simulation studying some statistical quantity such the energy density and specific heat varying the parameters $\Omega, \frac{\chi_{-1}}{\chi_0}, \alpha$ and gathering some informations on the various contributions of the fields to the action. The simulations are quite cumbersome due the  complexity of the action and the number of independent matrices to handle but  we are able to get an acceptable balance between the  computation precision and the computation time. For the simulations is applied a standard Metropolis-Monte Carlo algorithm \cite{Metro} with 
various estimators for the error and for the autocorrelation time of the samples. In general we chose the range of parameters in order to avoid problems with the thermalization process, obtaining numerical simulations where is enough to wait a relative small number  of Monte Carlo steps  to compute  independent results  from the initial conditions. we are interested on the continuous limit that correspond to matrices of infinite size. We will consider various size of the matrices expecting  a stabilization of the values of observables like the energy density,  increasing the matrix size. In order to find same possible phase transitions will be used the specific heat which is a measure of the dispersion of the energy. The  phase transitions are registered as peaks of the specific heat,  increasing the matrices size.

\section{8-dim spectral action }
In this section will be computed a spectral action starting from a \-non\--com\-mu\-ta\-ti\-ve spectral triple. The feature of this particular triple is the choice of a 4-dimension Harmonic  Dirac  operator. The idea behind this construction \cite{8-dim} is to relate  the Dirac operator with the oscillator Hamiltonian operator. Roughly speaking, we  look at the Dirac operator as a generalization of the Laplace operator  so we have $\mathcal{D}^2\approx H$. 
Considering the spectrum $\omega(2n+1)$ $n \in N$ of  the one-dimensional  harmonic oscillator Hamiltonian $H$, can be deduced  that  $H^{-1}$ is a non-commutative infinitesimal of order one. The non-commutative dimension of a spectral triple, equipped with the 4D Harmonic Dirac operator $ \mathcal{D}^2_4= H_4$, is fixed by the non-commutative order of the inverse operator $\mathcal{D}^{-1}$ which is eight not four. This occurrence connects the spectral dimension  to the 
phase space dimension instead the one of the configuration space \cite{con-phase}.
In order to construct such Harmonic Dirac operator and the spectral triple we will work in the framework of the generalized $n$-dimensional harmonic operators. Will be studied the 4-dimensional case in order to construct the non-commutative spectral triple which is starting point for the field theory we are interested in.
Having the 4-dimensional Harmonic Dirac operator with harmonic oscillator spectrum, to implement the Higgs mechanism we will consider 
the tensor product of the non-commutative triple with a finite Connes-Lott type spectral triple \cite{con-lot}.
We will fluctuate the total Dirac operator following the  standard machinery \cite{S-A,Connes3} of non-commutative geometry to get "Gauged" Dirac operator. Thus  we  will proceed to compute the spectral action in which are present two U(1)-Moyal Yang-Mills fields unified with a complex Higgs field.

\section{Harmonic Dirac operators }
\textbf{1}

The Harmonic Dirac operator  in $n$-dimensions  can be defined using the Clifford algebra of $\mathbb{R}^{2n}$ represented on the  Hilbert space $ \mathbb{C}^{2^n}$,  it is very useful to consider  $n$-dimensional fermionic annihilation and creation operators $b_\mu$, $b_\nu^\dagger$ and  $n$-dimensional bosonic annihilation and creation operators $a_\mu$, $a^\dagger_\nu$ satisfying for $\mu,\nu=1,\cdots,n$:
\begin{eqnarray}
&[a_\mu,a_\nu]=[a^\dagger_\mu,a^\dagger_\nu]=0, & [a_\mu,a^\dagger_\nu]=\delta_{\mu\nu} \label{com1} \\ 
&\{b_\mu,b_\nu\}=\{b_\mu^\dagger ,b_\nu^\dagger\}=0, & \{b_\mu,b_\nu^\dagger\}=\delta_{\mu\nu} \label{anti1}
\end{eqnarray}
Where $a_\mu =\frac{1}{\sqrt{2\omega}}  (\omega x_\mu + \partial_\mu ), \ a_\mu^\dagger = \frac{1}{\sqrt{2\omega}}(\omega x_\mu-\partial_\mu ) $.
Using this operators is possible to construct a Dirac operator as:
\begin{equation}
\mathcal{D}_n=-i\sqrt{2\omega}\delta^{\mu\nu}a_\mu^\dagger\otimes b_\nu + i\sqrt{2\omega}\delta^{\mu\nu}a_\mu\otimes b_\nu^\dagger=i\frac{d}{dx_\mu}\otimes(b_\mu+b_\mu^\dagger)+ i\omega x^\mu\otimes(b_\mu-b^\dagger_\mu) \label{dirac2}
\end{equation}
summed over repeated index. We can define the fermionic part of the Hilbert space on which the Dirac operator \eqref{dirac2} acts starting from the vacuum state by subsequent applications of the fermionic creation operators  $b_\nu^\dagger$  on the vacuum $b|0\rangle=0$, using the anti-commutation relations \eqref{anti1} defining $\Lambda(\mathbb{C}^n)$. The complete Hilbert  space is $\mathcal{H}_n=S(\mathbb{R}^n)\otimes\Lambda(\mathbb{C}^n)$. Beside, we can define a grading operator $\chi_n $ as:
 \begin{eqnarray}
\chi_n= \textbf{I}\otimes \prod^n_{\mu=1}(b_\mu b^\dagger_\mu-b_\mu b^\dagger_\mu )
\end{eqnarray}
Using the relations \eqref{com1}-\eqref{anti1} we can compute the square the Dirac operator \eqref{dirac2} as:
\begin{equation}
 \mathcal{D}_n^2=2\omega a^\dagger_\mu a^\mu\otimes \textbf{I} -2\omega \textbf{I} \otimes b^\dagger_\mu b^\mu =2\omega N_B\otimes \textbf{I} -2\omega \textbf{I}\otimes N_F
\end{equation}
Where $N_F$ and $N_B$ are the number operators. In this form it easy to see that $\mathcal{D}^2_n $, being a "difference" between fermionic and bosonic number operator, has only one zero mode corresponding to the vacuum state. For practical reasons it is convenient write $\mathcal{D}^2_n $ as:
\begin{equation}
 \mathcal{D}_n^2=\omega \delta^{\mu\nu}(a^\dagger_\mu a_\nu + a_\nu a^\dagger_\mu)\otimes \textbf{I} -2\omega \textbf{I} \otimes \delta^{\mu\nu}(b^\dagger_\mu b_\nu - b_\nu b^\dagger_\mu)=H_n\otimes \textbf{I} +\omega\otimes \Sigma_n 
\end{equation}
where in $H_n$ we can recognize the harmonic oscillator Hamiltonian and the spin operator $\Sigma_n$.
The universality property of the Clifford algebra grants the existence of an isomorphism between the 2$n$-dimensional Clifford algebra
and the Hilbert space $ \mathcal{H}_n=S(\mathbb{R}^n)\otimes \mathbb{C}^{2^n}$.  In this representation the Dirac operator is: 
\begin{equation}
\mathcal{D}_n=i\Gamma^\mu\partial_\mu +\omega\Gamma^{\mu+n}x_\mu  \label{Diracn}
\end{equation}
Where $\Gamma^\mu$ turns to be $\Gamma^\mu=(b_\mu+b^\dagger_\mu)$, $ \Gamma^{\mu+n}=i(b^\dagger_\mu-b_\mu)$  which satisfy the relations:
\begin{equation}
\Gamma_a\Gamma_b+\Gamma_b\Gamma_a=2\delta_{ab} \ \textrm{with} \ a,b=1,\cdots,2n \label{Gamman}
\end{equation}
Beside, the grading operator is represented as:
\begin{equation}
\chi_n=(-i)^n(-1)^{\frac{n(n-1)}{2}}\otimes \Gamma_1\cdots\Gamma_{2n}
\end{equation}

\section{An harmonic spectral triple for the Moyal plane}
In the framework of non-commutative field theories on 4-dimensional Moyal plane has been proved \cite{phi4-non,matrix-renorm} that the introduction of an harmonic oscillator term makes a $\phi^4$-model on 4-dimensional Moyal plane renormalizable. Such oscillator term can be written as:  
\begin{equation}
H_m=-\frac{\partial^2}{\partial x_\mu \partial x^\mu}+\Omega^2 \tilde{x}_\mu\tilde{x}^\mu +m^2
\end{equation}
where $\tilde{x}_\mu:=2(\Theta^{-1})_{\mu\nu} x^\nu$, $\Theta$ can be chosen as two copies of the  Pauli matrix $\Theta = i\theta\sigma_2  \otimes \textbf{I}_2$ or explicitly:
\begin{equation}
\Theta=\left(
\begin{array}{cccc}
0& \theta &0&0 \\
-\theta &0&0&0 \\
0&0&0&\theta \\
0&0&-\theta&0
\end{array}
\right), \ \theta\in\mathbb{R}
\end{equation}
With this choice we have $\Theta^{-1} =  -\frac{i\sigma_2  }{\theta}\otimes \textbf{I}_2 $.
Quantum mechanics tell us that in the Hilbert space $L^2(\mathbb{R}^4)$ exists
an orthonormal basis $\psi_s ,\ s\in\mathbb{R}^4$ of eigenfunctions of $H_m$ with eigenvalues
\begin{equation}
\lambda_s(m)=\left(\frac{4\Omega}{\theta} \left(s+2+ \frac{\theta m^2}{2\Omega}\right)\right), \ s\in\mathbb{N}
\end{equation}
The inverse $H^{-1}_m$ extends to a selfadjoint compact operator on $L^2(\mathbb{R}^4)$ with eigenvalues $\lambda^{-1}_s(m)$. 
If we look  at the trace the operator $H^{-4}_m$ we find:  
\begin{equation}
\operatorname{Tr}(H^{-s}_m)=\sum_{n=0}^\infty(n + 3)(n + 2)(n + 1)(\lambda_n (m))^s
\end{equation}
Which is derived simply from the number of possibilities to express $s$ as a sum of four ordered natural numbers. 
This means that   $H^{-4}$  belongs to the Dixmier trace ideal 
$L^{(1,\infty)}(L^2(\mathbb{R}^4))$ of compact operators and the relation $\mathcal{D}^2=H$  implies  that the 4-dimensional Moyal space has spectral dimension 8.
From the previous section, we can define a proper Dirac operator just considering  the 4-dimensional case obtaining a Dirac operator built from a 8-dimensional Cifford algebra:
\begin{equation}
\mathcal{D}_4 = i\Gamma_\mu \partial_\mu + \Omega \Gamma_{\mu+4} \tilde{x}_\mu
\end{equation}
Here, the $\Gamma_k \in M_{16}(\mathbb{C}), k = 1, . . . , 8$ are the generators of the 8-dimensional real Clifford
algebra, satisfying
\begin{equation}
\Gamma_k \Gamma_l + \Gamma_l\Gamma_k = 2\delta_{kl} \textbf{I} 
\end{equation}
We take the Hilbert space $\mathcal{H}_4 = S^2(\mathbb{R}^4,\mathcal{S})=L^2(\mathbb{R}^4)\otimes\mathbb{C}^{16} $ of Schwartz functions of spinors over
4-dimensional euclidean space. 
Accordingly with \eqref{Gamman} for $\psi \in \mathcal{H}_4$ we obtain:
\begin{equation}
\mathcal{D}^2_4 \psi= \left((-\Delta + \Omega^2 \tilde{x}_\mu \tilde{x}^\mu )\textbf{I} + \Sigma\right)\psi \ , \ \Sigma:= -i\Omega(\Theta^{-1})^{\mu\nu} [\Gamma_\mu ,\Gamma_{\nu+4}] \label{Sigma1}
\end{equation}
with $\Delta = \partial^\mu \partial_\mu$.
As algebra we chose the Moyal algebra $\mathbb{R}^4_\Theta$:
\begin{equation}
\mathcal{A}_4 = \mathbb{R}_\Theta^4=(\mathcal{S}(\mathbb{R}^4),\star)
\end{equation}
where $(\mathcal{S}(\mathbb{R}^4),\star)$ is the algebra of the Schwartz functions on $\mathbb{R}^4$,  with the Moyal product
\begin{equation}
(f \star g)(x) = \int \int d^4 y \frac{d^4 k}{(2\pi)^4} f(x+\frac{1}{2}\Theta \cdot k) g(x+y) e^{i\langle k,y\rangle} \ ,\ f,g \in \mathcal{A}_4 \label{star}
\end{equation}
The representation of the algebra $\mathcal{A}_4$ on $\mathcal{H}_4$ is by component-wise diagonal Moyal product \cite{moyal-triple} $\star : \mathcal{A}_4 \times \mathcal{H}_4 \to \mathcal{H}_4$. The Moyal product can be  extended to constant functions using   another representation of the product with the integral representation of the Dirac distribution.
Taking in account, for smooth spinors, the identity $2x^\mu\psi=x\star\psi+\psi\star x$ and the relation 
\begin{equation}
[x^\nu,f]_\star = i\Theta^{\nu\rho} \partial_\rho \label{comxf}
\end{equation}
we compute the commutator of that action with the Dirac operator
\begin{equation}
\begin{array}{l}
\mathcal{D}_4(f\star\psi)-f\star(\mathcal{D}_4 \psi) \\
= i\Gamma^\mu ((\partial_\mu f) \star\psi + f\star \partial_\mu \psi) + \frac{1}{2} \Omega\Gamma^{\mu+4} (\tilde{x}_\mu\star(f \star \psi) + (f \star \psi) \star \tilde{x}_\mu) \\
-i\Gamma^\mu f \star \partial_\mu \psi - \frac{1}{2} \Omega\Gamma^{\mu+4}(f\star( \tilde{x}_\mu\star\psi)+f\star(\psi \star \tilde{x}_\mu))\\
= i(\Gamma^\mu + \Omega\Gamma^{\mu+4} )(\partial_\mu f ) \star \psi .
\end{array} \label{d4-com}
\end{equation}
The previous commutator confirms that
$(\mathcal{A}_4,\mathcal{H}_4,\mathcal{D}_4)$ satisfy the main\footnote{Orientability axiom and Poincar\'e duality will be not considered} axioms of spectral triple, in fact the commutator is bounded and due to its commutation with
Moyal multiplication, order-one condition is fulfilled.
Now we introduce a very useful relation connected to the heat kernel  type expansion associated to a regular spectral 
triple taken from \cite{non-com-tri}. This relation will be used later in order to  compute the spectral action. Considering a regular non-unital spectral triple $(\mathcal{A},\mathcal{H},\mathcal{D},J)$
and two pseudo-differential operator  $A_0 \in \Psi_0(\mathcal{A}) $ $A_1\in \Psi_1(\mathcal{A})$ of order respectively 0 and 1.
We consider the following decomposition: 
\begin{equation}
e^{-t(\mathcal{D}^2+A_0+A_1)}=\sum_{j=0}^4(-1)^jE_j(t)-t^5R
\end{equation}
Using Duhamel principle \cite{Duhamel}
\begin{equation}
e^{-t(A+B)}= e^{-tA}-t\int_0^1 ds \ e^{-st(A+B)}Be^{-(1-s)tA}
\end{equation}
we can identify:
\begin{eqnarray}
E_0(t)&=&e^{-t\mathcal{D}^2}\nonumber \\ \ 
E_j(t)&=&\int_{\Delta_j} d^js \ e^{-s_1t\mathcal{D}^2}(A_0+A_1)e^{-(s_2-s_1)t\mathcal{D}^2}\cdots(A_0+A_1)e^{-(1-s_j)t\mathcal{D}^2} \nonumber \\ \ 
&&
\end{eqnarray}
and
\begin{eqnarray}
&&R=\int_{\Delta_5}ds_1ds_2ds_3ds_4ds_5 e^{-s_1t(\mathcal{D}^2+A_0+A_1)}(A_0+A_1)e^{-(s_2-s_1)t\mathcal{D}^2}(A_0+A_1)    \nonumber\\
&& \times e^{-(s_3-s_2)t\mathcal{D}^2}(A_0+A_1)e^{-(s_4-s_3)t\mathcal{D}^2}(A_0+A_1)e^{-(s_5-s_4)t\mathcal{D}^2}(A_0+A_1)e^{-(1-s_5)t\mathcal{D}^2}\nonumber\\
&&
\end{eqnarray}
The domains of the integrals $\Delta_j$ are the $j$-simplex:
\begin{equation}
\Delta_j=\{s\in\mathbb{R}^j;0\leq s_1\leq s_2\leq s_1\leq\cdots\leq s_j\leq 1\} \simeq \{s\in\mathbb{R}^{j+1};s_i\geq 0,\sum_{i=0}^j s_i=1\}
\end{equation}
Taking in account the relation:
\begin{equation}
\left[ e^{-t\mathcal{D}^2},A \right]=\int_0^1 ds \frac{d}{ds}\left(e^{-ts\mathcal{D}^2}Ae^{-t(1-s)\mathcal{D}^2}\right) =-t\int_0^1 ds e^{-ts\mathcal{D}^2}\left[\mathcal{D}^2,A\right]e^{-t(1-s)\mathcal{D}^2}
\end{equation}

and considering the trace in \cite{non-com-tri} is computed the leading term of $e^{-t(\mathcal{D}^2+A_0+A_1)}$ for $t\to 0$:
\begin{eqnarray}
&&\operatorname{Tr}(e^{-t(\mathcal{D}^2+A_0+A_1)})  \nonumber \\
&&=\textrm{Tr}\Big(\Big\{1-t(A_0+A_1)+\frac{t^2}{2}(A^2_0+A_1A_0+A_0A_1+A^2_1) \nonumber \\
&&-\frac{t^3}{6}(A_0\left[\mathcal{D}^2,A_1\right]+A_1\left[\mathcal{D}^2,A_0\right]+A_1\left[\mathcal{D}^2,A_1\right]+A_0A^2_1+A_1A_0A_1+A_1^2A_0\nonumber \\
&&+A^3_1)+\frac{t^4}{24}(A_1\left[\mathcal{D}^2 \left[\mathcal{D}^2,A_0\right]\right]+2A^2_1\left[\mathcal{D}^2 ,A_1\right]+A_1\left[\mathcal{D}^2 ,A_1\right]A_1+A_1^4\Big)e^{-t\mathcal{D}^2}\Big\}\nonumber \\
&&+\mathcal{O}(\sqrt{t}) \label{heat1}
\end{eqnarray}

\section{4-dimensional harmonic Yang-Mills model}
Following the Connes-Lott models, in order to implement the Higgs mechanism,  we consider the total spectral triple 
as the tensor product of the 8-dimensional spectral triple $(\mathcal{A}_4,\mathcal{H}_4,\mathcal{D}_4,\Gamma_9)$ with the two point Connes-Lott like spectral triple $(\mathbb{C}\otimes \mathbb{C},\mathbb{C}^2, M\sigma_1 )$. The total Dirac operator of the product triple is:
\begin{equation}
\mathcal{D}_T = \mathcal{D}_4 \otimes \textbf{I}  + \Gamma_9 \otimes M\sigma_1
\end{equation}
Or explicitly:
\begin{equation}
\mathcal{D}_T= \left(\begin{array}{cc}
    \mathcal{D}_4& M\Gamma_9 \\
    M\Gamma_9& \mathcal{D}_4
\end{array}\right)
\end{equation}
The algebra becomes $ \mathcal{A}_T=\mathcal{A}_4 \oplus \mathcal{A}_4 $ and acts by diagonal star multiplication \eqref{star} on $\mathcal{H}_T= \mathcal{H}_4 \oplus \mathcal{H}_4$. The fluctuated Dirac operator is found  using 
$\mathcal{D}_A = \mathcal{D}_T + \Sigma_i a_i[\mathcal{D}_T,b_i]$ with $a_i,b_i\in\mathcal{A}_T $ of the form $(f,g)$, the computation of the commutator $\mathcal{D}_T$ with $(f,g)$ gives:
\begin{equation}
[\mathcal{D}_T,(f, g)] = 
\left(\begin{array}{cc}
    i(\Gamma^\mu + \Omega \Gamma^{\mu+4} )L_\star(\partial_\mu f) & M\Gamma_9 L_\star(f-g) \\
    M\Gamma_9 L_\star(g-f) & i(\Gamma^\mu + \Omega\Gamma^{\mu+4} )L_\star(\partial_\mu g)
\end{array}\right)
\end{equation}
$L_\star(f)\psi = f \star \psi$ is the left Moyal multiplication. From the commutator we deduce that the form of selfadjoint fluctuated Dirac has to be: 
\begin{equation}
\mathcal{D}_A =\left(\begin{array}{cc} 
\mathcal{D}_4 + (\Gamma_\mu + \Omega \Gamma_{\mu+4})L_\star(A^\mu )& \Gamma_9 L_\star(\phi) \\
\Gamma_9 L_\star (\bar{\phi}) &\mathcal{D}_4 + (\Gamma_\mu + \Omega \Gamma_{\mu+4} )L_\star (B^\mu ) \end{array}\right)
\end{equation}
Where $\phi \in \mathcal{A}_4$ is the Higgs complex field and $A_\mu, B_\mu \in \mathcal{A}_4$ are real fields. The spectral action computation needs the square of $\mathcal{D}_A$:
\begin{equation}
\mathcal{D}^2_A =\left(\begin{array}{cc}
(H^2_0 + L_\star (\phi \star \varphi))1 + \Sigma + F_A
&i(\Gamma_\mu + \Omega\Gamma_{\mu+4})\Gamma_9 L_\star (D^\mu \phi)\\
i(\Gamma_\mu + \Omega \Gamma_{\mu+4} )\Gamma_9 L_\star (\overline{D^\mu \phi})&
(H^2 + L_\star (\phi \star \phi))1 + \Sigma + F_B
\end{array}\right)
\end{equation}
with
\begin{eqnarray}
D_\mu \phi &=& \partial_\mu\phi - iA_\mu \star \phi + i\phi\star B_\mu\\ 
F_A &=& \{\mathcal{D}_4 , (\Gamma_\mu + \Omega \Gamma_{\mu+4} )L_\star (A^\mu )\} + (\Gamma_\mu + \Omega\Gamma_{\mu+4} )(\Gamma_\nu + \Omega \Gamma_{\nu+4} )L_\star (A^\mu \star A^\nu)\nonumber \\
&=& \{ L_\star(A^\mu ), i\partial_\mu + \Omega^2 M_\bullet(x_\mu)\} + (1 +\Omega^2)L_\star (A_\mu \star A^\mu )  \nonumber \\
&+& i \left(\frac{1}{4} [\Gamma_\mu, \Gamma_\nu]+ \frac{1}{4} \Omega^2 [\Gamma_{\mu+4} , \Gamma_{\nu+4} ] + \Omega\Gamma_\mu \Gamma_{\nu+4} \right) L_\star (F^{\mu\nu}_A ) ,
\end{eqnarray}
$(M_\bullet(\tilde{x}_\mu)\psi)(x) = \tilde{x}_\mu\psi(x)$ is ordinary pointwise multiplication and $F_B$  is obtained just replacing $A$ with $B$. We can recognize in  previous expression the field strength $F^A_{\mu\nu} = \partial_\mu A_\nu - \partial_\nu A_\mu -i(A_\mu\star A_\nu - A_\nu\star A_\mu )$

\subsection{Spectral action}
Recalling  the spectral action principle, the bosonic action can be defined exclusively by the
spectrum of the Dirac operator. The general form for such bosonic action is:
\begin{equation}
 S(\mathcal{D}_A) = \textrm{Tr}\chi(\mathcal{D}^2_A) \label{Action}
\end{equation}
Where $\chi$ is a regularization function $\chi: R_+\to R_+$ for which trace exists. \\
The trace in \eqref{Action}  is  defined on $\mathcal{B}(L^2(\mathbb{R}^4))$ by 
\begin{equation}
 \operatorname{Tr}(A) =\int_{\mathbb{R}^4} dx \ A(x,x)
\end{equation}
together with the matrix trace including  the Clifford algebra.
By Laplace transformation one has
\begin{equation}
S(\mathcal{D}_A) =\int_0^\infty dt \operatorname{Tr}(e^{-t\mathcal{D}^2_A})\tilde{\chi}(t)  \label{Action1}
\end{equation}
where $\tilde{\chi}$ is the inverse Laplace transform of $\chi(s)$, 
\begin{equation}
\chi(s) = \int_0^\infty dt e^{-st}\tilde{\chi}(t). 
\end{equation}
The trace in \eqref{Action1} is given by:
\begin{equation}
 \operatorname{Tr}(e^{-t\mathcal{D}^2_A}) =\int_{\mathbb{R}^4} dx \ \textrm{tr}(e^{-t\mathcal{D}^2_A})(x,x)
\end{equation}
Assuming the trace of the heat kernel $e^{-t\mathcal{D}^2_A}$ has an asymptotic expansion
\begin{equation}
 \operatorname{Tr}(e^{-t\mathcal{D}^2_A})= \sum^\infty_{n=-\delta} a_n(\mathcal{D}^2_A)t^{n} \ , \ \delta \in \mathbb{N}  \label{heat}
\end{equation}
we obtain replacing the previous expansion into \eqref{Action1}
\begin{equation}
S(\mathcal{D}_A) =\sum_{n=-\delta}^\infty  a_n(\mathcal{D}^2_A)\int_0^\infty dt \ t^{n} \tilde{\chi}(t) \label{heatd}
\end{equation}
To compute the integrals we have to consider separately the cases $ n\notin\mathbb{N}$ and $ n\in\mathbb{N}$:
\begin{equation}
\chi_n =\Bigg\{
\begin{array}{ll}
\frac{1}{\Gamma(-n)}\int_0^\infty ds \ s^{-n-1} \chi(s) & \textrm{for} \ n\notin\mathbb{N}  \\
(-1)^{n-\delta} \chi^{(n)}(0) & \textrm{for} \ n\in\mathbb{N}  
\end{array}\label{chi}
\end{equation}
Due to the nature of the $\chi(t)$ function (usually one chose a characteristic function), we can assume  $\chi(0)$ much bigger then the derivatives $\chi^{(m)}(0)$ for any $m>0$ appearing in \eqref{chi}. Consequently in  the expansion \eqref{heat}  we will  take in account only the finite or singular part for $t\to 0$   

Our strategy to compute the action is to use the relation \eqref{heat1}, therefore after explicitly	expressed  $A_0$ and $A_1$ we proceed to the calculus of the traces and in the end we will identify the leading part of the action comparing the result with the expansions \eqref{heat}-\eqref{heatd}. We can identify the operators $A_0$ and $A_1$ appearing in the \eqref{heat1} as follow:
\begin{equation}
A_0 =\left(\begin{array}{cc}
L_\star (V_{A,\phi})\textbf{I} + L_\star(F_A^{\mu\nu})\Gamma^\Omega_{\mu\nu}
&i(\Gamma^\mu + \Omega\Gamma^{\mu+4})\Gamma_9 L_\star (D_\mu \phi)\\
i(\Gamma^\mu + \Omega \Gamma^{\mu+4} )\Gamma_9 L_\star (\overline{D_\mu \phi})&
L_\star (V_{B,\phi})\textbf{I} + L_\star(F_B^{\mu\nu})\Gamma^\Omega_{\mu\nu}
\end{array}\right) \label{A0}
\end{equation}
\begin{equation}
A_1 =\left(\begin{array}{cc}
2i(1+\Omega^2)L_\star(A^\mu)\nabla_\mu^{(\Omega)} &0\\
0& 2i(1+\Omega^2)L_\star(B^\mu)\nabla_\mu^{(\Omega)}
\end{array}\right)\label{A1}
\end{equation}
with
\begin{equation}
 V_{A,\phi}= \phi\star\bar{\phi}+(1+\Omega^2)(i \partial_\mu A^\mu +A_\mu\star A^\mu) , \  V_{B,\phi}= \bar{\phi}\star\phi+(1+\Omega^2)(i\partial_\mu B^\mu+B_\mu\star A^\mu)
\end{equation}
$\nabla_\mu^{(\Omega)}$  are define as $\nabla_\mu^{(\Omega)}=\frac{1}{1+\Omega^2}\left(\partial_\mu-i\Omega^2 M_\bullet(\tilde{x}_\mu)\right) $.
We are allowed to split the traces in two parts a matrix trace and the continuous one. 
After the matrices trace computations we obtain \cite{non-com-tri} for the $A$ field:
\begin{eqnarray}
&&\textrm{Tr}(e^{-t\mathcal{D}_A^2})\nonumber \\
&&=\Bigg\{16\cosh^4(t\Omega)\textrm{tr}(e^{-tH_4^2})-tT(16V_{A,\phi})-t\mathcal{T}_\mu(32i(1+\Omega^2)A^\mu)\nonumber \\
&&+\frac{t^2}{2}\mathcal{T}\Big(16V_{A,\phi}\star V_{A,\phi}+16(1+\Omega^2)D_\mu\phi\star\overline{D^\mu\phi}+8(1+\Omega^2)F_{\mu\nu}^A F^{\mu\nu}_A\nonumber \\
&&+ 32i(1+\Omega^2)A^\mu\star\partial_\mu V_{A,\phi}\Big)\nonumber \\
&&+ \frac{t^2}{2}\mathcal{T}_\mu\Big(32i(1+\Omega^2)A^\mu\star V_{A,\phi}+32i(1+\Omega^2)V_{A,\phi}\star A^\mu\nonumber \\
&&-64(1+\Omega^2)^2A^\nu\star\partial_\nu A^\mu\Big)+\frac{t^2}{2}\mathcal{T}_{\mu\nu}\left(-64(1+\Omega^2)^2 A^\mu\star A^\nu \right)\nonumber \\
&&- \frac{t^3}{6}\mathcal{T}_{\mu\nu}\Big(-64(1+\Omega^2)^2V_{A,\phi}\star \partial^\mu A^\nu -64i(1+\Omega^2)^2A^\nu\star\partial^\nu V_{A,\phi}\nonumber \\
&&+64(1+\Omega^2)^3A_\rho\star(\delta^{\rho\mu}\triangle A^\nu +2\partial^\rho\partial^\mu A^\nu)\nonumber \\
&&- 64(1+\Omega^2)^2(V_{A,\phi}\star A^\mu\star A^\nu + A^\mu\star V_{A,\phi}\star A^\nu +A^\mu\star A^\nu\star V_{A,\phi})\nonumber \\
&&-128i(1+\Omega^2)^3\left(A^\rho\star(\partial_\rho A^\mu)\star A^\nu +A^\rho\star A^\mu(\partial_\rho A^\nu\right)\Big)\nonumber \\
&&- \frac{t^3}{6}\mathcal{T}_{\mu\nu\rho}\left(128(1+\Omega^2)^3A^\mu\star\partial^\nu A^\rho -128i(1+\Omega^2)^3A^\nu\star A^\mu\star A^\rho\right)\nonumber \\
&&+ \frac{t^3}{6}\tilde{\mathcal{T}}_{\mu\nu}\left(512i\Omega^2(1+\Omega^2)(\Theta^{-1})^{\rho\nu}A^\mu\star A_\rho\right)\nonumber \\
&&+ \frac{t^4}{24}\mathcal{T}_{\mu\nu\rho\sigma}\Big(-256(1+\Omega^2)^4A^\mu\star\partial^\nu\partial^\rho A^\sigma +512i(1+\Omega^2)^3 A^\mu\star A^\nu\star \partial^\rho A^\sigma\nonumber \\
&&+256i(1+\Omega^2)^4 A^\mu\star(\partial^\nu A^\rho)\star A^\sigma +256(1+\Omega^2)^4 A^\mu\star A^\nu\star A^\rho\star A^\sigma\Big)\Bigg\}\nonumber \\
&& + \mbox{$B$ field contribution} \ +\mathcal{O}(\sqrt{t})
\end{eqnarray} 
Where in order to simplify the notation we introduce the functions:
\begin{eqnarray}
 \mathcal{T}(f)&=&\textrm{Tr}_{L^2(\mathbb{R}^4)}\left(L_\star(f)e^{-tH_4}\right) \nonumber \\
 \mathcal{T}_{\mu_1\cdots\mu_k}(f)&=&\textrm{Tr}_{L^2(\mathbb{R}^4)}\left(L_\star(f)\nabla^{(\Omega)}_{\mu_1}\cdots\nabla^{(\Omega)}_{\mu_k}e^{-tH_4}\right)  \nonumber \\
 \tilde{\mathcal{T}}_{\mu\nu}(f)&=&\textrm{Tr}_{L^2(\mathbb{R}^4)}\left(L_\star(f)\nabla^{(\Omega)}_{\mu_1}\nabla^{(1)}_{\mu_k}e^{-tH_4}\right) 
\end{eqnarray}

The contributions for the $B$ fields are obtained operating the following substitutions:
\begin{equation}
\left\{ A_\mu \to B_\mu, \ F^A_{\mu\nu}\to F^B_{\mu\nu}, \ V_{A,\phi} \to V_{B,\phi}, \ D_\mu\phi \leftrightarrow \overline{D^\mu\phi}\right\}
\end{equation}
After the computation of the $\mathcal{T},\mathcal{T}_{\mu_1\cdots\mu_k}$ \footnote{$\tilde{\mathcal{T}}_{\mu\nu}$ can be neglected}, we have all the ingredients required to compute the leading part of the action \eqref{Action} replacing all the traces into the \eqref{heat1}.  
Using the trace property of the star product and the identities 
\begin{eqnarray}
\tilde{z}_\mu\star f=\frac{1}{2}\{\tilde{z}_\mu,f\} +\frac{1}{2}[\tilde{z}_\mu,f]=\frac{1}{2}\{\tilde{z}_\mu,f\}+i\partial_\mu f \\
\{\tilde{z}_\mu , f\star g \}_\star=\{\tilde{z}_\mu,f \}\star g -if\star \partial_\mu g 
\end{eqnarray}
we get after some manipulations:   
\begin{eqnarray}
&&\textrm{Tr}(e^{-t\mathcal{D}^2})=2 \cosh^4(\tilde{\Omega}t) \nonumber \\
&&+\frac{1}{\pi^2(1+\Omega^2)^2} \int d^4z \Bigg\{ -\frac{1}{t}\Bigg(\left(\phi\star\bar{\phi}+ \frac{4\Omega^2}{1+\Omega^2}(\tilde{X}^A_\mu\star\tilde{X}_A^\mu -\tilde{X}^0_{\mu}\star\tilde{X}_0^\mu)\right)\nonumber \\
&&+\left(\bar{\phi}\star\phi+\frac{4\Omega^2}{1+\Omega^2}(\tilde{X}^B_\mu\star\tilde{X}_B^\mu -\tilde{X}^0_{\mu}\star\tilde{X}_0^\mu)\right)\Bigg)\nonumber \\
&&+\frac{1}{2}\left( \phi\star\bar{\phi}+\frac{4\Omega^2}{1+\Omega^2}\tilde{X}^A_\mu\star\tilde{X}_A^\mu\right)^2-\frac{1}{2}\left(\frac{4\Omega^2}{1+\Omega^2}\tilde{X}^\mu_0\star\tilde{X}^0_\mu\right)^2\nonumber \\
&&+\frac{1}{2}\left( \bar{\phi}\star\phi+\frac{4\Omega^2}{1+\Omega^2}\tilde{X}^B_\mu\star\tilde{X}_B^\mu\right)^2-\frac{1}{2}\left(\frac{4\Omega^2}{1+\Omega^2}\tilde{X}^\mu_0\star\tilde{X}^0_\mu\right)^2\nonumber \\
&&+\frac{1}{2}\left(\frac{(1-\Omega^2)^2}{2}-\frac{(1-\Omega^2)^4}{6(1+\Omega^2)^2}\right)(F^A_{\mu\nu}\star F_A^{\mu\nu}+F^B_{\mu\nu}\star F_B^{\mu\nu})\nonumber \\
&&+\frac{1+\Omega^2}{2}\left(D_\mu\phi\star\overline{D^\mu\phi}+\overline{D_\mu\phi}\star D^\mu\phi\right)\Bigg\}(z)+\mathcal{O}(\sqrt{t})
\end{eqnarray}
where 
\begin{equation}
 \tilde{X}^A_\mu=\frac{\tilde{z}_\mu}{2}+A_\mu, \ \tilde{X}^B_\mu=\frac{\tilde{z}_\mu}{2}+B_\mu, \ \tilde{X}^0_\mu=\frac{\tilde{z}_\mu}{2}
\end{equation}
Using the Laurent expansion of $ \coth^4(t^\prime)=t^{\prime-4}+ \frac{4}{3}t^{\prime-2}+\frac{26}{45}+\mathcal{O}(t^{\prime 2})$
and comparing the previous expression to the expansion \eqref{heatd} and putting  $\chi_0=\chi(0)$  we  are finally able to write the spectral action \eqref{Action} as:
\begin{eqnarray}
&&S(\mathcal{D}_A)= \frac{\theta^4\chi_{-4}}{8\Omega^4}+\frac{2\theta^2\chi_{-2}}{3\Omega^2}+\frac{52\chi_{0}}{45} \nonumber \\
&&+\frac{\chi_0}{\pi^2(1+\Omega^2)^2}+\frac{1}{\pi^2(1+\Omega^2)^2} \int d^4z \Bigg\{\left(\frac{(1-\Omega^2)^2}{2}-\frac{(1-\Omega^2)^4}{6(1+\Omega^2)^2}\right)(F^A_{\mu\nu}\star F_A^{\mu\nu}+F^B_{\mu\nu}\star F_B^{\mu\nu})\nonumber \\
&&+\left(\bar{\phi}\star\phi+ \frac{4\Omega^2}{1+\Omega^2}\tilde{X}^A_\mu\star\tilde{X}_A^\mu-\frac{\chi_{-1}}{\chi_0}\right)^2\nonumber \\
&&+\left(\phi\star\bar{\phi}+\frac{4\Omega^2}{1+\Omega^2}\tilde{X}^B_\mu\star\tilde{X}_B^\mu-\frac{\chi_{-1}}{\chi_0}\right)^2\nonumber \\
&&-2\left(\frac{4\Omega^2}{1+\Omega^2}\tilde{X}^\mu_0\star\tilde{X}^0_\mu-\frac{\chi_{-1}}{\chi_0} \right)^2+ 2(1+\Omega^2)D_\mu\phi\star\overline{D^\mu\phi}\Bigg\}(z)+\mathcal{O}(\chi_1)\label{faction}
\end{eqnarray}
We notice that Higgs mechanism introduces an extension of the standard Higgs potential in the commutative case, in fact the Higgs scalar field $\phi$ and the $\tilde{X}_A^\mu$, $\tilde{X}_B^\mu$ fields are present together in the potential. In this way the  gauge field  takes part in the definition of the vacuum. Another important property of the action,  considering the $\tilde{X}_A^\mu$, $\tilde{X}_B^\mu$ as independent, is the invariance under the translations:
\begin{equation}
\phi(x)\to\phi(x+a),\ X^\mu_A(x)\to X_A^\mu(x+a), \  X^\mu_B(x)\to X^\mu_B(x+a) , \ X^\mu_0(x)\to X^\mu_0(x+a)
\end{equation} 
which in other $\phi^4$-renormalizable theory is broken.
Beside, the action is invariant under $U(1) \times U(1)$ transformations:
\begin{equation}
\phi \to u_A\star\phi\star\overline{u_B}, \ \tilde{X}\to u_A\star \tilde{X}^\mu_A\star \overline{u_A}, \ \tilde{X}^\mu_B\to u_B\star\tilde{X}^\mu_B\star \overline{u_B} \label{gauge}
\end{equation}
In field theory the ground state can be defined through the minimum of the action, the relevant part of the \eqref{faction} for the minimization is:
\begin{eqnarray}
&&S(\mathcal{D}_A)=  \nonumber \\
&&+\frac{1}{(1+\Omega^2)^2} \int d^4z \Bigg\{\left(\frac{(1-\Omega^2)^2}{2}-\frac{(1-\Omega^2)^4}{6(1+\Omega^2)^2}\right)(F^A_{\mu\nu}\star F_A^{\mu\nu}+F^B_{\mu\nu}\star F_B^{\mu\nu})\nonumber \\
&&+\left(\bar{\phi}\star\phi+ \frac{4\Omega^2}{1+\Omega^2}\tilde{X}^A_\mu\star\tilde{X}_A^\mu-\frac{\chi_{-1}}{\chi_0}\right)^2\nonumber \\
&&+\left(\phi\star\bar{\phi}+ \frac{4\Omega^2}{1+\Omega^2}\tilde{X}^B_\mu\star\tilde{X}_B^\mu-\frac{\chi_{-1}}{\chi_0}\right)^2\nonumber \\
&& -2\left(\frac{4\Omega^2}{1+\Omega^2}\tilde{X}^\mu_0\star\tilde{X}^0_\mu-\frac{\chi_{-1}}{\chi_0} \right)^2+2(1+\Omega^2)D_\mu\phi\star\overline{D^\mu\phi}\Bigg\}(z)+\mathcal{O}(\chi_1) \label{faction1}
\end{eqnarray}
Where we have omitted the constant part and we have rescaled the coefficient in front of the integral.
Considering the fields $X_A^\mu$, $X_B^\mu$ as fields variables instead $A^\mu$, $B^\mu$ we can state that each terms of the action is semi-positive defined, so in order to find the minimum it is sufficient to minimize them separately.
There are the two possible minimum for the field strength part and for the covariant derivative part: the trivial solution with $\phi$ and $\tilde{X}_A^\mu$, $\tilde{X}_B^\mu$ equal to the null fields and the solution with  $\phi$, $X_A^\mu$, $X_B^\mu$ proportional to the identity. In each cases both the field strength part and the covariant derivative part 	disappear. For the potential parts we have:
\begin{eqnarray*}
V_A=V_B=\left(\frac{\chi_{-1}}{\chi_0}\right)^2 &&  \textrm{for} \ \phi=\tilde{X}^\mu_A=\tilde{X}^\mu_B=0 \\
V_A=(\alpha^2_\phi+4\frac{4\Omega^2}{1+\Omega^2}\alpha^2_A-\frac{\chi_{-1}}{\chi_0})^2,  &&  \textrm{for} \ \phi=\alpha_\phi \textbf{I}, \ \tilde{X}^\mu_A=\alpha_A \textbf{I}^\mu, \  \tilde{X}^\mu_B=\alpha_B \textbf{I}^\mu \\
V_B=(\alpha^2_\phi+4\frac{4\Omega^2}{1+\Omega^2}\alpha^2_B-\frac{\chi_{-1}}{\chi_0})^2 &&
\end{eqnarray*} 
Referring to the second case and minimizing the potentials, the minimum seems to be  for 
\begin{equation}
\phi_0=\frac{\chi_{-1}}{\chi_0}\cos\alpha \textbf{I},\ \tilde{X}^\mu_{A0}=\tilde{X}^\mu_{B0}= \frac{1}{2}\sqrt{\frac{\chi_{-1}}{\chi_0}}\sqrt{ \frac{1+\Omega^2}{4\Omega^2}} \sin\alpha \textbf{I}^\mu \label{Vacuum}
\end{equation}
However, the previous position is not allowed because the identity does not belong to the algebra under consideration. In general the non-triviality of the vacuum makes very difficult to explicit the vacuum configuration of the system in \cite{Goursac1} A. de Goursac, J.C. Wallet, and R. Wulkenhaar, using the matrix base formalism, have found an expressions from vacuum solutions deriving them from the relevant solutions equations of motion.
Although, the  complexity  of the vacuum configuration makes the perturbative approach very complicated, in order to conduct some investigation in the next section will be consider a non-perturbative approach using a discretized matrix model of the action \eqref{faction1} obtained using a  Moyal base. In this setting the action reduces to  
\begin{eqnarray}
&&S(\mathcal{D}_A)=  \nonumber \\
&&+\frac{1}{(1+\Omega^2)^2} \int d^4z \Bigg\{\left(\frac{(1-\Omega^2)^2}{2}-\frac{(1-\Omega^2)^4}{6(1+\Omega^2)^2}\right)(F^A_{\mu\nu}\star F_A^{\mu\nu}+F^B_{\mu\nu}\star F_B^{\mu\nu})\nonumber \\
&&+\left(\bar{\phi}\star\phi+\frac{4\Omega^2}{1+\Omega^2}\tilde{X}^A_\mu\star\tilde{X}_A^\mu-\frac{\chi_{-1}}{\chi_0}\right)^2 \nonumber \\
&&+\left(\phi\star\bar{\phi}+ \frac{4\Omega^2}{1+\Omega^2}\tilde{X}^B_\mu\star\tilde{X}_B^\mu-\frac{\chi_{-1}}{\chi_0}\right)^2\nonumber \\
&&+ 2(1+\Omega^2)D_\mu\phi\star\overline{D^\mu\phi}\Bigg\}(z)+\mathcal{O}(\chi_1) \label{S}
\end{eqnarray}
The omitted factor for the finite matrix model of size $N$ becomes constant so can be ignored. The minimum is obtained like before and formally is \eqref{Vacuum}  in this case the identity,  of course, belongs to the matrix space. It is interesting to notice that the vacuum of the finite model, due to the Higgs field,  is no longer invariant under the transformations \eqref{gauge}, but is invariant under a subgroup of $U(N)\times U(N)$:
\begin{equation}
u_A=\overline{u_B} \longrightarrow   u_A\star \textbf{I} \star\overline{u_B}=u_A\star \overline{u_A}=\textbf{I}
\end{equation}
Having discretized the model will be performed a Monte Carlo simulation studying some statistical quantity such the energy density, specific heat, varying the parameters $\Omega, \frac{\chi_{-1}}{\chi_0}, \alpha$ and gathering some informations on the various contributions of the fields to the action. The simulations are quite cumbersome due to the 
complexity of the action and to the number of independent matrix to handle.

\section{Discretization of the action}
The first step across the numerical analysis is to apply a discretization scheme. Various schemes can be used like lattice approximation, but the nature of star product due to its oscillator exponential,  makes the lattice approach not suitable without adaptations. We will use  another discretization scheme in which our fields are approximated  by finite matrices and the star product becomes the standard matrix multiplication.
Using the identity $D_\mu\phi =\phi\star\tilde{X}_{B\mu}-\tilde{X}_{A\mu}\star\phi$   we can recast the action \eqref{S} in the following form:
\footnotesize
\begin{eqnarray}
&S(\phi,\tilde{X}_A,\tilde{X}_B)&= \frac{1}{(1+\Omega^2)^2}\int d^4z\Bigg\{\left(\frac{\left(1-\Omega^2\right)^2}{2}-\frac{\left(1-\Omega^2\right)^4}{6\left(1+\Omega^2\right)^2}\right)\Big(\left[\tilde{X}_{A\mu},\tilde{X}_{A\nu}\right]_\star\left[\tilde{X}_{A}^\mu,\tilde{X}_{A}^\nu\right]_\star  \nonumber \\
& &+\left[\tilde{X}_{B\mu},\tilde{X}_{B\nu}\right]_\star\left[\tilde{X}_{B}^\mu,\tilde{X}_{B}^\nu\right]_\star\Big)
+ \left(\phi\star\bar{\phi}+\frac{4\Omega^2}{1+\Omega^2}\tilde{X}_A^\mu\star\tilde{X}_{A\mu}-\frac{\chi_{-1}}{\chi_0}\right)^2 \nonumber \\
& & +\left(\bar{\phi}\star\phi +\frac{4\Omega^2}{1+\Omega^2}\tilde{X}_B^\mu\star\tilde{X}_{B\mu}-\frac{\chi_{-1}}{\chi_0} \right)^2  \nonumber \\
& &+2(1+\Omega^2)\left(\phi\star\tilde{X}_{B\mu}-\tilde{X}_{A\mu}\star\phi\right)
\left(\bar{\phi}\star\tilde{X}_{A}^\mu-\tilde{X}_{B}^\mu\star\bar{\phi}\right)\Bigg\}(z) \nonumber \\
& &+\mathcal{O}(\chi_1)\label{S0}
\end{eqnarray}\normalsize
As a first approach to the numerical simulation and forced by limited computation resource,  we will consider the Monte Carlo simulation of the previous action around  its minimum and the simulation will take $\sqrt{\frac{\chi_{-1}}{\chi_0}} $  as a positive parameter. In this setting  the behavior of the simulations will be identical for the negative case and avoiding $\frac{\chi_{-1}}{\chi_0}$ to be negative we  have not any problems about the thermalization. In order to define the previous action around the minimum we translate the fields $\phi$, $\tilde{X}_{A\mu}$, $\tilde{X}_{B\mu}$ using the following translated fields:
\begin{eqnarray}
\phi &=& \psi +\sqrt{\frac{\chi_{-1}}{\chi_0}}\cos\alpha\textbf{I} \ \\
\tilde{X}_{A\mu}&=& Y_{A\mu} +\frac{1}{2}\sqrt{\frac{\chi_{-1}}{\chi_0}}\sqrt{\frac{2\Omega^2}{(1+\Omega^2)}}\textbf{I}_\mu\sin\alpha  \\
\tilde{X}_{B\mu}&=& Y_{B\mu} +\frac{1}{2}\sqrt{\frac{\chi_{-1}}{\chi_0}}\sqrt{\frac{2\Omega^2}{(1+\Omega^2)}}\textbf{I}_\mu\sin\alpha  
\end{eqnarray}
Substituting the previous fields into \eqref{S0} we get a positive action with minimum in zero:
\begin{eqnarray}
&S(\psi,Y_A,Y_B)&= +\frac{1}{(1+\Omega^2)^2}\int d^4z\Bigg\{D\Big(\left[Y_{A\mu},Y_{A\nu}\right]_\star\left[Y_{A}^\mu,Y_{A}^\nu\right]_\star 
+ \left[Y_{B\mu},Y_{B\nu}\right]_\star\left[Y_{B}^\mu,Y_{B}^\nu\right]_\star\Big) \nonumber \\
& &+ \left(\psi\star\bar{\psi}+\mu\cos\alpha(\psi+\bar{\psi})+ CY_A^\mu\star Y_{A\mu} +\mu \textbf{I}^\mu Y_{A\mu}\sin\alpha \right)^2  \nonumber \\
& &+ \left(\bar{\psi}\star\psi +\mu\cos\alpha(\psi+\bar{\psi})+ CY_B^\mu\star Y_{B\mu} +\mu \textbf{I}^\mu Y_{B\mu}\sin\alpha  \right)^2 \nonumber \\ \nonumber \\
& &+2(1+\Omega^2)\left((Y_{B\mu}-Y_{A\mu})\mu\cos\alpha +\psi\star Y_{B\mu}-Y_{A\mu}\star\psi\right)\nonumber \\
& &\star\left((Y_{A}^\mu-Y_{B}^\mu)\mu\cos\alpha+\bar{\psi}\star Y_{A}^\mu -Y_{B}^\mu\star\bar{\psi}\right)\Bigg\}(z)  +\mathcal{O}(\chi_1)
\label{Sf} \end{eqnarray}
Where for simplicity we put:
\begin{eqnarray}
C=\frac{1+\Omega^2}{4\Omega^2},& D=\frac{\left(1-\Omega^2\right)^2}{2}-\frac{\left(1-\Omega^2\right)^4}{6\left(1+\Omega^2\right)^2} , & \frac{\chi_{-1}}{\chi_0}=\mu^2
\end{eqnarray}

\subsection{Discretization by Moyal base}
The following treatment is mainly taken from \cite{moyal-triple,bondia2} as introduction to the Moyal base which will be used later.
We can define on the algebra $\mathbb{R}_\Theta^2$  a natural basis of eigenfunctions $f_{mn}$ of the harmonic
oscillator, where $m, n \in \mathbb{N}$. 
This base satisfy the $\star$-multiplication rule: 
\begin{equation}
(f_{mn} \star f_{kl} )(x) = \delta_{nk} f_{ml} (x) \label{star-rule}
\end{equation}
and  this useful property:
\begin{eqnarray}
\int d^2 x f_{mn}(x)&=& \delta_{mn} \int d^2x  f_0 =2\pi\theta\delta_{mn}
\end{eqnarray}
The previous multiplication rule  associates the $\star$-product between $f_{ml}$ with the ordinary matrix product:
In this base we can write any elements of $ \mathbb{R}_\Theta^2$ but we have to require the rapid decay \cite{bondia2} of the sequences of coefficients $\{a_{mn}\}$:
\begin{equation}
\sum^\infty_{m,n=0} a_{mn} f_{mn}(x) \in \mathbb{R}_\Theta^2, \ \textrm{if}, \ \sum^\infty_{m,n=0} \left((2m+1)^{2k} (2n+1)^{2k} |a_{mn}|^2 \right)^\frac{1}{2} < \infty \ \textrm{for all}  \ k .
\end{equation}
The eigenfunctions $f_{nm}$ can be expressed with the help of
Laguerre functions \cite{bondia2,phi4-non,moy-base}:
\begin{equation}
f_{mn}(\rho,\varphi)= 2(-1)^m \sqrt{\frac{m!}{n!}} e^{i\varphi(n-m)} \left(\sqrt{\frac{2}{\theta}}\rho\right)^{n-m} e^{-\frac{\rho^2}{\theta}}L_{m}^{n-m}\left(\frac{2}{\theta}\rho^2\right) \label{mb}
\end{equation}
Our fields can be expanded in this base as:
\begin{equation}
X^\mu(x) =\sum_{m_i,n_i \in \mathbb{N}} X^\mu_{\genfrac{}{}{0pt}{}{m_1n_1}{m_2n_2}} f_{m_1n_1}(x_0,x_1)f_{m_2n_2}(x_2,x_3 )\label{X-exp}
\end{equation}
and
\begin{equation}
\psi(x) =\sum_{m_i,n_i \in  \mathbb{N}} \psi_{\genfrac{}{}{0pt}{}{m_1n_1}{m_2n_2}} f_{m_1n_1}(x_0,x_1)f_{m_2n_2}(x_2,x_3 )\label{psi-exp}
\end{equation}
Using this base we can forget the Moyal product in this way the model becomes to 9-matrix model. A $\star$-product between two fields using \eqref{star-rule} can be written as
\begin{eqnarray}
\Psi(x)\star\Phi(x)&=&\sum_{m_i,n_i,k_1,l_1\in  \mathbb{N}}\Psi_{\genfrac{}{}{0pt}{}{m_1n_1}{m_2n_2}}\Phi_{\genfrac{}{}{0pt}{}{k_1l_1}{k_2l_2}} f_{m_1n_1}(x_0,x_1)\star f_{k_1l_1}(x_0,x_1) \nonumber \\
&\times & f_{m_2n_2}(x_2,x_3)\star f_{k_2l_2}(x_2,x_3) \nonumber \\
&=& \sum_{m_i,n_i,k_1,l_1\in  \mathbb{N}}\Psi_{\genfrac{}{}{0pt}{}{m_1n_1}{m_2n_2}}\Phi_{\genfrac{}{}{0pt}{}{k_1l_1}{k_2l_2}}\delta_{n_1k_1}\delta_{n_2k_2}f_{m_1l_1}(x_0,x_1)f_{m_2l_2}(x_2,x_3)\nonumber \\
&=& \sum_{m_i,l_1\in  \mathbb{N}}\Psi\Phi_{\genfrac{}{}{0pt}{}{m_1l_1}{m_2l_2}} f_{m_1l_1}(x_0,x_1)f_{m_2l_2}(x_2,x_3)
\end{eqnarray}
\normalsize
where 
\begin{equation}
\Psi\Phi_{\genfrac{}{}{0pt}{}{m_1l_1}{m_2l_2}}=\sum_{n_1,n_2 \in  \mathbb{N}}\Psi_{\genfrac{}{}{0pt}{}{m_1n_1}{m_2n_2}}\Phi_{\genfrac{}{}{0pt}{}{n_1l_1}{n_2l_2}}
\end{equation}
So the star product became a "double" matrix multiplication, the action, the equations of field and all treatments can be conducted on the infinite matrices instead directly on the continues fields.

Beside, for finite matrices, the $\mathbb{N}^2$-indexed double sequences
  can be written as tensor products of ordinary matrices, 
\begin{equation}
X_{\genfrac{}{}{0pt}{}{m_1n_1}{m_2n_2}}= \sum_{i=1}^K 
X^i_{m_1n_1} \otimes X^i_{m_2n_2} \;.
\label{tensorproduct}
\end{equation}
Since the matrix product and trace also
factor into these independent components, the action factors into 
$S=\sum_{i=1}^K
S(\psi^{1i},Y_A^{1i},Y_B^{1i})S(\psi^{2i},Y_A^{2i},Y_B^{2i})$. 
Then, regarding all 
$\psi^{1i},Y_A^{1i}$, $Y_B^{1i},\psi^{2i},Y_A^{2i},Y_B^{2i}$
as random variables over which to integrate in the partition function,
the partition function factors, too:
\begin{align}
&\int
\mathcal{D}(\psi^{11},Y_A^{11},Y_B^{11},\psi^{21},Y_A^{21},Y_B^{21})
\cdots 
\mathcal{D}(\psi^{1K},Y_A^{1K},Y_B^{1K},\psi^{2K},Y_A^{2K},Y_B^{2K})
\;e^{-S}
\nonumber
\\
&=\bigg(
\int \mathcal{D}(\psi^{1i},Y_A^{1i},Y_B^{1i},\psi^{2i},Y_A^{2i},Y_B^{2i})
\;e^{-S(\psi^{1i},Y_A^{1i},Y_B^{1i})\cdot S(\psi^{2i},Y_A^{2i},Y_B^{2i})}
\bigg)^K\;.
\end{align}
We may therefore restrict ourselves to $K=1$.
Using this approximations the calculus will be performed just on standard  infinite matrix, but is not enough to be handled numerically. We have to perform a truncation in order to obtain finite matrices, this truncation will consist in a maximum $m,n<N$ in the expansion \eqref{psi-exp}-\eqref{X-exp}. It is easy to verify that this kind of approximation corresponds in a cut in energy, in fact from the definition  of  $f_{mn}$ we have:
\begin{equation}
\{H, f_{mn}\}_\star= \theta(m+n+\frac{1}{2} )f_{mn}(\varphi,\rho)
\end{equation}
Beside, can be proved \cite{moy-base} that the functions $f_{mn}$ with $m, n < N$ induce a cut-off  in position space and momentum space:
\begin{equation}
\rho_{max} \sim \sqrt{2\theta}  \ ,\ \textrm{for} \ m, n < N 
\end{equation}
and
\begin{equation}
p_{max} \sim \sqrt{\frac{8N}{\theta}}  \,\ \textrm{for} \ m, n < N 
\end{equation}
Summarizing, to operate the discretization we have the following correspondences:
\begin{eqnarray}
\phi(x) \in \mathbb{R}_\Theta^4  &\rightarrow & \hat{\phi} \in \mathbb{M}_N \\
Y^A_\mu(x) \in \mathbb{R}_\Theta^4 &\rightarrow & \hat{Y}^A_\mu \in  \mathbb{M}_N \\
Y^B_\mu(x) \in \mathbb{R}_\Theta^4 &\rightarrow & \hat{Y}^B_\mu \in  \mathbb{M}_N \\
 \int a(x)dx &\rightarrow & \textrm{Tr}(\hat{a}) 
\end{eqnarray}
After truncating the representative matrices is convenient to operate  another substitution \cite{Goursac1}:
\begin{eqnarray}
 Z_0=\hat{Y}^A_0+i\hat{Y}^A_1, & \bar{Z}_0=\hat{Y}^A_0-i\hat{Y}^A_1  \nonumber \\
 Z_1=\hat{Y}^B_0+i\hat{Y}^B_1, & \bar{Z}_1=\hat{Y}^B_0-i\hat{Y}^B_1 \nonumber \\
 Z_2=\hat{Y}^A_2+i\hat{Y}^A_2, & \bar{Z}_2=\hat{Y}^A_2-i\hat{Y}^A_3  \nonumber \\
 Z_3=\hat{Y}^B_2+i\hat{Y}^B_3, & \bar{Z}_3=\hat{Y}^B_2-i\hat{Y}^B_3  \label{Z-sub}
 \end{eqnarray}
In the end the discretized action is:
\begin{equation}
S_4=\frac{1}{1+\Omega^2}\operatorname{Tr}\left(\mathcal{L}_F+\mathcal{L}_{V_0}+\mathcal{L}_{V_1}+\mathcal{L}_{D_0}\bar{\mathcal{L}}_{D_0} +\mathcal{L}_{D_1}\bar{\mathcal{L}}_{D_1}+\mathcal{L}_{D_2}\bar{\mathcal{L}}_{D_2}+\mathcal{L}_{D_3}\bar{\mathcal{L}}_{D_3}\right) \label{S4}
\end{equation}
With
\footnotesize
\begin{eqnarray}
\mathcal{L}_{4F}&=&\frac{D}{2}\Big(\left[\bar{Z}_0,Z_0\right]^2 +\left[\bar{Z}_1,Z_1\right]^2 + 
\frac{1}{4}\Big(\left[Z_0+\bar{Z}_0,Z_2-\bar{Z}_2\right]^2-\left[Z_0+\bar{Z}_0,Z_2+\bar{Z}_2\right]^2\nonumber \\
&+& \left[Z_0-\bar{Z}_0,Z_2+\bar{Z}_2\right]^2- \left[Z_0-\bar{Z}_0,Z_2-\bar{Z}_2\right]^2 -\left[Z_1+\bar{Z}_1,Z_3+\bar{Z}_3\right]^2\nonumber \\
&+& \left[Z_1+\bar{Z}_1,Z_3-\bar{Z}_3\right]^2 +\left[Z_1-\bar{Z}_1,Z_3+\bar{Z}_3\right]^2 -\left[Z_1-\bar{Z}_1,Z_3-\bar{Z}_3\right]^2\Big)\Big)\nonumber \\
\mathcal{L}_{4V_0}&=&\big(\psi\bar{\psi}+\mu\cos\alpha(\psi+\bar{\psi})+ \frac{1}{2}\left(\left\{\bar{Z}_0,Z_0\right\} +\left\{\bar{Z}_2,Z_2\right\}\right)\nonumber \\
&+&\frac{\mu\sin\alpha}{2\sqrt{C}}((-1+i)(Z_0+Z_2)+(1+i)(\bar{Z}_0+\bar{Z}_2))\big)^2    \nonumber \\
\mathcal{L}_{4V_1}&=&\big(\bar{\psi}\psi+\mu\cos\alpha(\psi+\bar{\psi})+ \frac{1}{2}\left(\left\{\bar{Z}_1,Z_1\right\} +\left\{\bar{Z}_3,Z_3\right\}\right)\nonumber \\
&+& \frac{\mu\sin\alpha}{2\sqrt{C}}((-1+i)(Z_1+Z_3)+(1+i)(\bar{Z}_1+\bar{Z}_3))\big)^2    \nonumber \\
\mathcal{L}_{4D_0}&=&  \sqrt{2(1+\Omega^2)}\left(\mu\cos\alpha(Z_1+\bar{Z}_1-Z_0-\bar{Z}_0 ) + \psi(Z_1+\bar{Z}_1)-(Z_0+\bar{Z}_0)\psi\right)\nonumber \\
\mathcal{L}_{4D_1}&=&  \sqrt{2(1+\Omega^2)}\left(\mu\cos\alpha(Z_1-\bar{Z}_1-Z_0+\bar{Z}_0 ) + \psi(Z_1-\bar{Z}_1)-(Z_0-\bar{Z}_0)\psi\right)\nonumber \\
\mathcal{L}_{4D_2}&=&  \sqrt{2(1+\Omega^2)}\left(\mu\cos\alpha(Z_3+\bar{Z}_3-Z_2-\bar{Z}_2 ) + \psi(Z_3+\bar{Z}_3)-(Z_2+\bar{Z}_2)\psi\right)\nonumber \\
\mathcal{L}_{4D_3}&=&  \sqrt{2(1+\Omega^2)}\left(\mu\cos\alpha(Z_3-\bar{Z}_1-Z_2+\bar{Z}_2 ) + \psi(Z_3-\bar{Z}_3)-(Z_2-\bar{Z}_2)\psi\right)\nonumber \\
\nonumber 
\end{eqnarray}
\normalsize
Where for simplicity we have omitted the hat on the matrices and the bars stand for the hermitian conjugate. In this case the action \eqref{S4} becomes 5 complex matrix model instead eight real matrices $Y_{A\mu}$, $Y_{B\mu}$ and one complex matrix $\psi$. This form  may seem cumbersome but it is  more comfortable for numerical simulations. The next step is to define the estimator for the average values of interest and to develop some numerical parameters in order to analyze  the numerical results.  

\section{Definition of the observables}

Following  Monte Carlo methods, will be produced a  sequence of configurations
$\{(\psi,Z_i)_j \}, j = 1, 2,\cdots,T_{MC}$  and evaluated the average of the observables over that set
of configurations. The sequences of configurations obtained,  a Monte Carlo chain,  are  representations of 
the configuration space at the given parameters. 
In this frame the expectation value is approximated as
\begin{equation}
\langle O\rangle \approx \frac{1}{T_{MC}}\sum_{j=1}^{T_{MC}}O_j
\end{equation}
where $O_j$ is the value of the observable $O$ evaluated in the $j$-sampled configuration,
$(\psi,Z_i)_j$, $O_i= O[(\psi,Z_i)_j]$.
The internal energy is defined as:
\begin{equation}
E(\Omega,\mu,\alpha)= \langle S\rangle
\end{equation}
and the specific heat takes the form
\begin{equation}
C(\Omega,\mu,\alpha)= \langle S^2\rangle - \langle S\rangle^2
\end{equation}
These terms correspond to the usual definitions for energy
\begin{equation}
E(\Omega,\mu,\alpha) = -\frac{1}{\mathcal{Z}}\frac{\partial\mathcal{Z}}{\partial\beta}
\end{equation}
and specific heat
\begin{equation}
C(\Omega,\mu,\alpha) =\frac{\partial E}{\partial\beta}
\end{equation} where $\mathcal{Z}$ is the partition function. 
It is very useful to compute separately the average values of the four contributions:
\begin{eqnarray}
F(\Omega,\mu,\alpha)&=& \langle \operatorname{Tr} \mathcal{L}_F \rangle \\
V_0(\Omega,\mu,\alpha) &=& \langle \operatorname{Tr} \mathcal{L}_{V_0} \rangle \\
V_1(\Omega,\mu,\alpha) &=& \langle \operatorname{Tr} \mathcal{L}_{V_1} \rangle \\
D(\Omega,\mu,\alpha)&=& \langle \operatorname{Tr}\left(\mathcal{L}_{D_j}\bar{\mathcal{L}}_{D^j}\right) \rangle \\
V&=&V_0+V_1
\end{eqnarray}
Where $i,j=1,2$ or $i,j=1,\cdots,4$. 

\subsection{Order parameters}
The previous quantities are not enough if we want to measure the various contributions of different modes of the fields 
to the configuration $\psi,Z_i$. Therefore we need  some control parameters usually called order parameters. As a first idea we can think about a quantity related to the norms of the fields for example the sums $\sum_{nm} |\psi_{nm}|^2$, $\sum_{nm} |Z_{inm}|^2$ of all squared entries of our matrices, this quantity is called the full power of the field \cite{order-par,order-par1} and it can be computed as  the trace of the square:
\begin{eqnarray}
\varphi^2_a &=& \operatorname{Tr}(|\psi|^2) \label{vara} \\
Z^2_{ia} &=& \operatorname{Tr}(|Z_i|^2) 
\end{eqnarray}
However, $\langle\varphi_a\rangle$ alone  is not a real order parameter because does not distinguish contributions from the different modes but we can use it as a reference to  define the quantities:
\begin{eqnarray}
\varphi^2_0 &=& \sum^N_{n=0} |a_{nn}|^2  \nonumber \\
Z^2_{i0} &=& \sum^N_{n=0} |z_{inn}|^2 \label{var0}
\end{eqnarray} 
Referring to the base \eqref{mb} it is easy to see that  such parameters are connected to the pure spherical contribution.  This quantity will be used to analyze the spherical contribution to the full power of the field.  
We can generalize the previous quantity defining some parameters $\varphi_l$ in such a way they form a decomposition of the full power of the fields. 
\begin{equation}
\varphi^2_a=\varphi^2_0+\sum_l \varphi^2_l , \  Z_{ia}^2=Z_{i0}^2+\sum_l Z_{il}^2
\end{equation}
Following this prescription the other quantity for $l>0$ can be  defined as:
\begin{equation}
\varphi^2_l =\sum^l_{n,m=0} |a_{nm}(1-\delta_{nm})|^2 , \  Z_{il}^2 =\sum^l_{n,m=0} |z_{lnm}(1-\delta_{nm})|^2  \label{varl}
\end{equation}
If the contribution is dominated from  the spherical symmetric parameter we expect to have  $\langle\varphi^2_a\rangle \sim \langle\varphi^2_0\rangle$, $\langle Z_{ia}^2\rangle \sim \langle Z^2_{i0}\rangle$. With \eqref{varl}  we can define the order parameters $\varphi^2_1$, $Z_{i1}^2$ as a particular case
$l = 1$ we have
\begin{equation}
\varphi^2_1 =|a_{10}|^2+ |a_{01}|^2, \  Z_{i1}^2 =|z_{i10}|^2+ |z_{i01}|^2
\end{equation}
In the next simulations we be evaluated  the quantities related to $l = 0$ and to $l = 1$ as representative of
those contribution where the rotational symmetry is broken.
Using higher $l$ in \eqref{varl} we can analyze  the contributions of the remaining modes, turns out that the measurements 
of the first two modes are  enough to characterize the behavior of the system. 

\section{Numerical results }

Now we  discuss the results of the Monte Carlo simulation on the approximated spectral model. As a first approach  we use some restrictions on the parameters. Will be considered the spectral action around its minimum \eqref{S4} in order to simplify the calculation, in this frame in the action  is symmetric under the transformation $\mu\to-\mu$ so  will be used the condition $\mu\geq0 $ and $\mu^2\geq0$. Will be explored the range $\mu\in[0,3.1]$, this interval was chosen to show a particular behavior of the system for fixed $\Omega$. The parameter $\Omega$  appears only with its square and is defined as a real parameter, therefore for the $\Omega$ too we require $\Omega\geq0$. Beside, if we refer to the scalar model, is possible to prove \cite{phi4-non} using the Langmann-Szabo duality \cite{L-S-dual}, that the model can be fully described varying $\Omega$ in the range $[0,1]$, for higher $\Omega$ the system can be remap inside the previous interval. In the present model the L-S duality does not hold any more, but forced by limited resource, we conjecture that the interval $\Omega\in[0,1]$ still enough to describe the system. The last parameter to consider is  $\alpha$,  it is connected to the choice of the vacuum state, the range of $\alpha$ is $[0,2\pi]$. The study of the system varying this parameter is quite important from a theoretical point of view because is related to the vacuum invariance. Beside, in the action  appear some contributions proportional to $(\sin\alpha)/\Omega$ which seem to diverge for $\Omega=0$, numerically we have verified that is an eliminable divergence and the curves of the observables can be extended in zero by continuity. Studying the dependence on $\alpha$ we can conclude that in the limit $N\to\infty$ the observables are independent from $\alpha$, therefore for our purposes $\alpha$ will be fixed equal to zero avoiding the annoying terms. In general for each observable  are computed the graphs for $N=$5, 10, 15, 20 matrix size.

\subsection{Varying $\alpha$ }
We start looking at the variation of energy density and the full power of the fields density for fixed $\mu$ and $\Omega$, varying $\alpha \in [0,2\pi]$. As representative here will be presented the graphs for $\mu=1$, $\Omega=\{1,0.5\}$ but we obtain the same behaviors for any other choice of the parameters allowed in the  considered range.
\begin{figure}[htb]
\begin{center}
\includegraphics[scale=0.45]{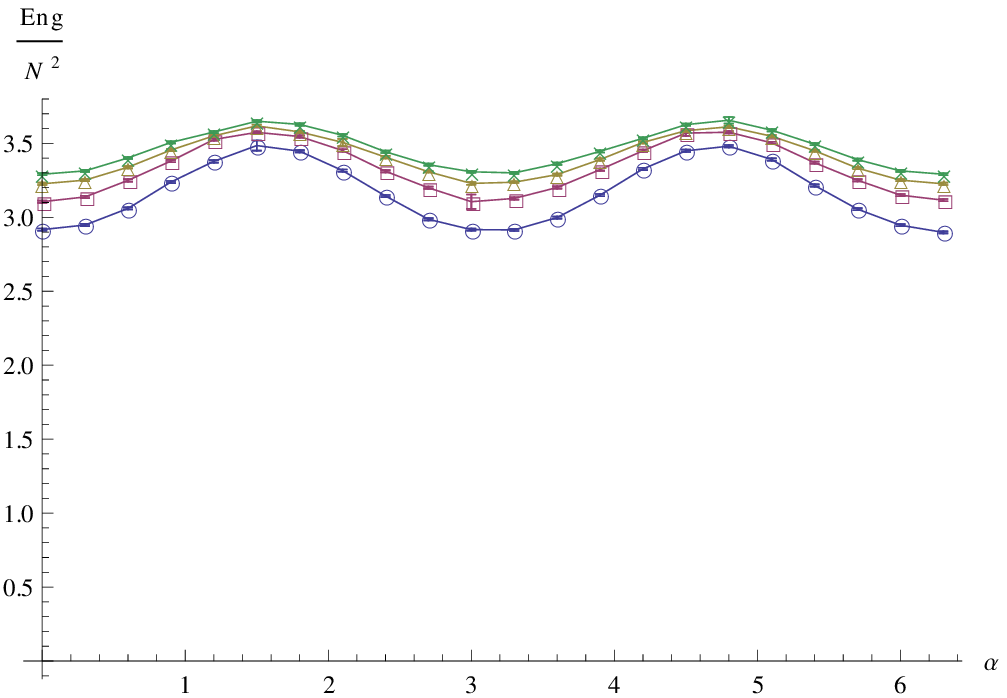}
\includegraphics[scale=0.45]{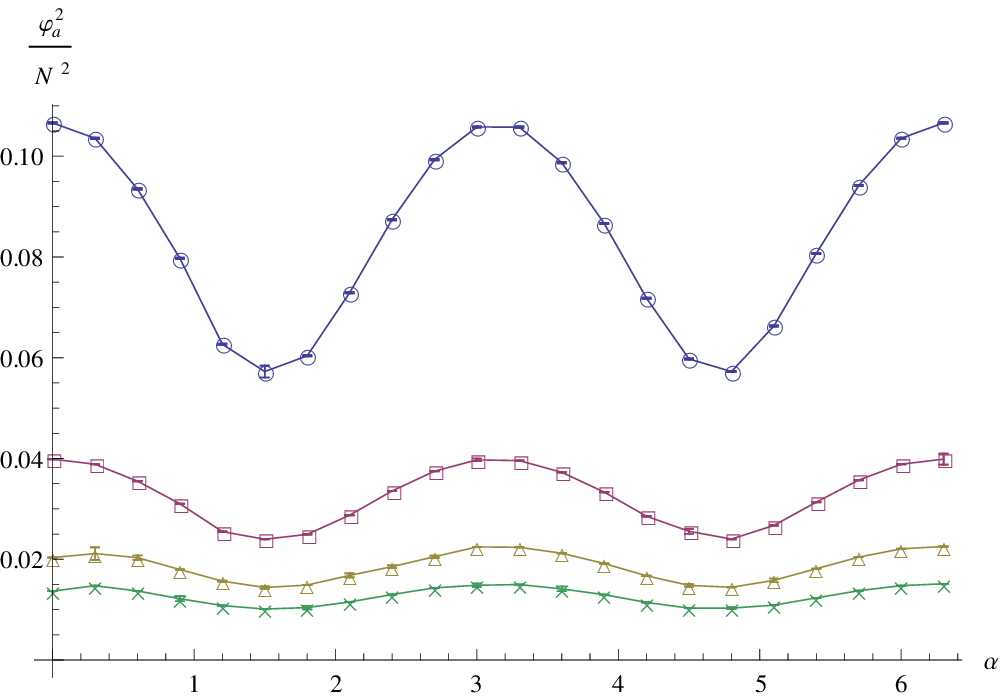}
\includegraphics[scale=0.45]{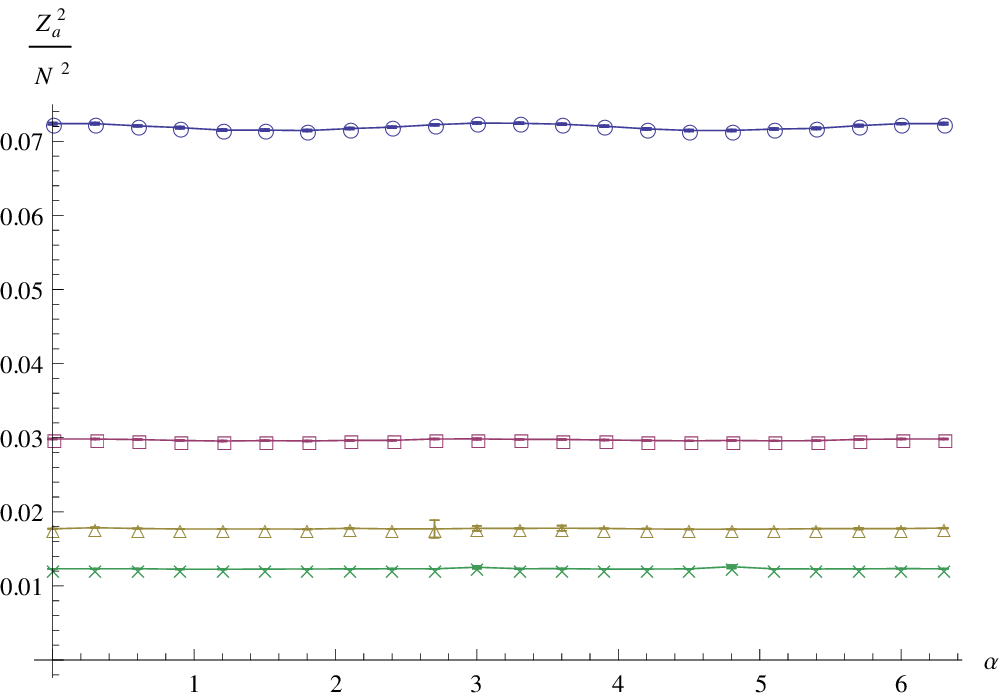}
\end{center}
\caption{\footnotesize Total energy density and full power of the fields density for $\varphi^2_a$, $Z_{0a}^2$ (from the left to the right) fixing $\mu=1$, $\Omega=1$, varying $\alpha$ and $N$. $N=5$ (circle), $N=10$ (square), $N=15$ (triangle), $N=20$ (cross).  \normalsize}\label{Figure 1}\end{figure}
All tree graphs show an oscillating behavior of the values, this oscillation is present in all other quantities measured. The amplitude of this oscillation becomes smaller and smaller increasing the size of the matrix and this is true for all the quantities measured. The same  trend  is described in fig.\ref{Figure 2} in which  position of the maximum are different but the amplitudes becomes smaller increasing $N$ 
\begin{figure}[htb]
\begin{center}
\includegraphics[scale=0.45]{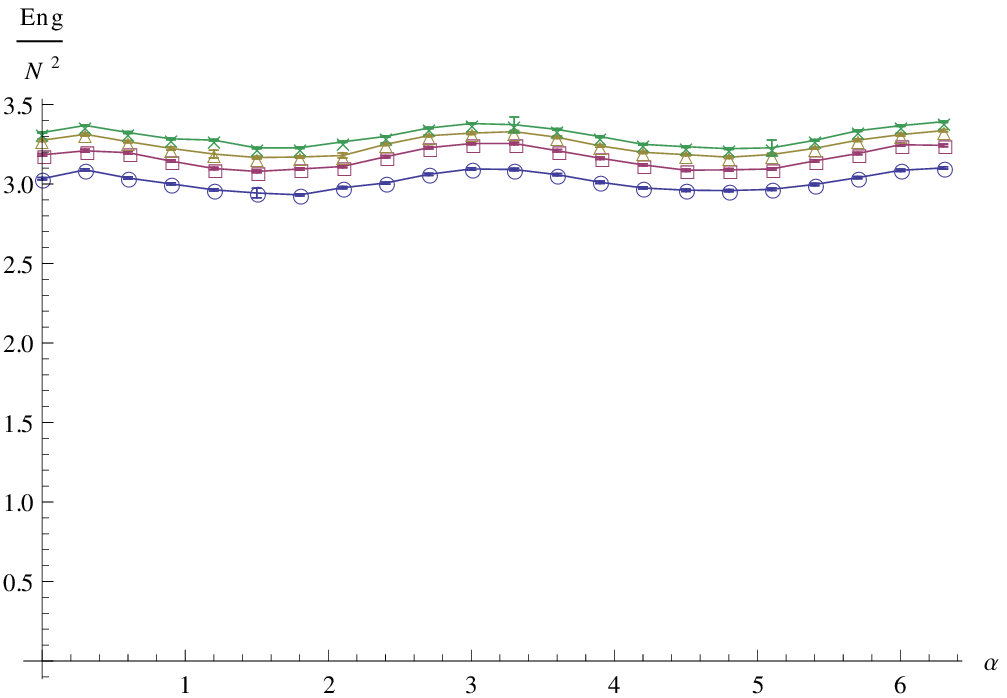}
\includegraphics[scale=0.45]{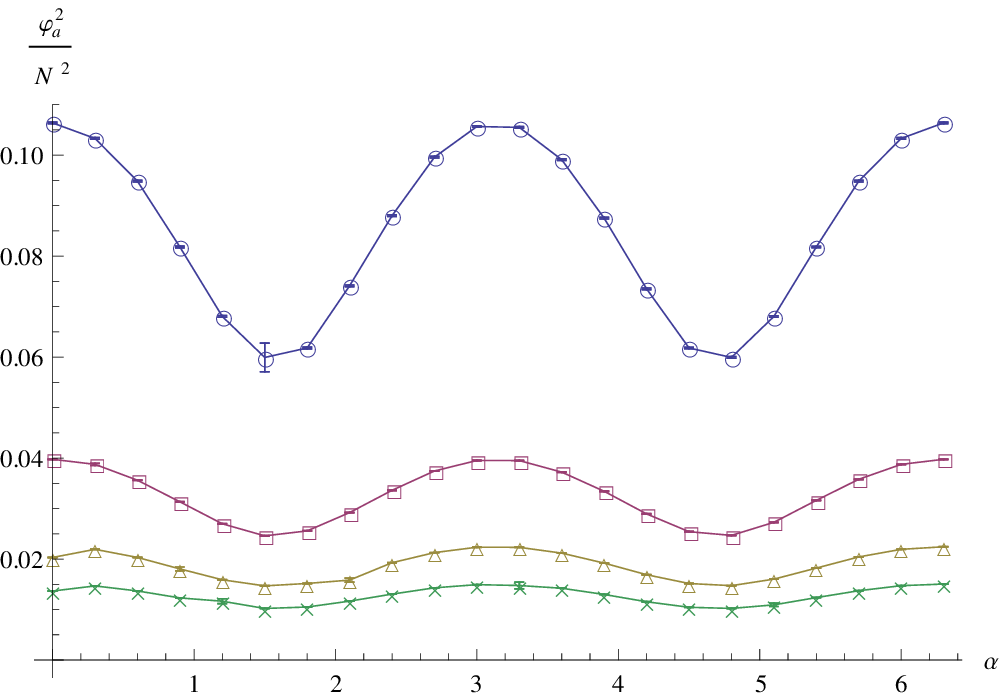}
\includegraphics[scale=0.45]{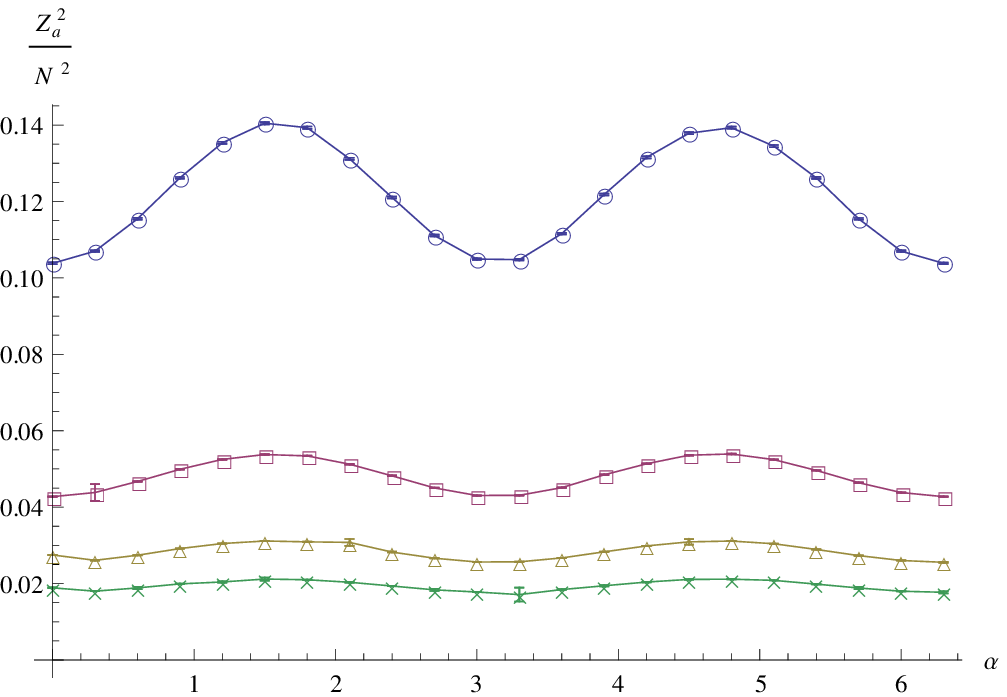}
\end{center}
\caption{\footnotesize Total energy density and full power of the fields density for $\varphi^2_a$, $Z_{0a}^2$  (from the left to the right) fixing $\mu=1$, $\Omega=0.5$, varying $\alpha$ and $N$. \normalsize}\label{Figure 2}\end{figure}
This results allow us to consider $\alpha=0$ for all next graphs, since we are interested in the behavior of the system for $N \to \infty $. This occurrence simplify all the next simulations thanks to the vanishing the of terms  $\sim(\sin\alpha)/\Omega $  appearing in the discretized action. Beside, such results induce us to reckon the parameter $\alpha$ as connected to the remaining invariance of the vacuum state for the exact model. 
\subsection{Varying $\Omega$ }
\begin{figure}[htb]
\begin{center}
\includegraphics[scale=0.4]{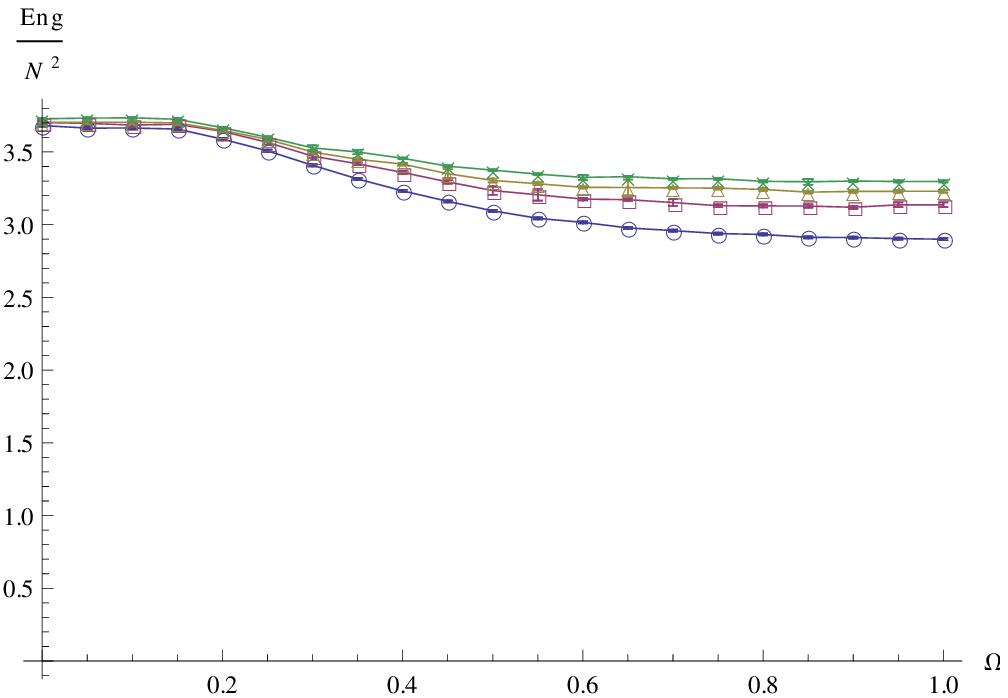}
\includegraphics[scale=0.4]{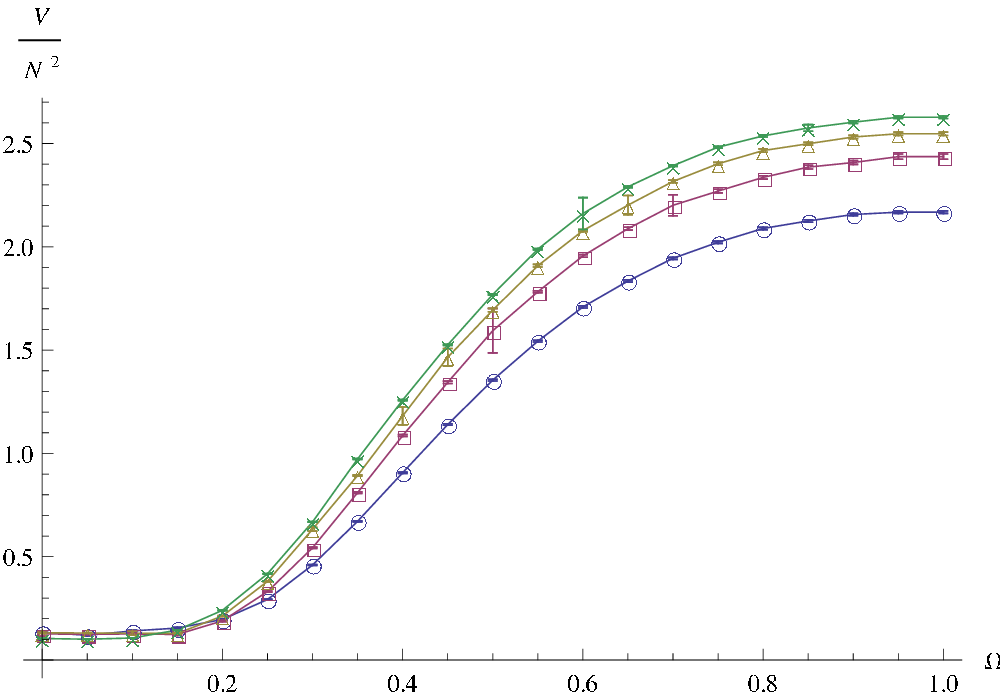}
\includegraphics[scale=0.4]{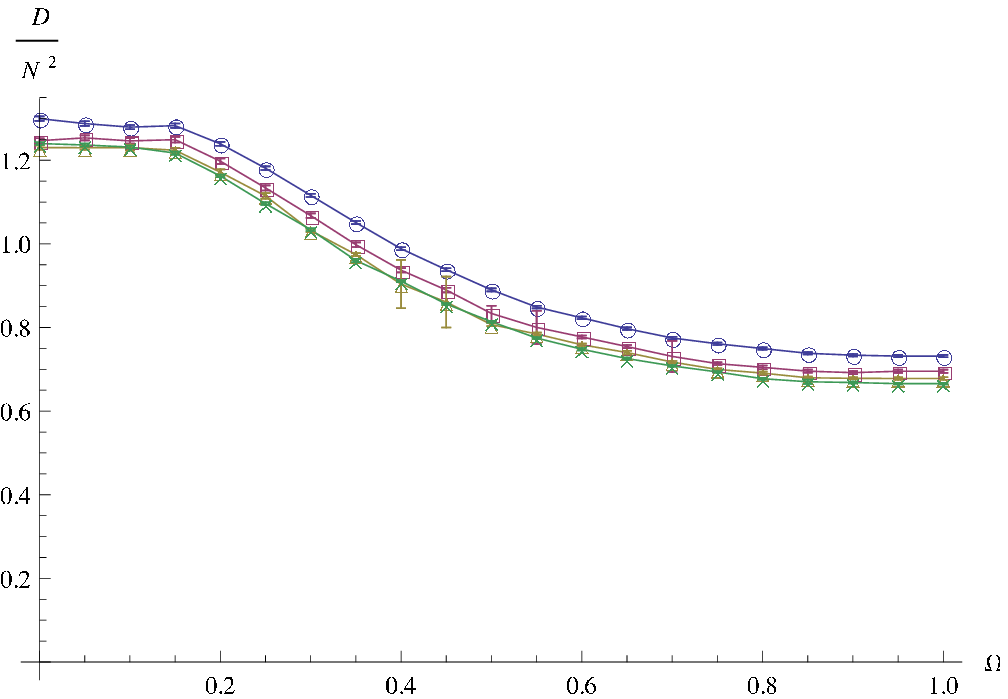}
\includegraphics[scale=0.4]{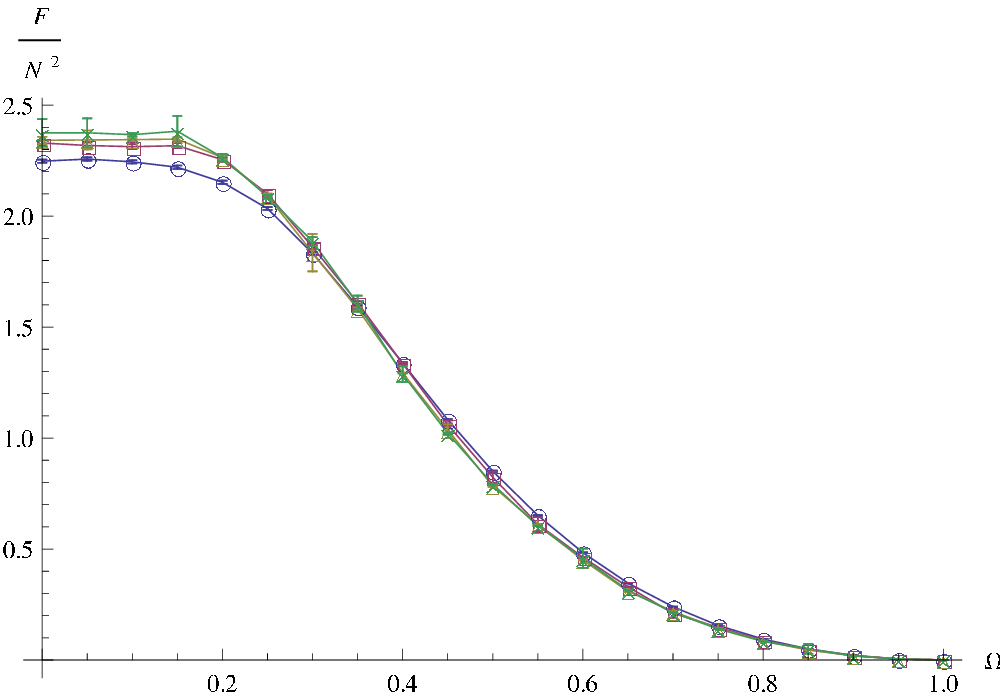}
\end{center}
\caption{\footnotesize Total energy density and the various contributions  for $\mu=1$, $\alpha=0$ varying $\Omega$ and $N$. From the left to the right  $E$, $V$, $D$, $F$ with $N=5$ (circle), $N=10$ (square), $N=15$ (triangle), $N=20$ (cross).\normalsize}\label{Figure 4}\end{figure}

\begin{figure}[htb]
\begin{center}
\includegraphics[scale=0.55]{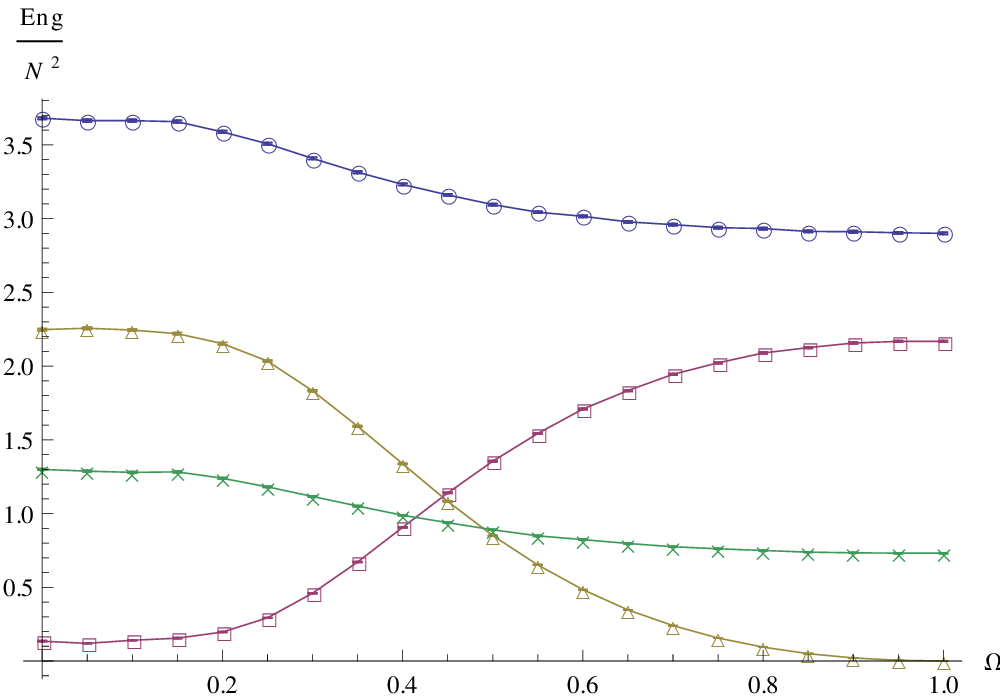}
\includegraphics[scale=0.55]{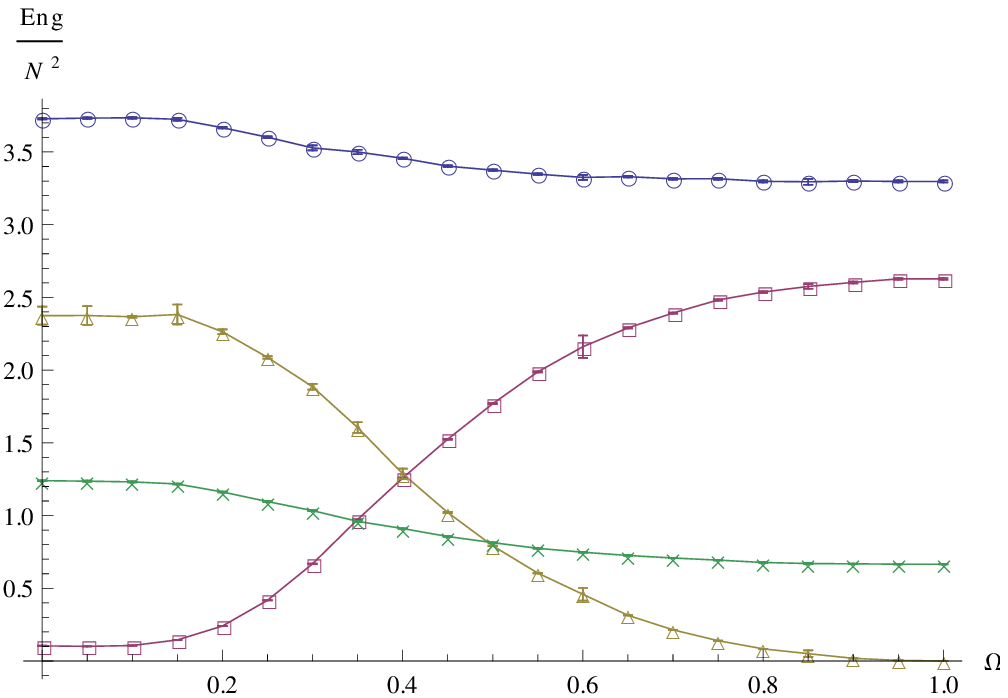}
\end{center}
\caption{\footnotesize Comparison of the total energy density and the various contributions for $\mu=1$, $\alpha=0$. $E$ (circle), $F$ (triangle), $D$ (cross), $V$ (square). With $N=5$ (left) and $N=20$ (right).  \normalsize}\label{Figure 5}
\end{figure}

Now we will analyze three cases in which $\mu$ is fixed to $0,1,3$, $\alpha$ is zero and we vary $\Omega\in[0,1]$. 
In the rest of this section we ignore for the computation of
$E,D,V,F$ the prefactor $(1+\Omega^2)$. In this way we focus our
attention to the integral as the source of possible phase transitions. 
The graphs in fig.\ref{Figure 4} show the total energy density $\langle S \rangle / N^2$ and the various contributions: the potential $V / N^2$, the Yang-Mills part $F / N^2$ and the covariant derivative part $D / N^2$, for $\mu=1$. There is no evident discontinuity or peak and increasing the size of the matrices the curves remain smooth. 
Comparing the energy density and the various contributions fig.\ref{Figure 5} we notice that the contributions between $F$ and $V$ balance each other and the total energy follows the slope of $D$, this  behavior continues increasing the size of the matrices fig.\ref{Figure 5}. \\
\newpage
\begin{figure}[htb]
\begin{center}
\includegraphics[scale=0.6]{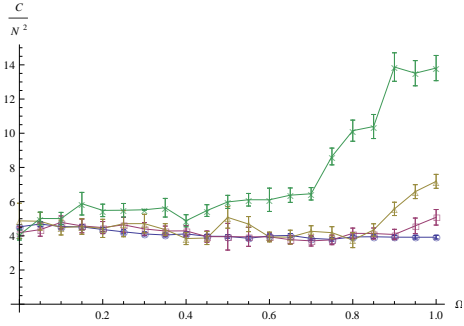}
\end{center}
\caption{\footnotesize Specific heat for $\mu=1$.\normalsize}\label{Figure 6}
\end{figure}
The specific heat density  shows fig.\ref{Figure 6} a peak in $\Omega=1$, this peak increases as $N$ increase. This behavior is typical of a phase transition, the peak is not clear for small $N$ due to the finite volume effect. 
In order to gain some informations on the composition  of the fields we look at the order parameters defined in the previous chapter. Starting from $\psi$ field, in the figure \ref{Figure 6} it is showed the graphs for $\varphi_a^2 $, $\varphi^2_0 $  and $\varphi^2_1 $ for $N=5$. \\
The three values  $\varphi_a^2 $, $\varphi^2_0 $  and $\varphi^2_1 $ seem essentially constant, comparing the three graphs fig.\ref{Figure 8} it is easy to see the dominance of the spherical contribution $\varphi^2_0 $ to the full power of the field.   
\begin{figure}[htb]
\begin{center}
\includegraphics[scale=0.55]{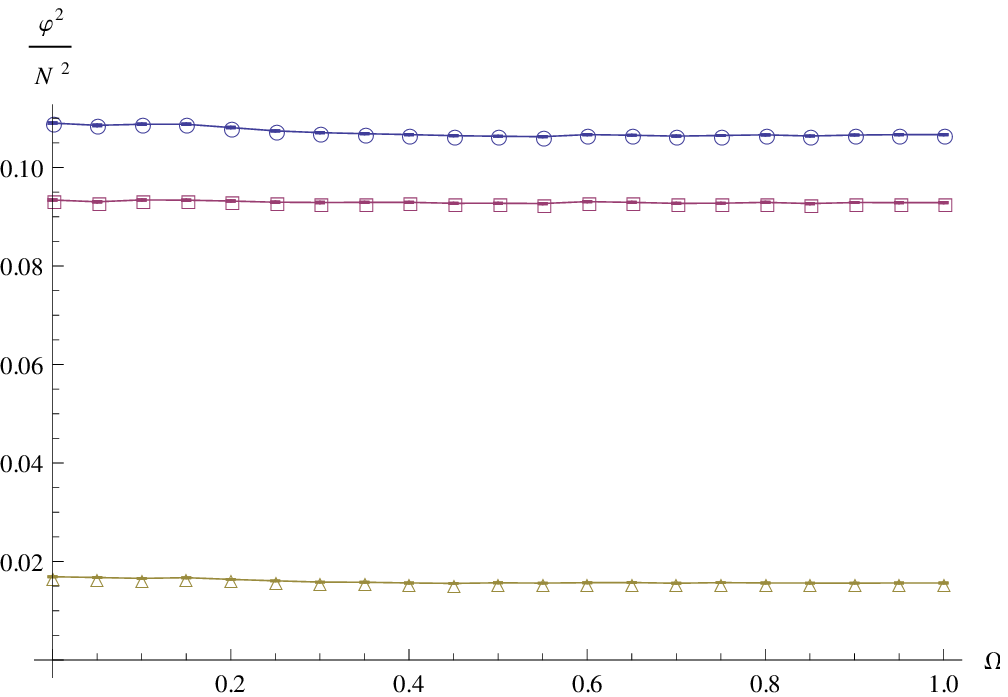}
\includegraphics[scale=0.55]{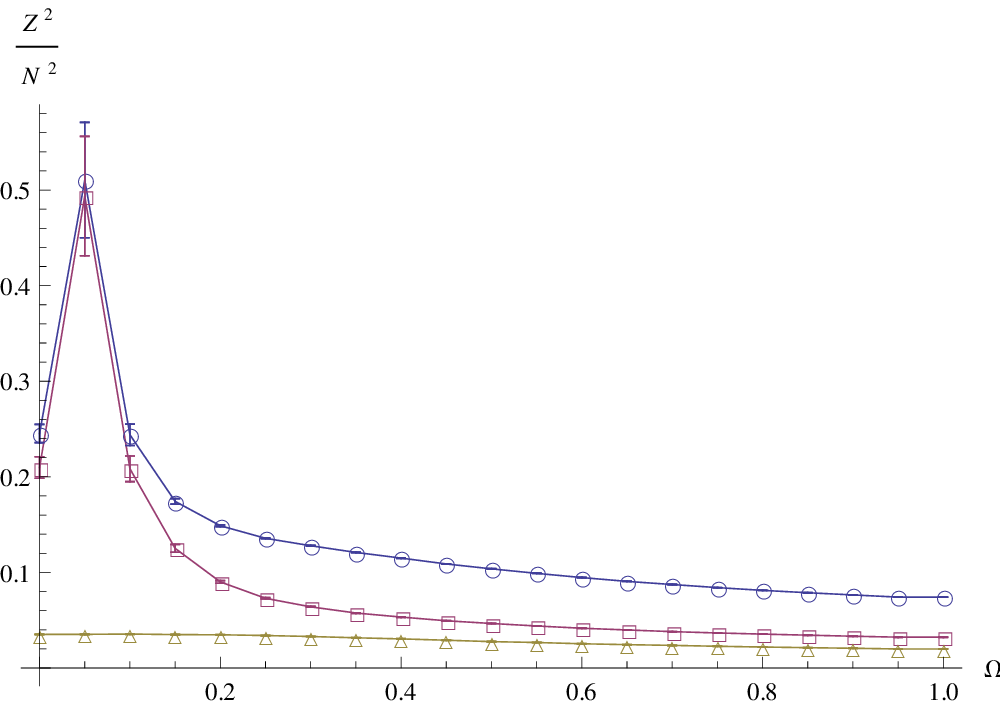}
\end{center}
\caption{\footnotesize On the left comparison of  $\varphi_a^2 $ (circle), $\varphi^2_0 $ (square)  and $\varphi^2_1 $ (triangle) density. On the right comparison of $Z_{0a}^2$ (circle), $Z_{00}^2$ (square) and $Z_{01}^2$ (triangle) density.\normalsize}\label{Figure 8}
\end{figure}
The behavior of the $Z_0$ fields is different,  the spherical contribution becomes dominant approaching to $\Omega=0$ starting from a zone in which the contribution of $Z_{00}^2$ and $Z_{01}^2 $ are comparable. 
For brevity  will not be showed only the graphs for $Z_{0a}^2 $, $Z_{00}^2$  and $Z_{01}^2 $ but taking in account the statistical errors the other $Z_i$ related graphs appear compatible to the $Z_0$ case. The dependence of the previous quantities on $N$ are showed in the following graphs fig.\ref{Figure 9}. 
\begin{figure}[htb]
\begin{center}
\includegraphics[scale=0.4]{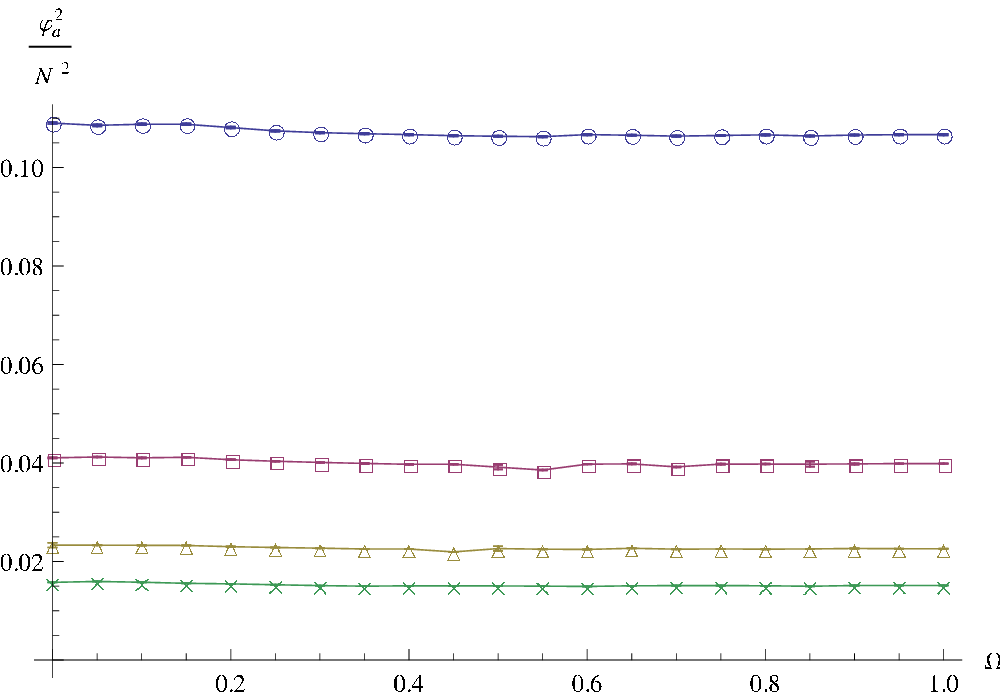}
\includegraphics[scale=0.4]{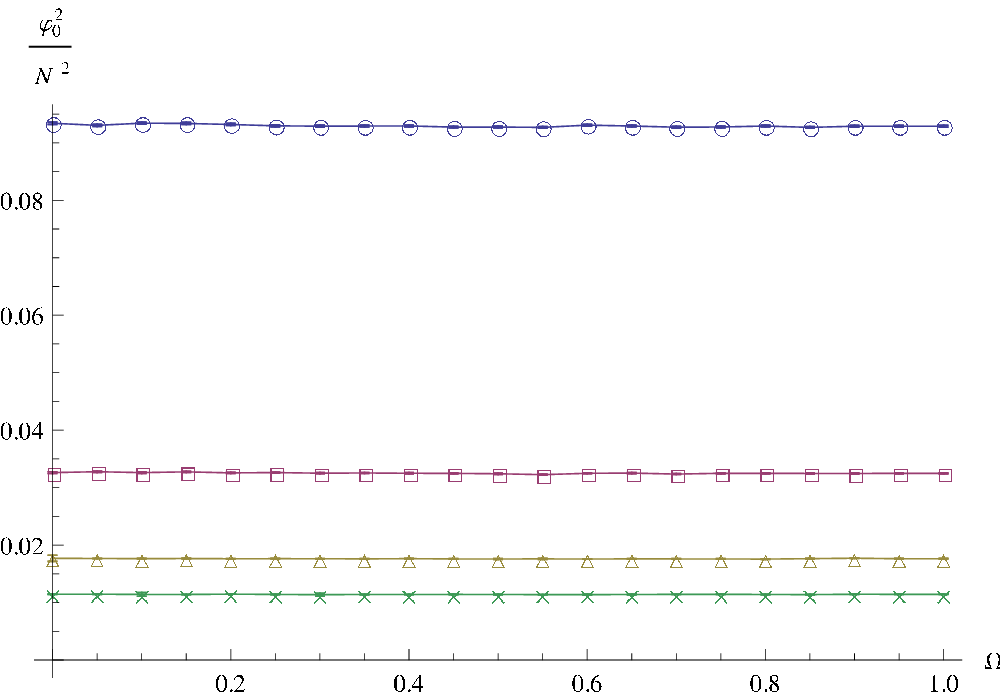}
\includegraphics[scale=0.4]{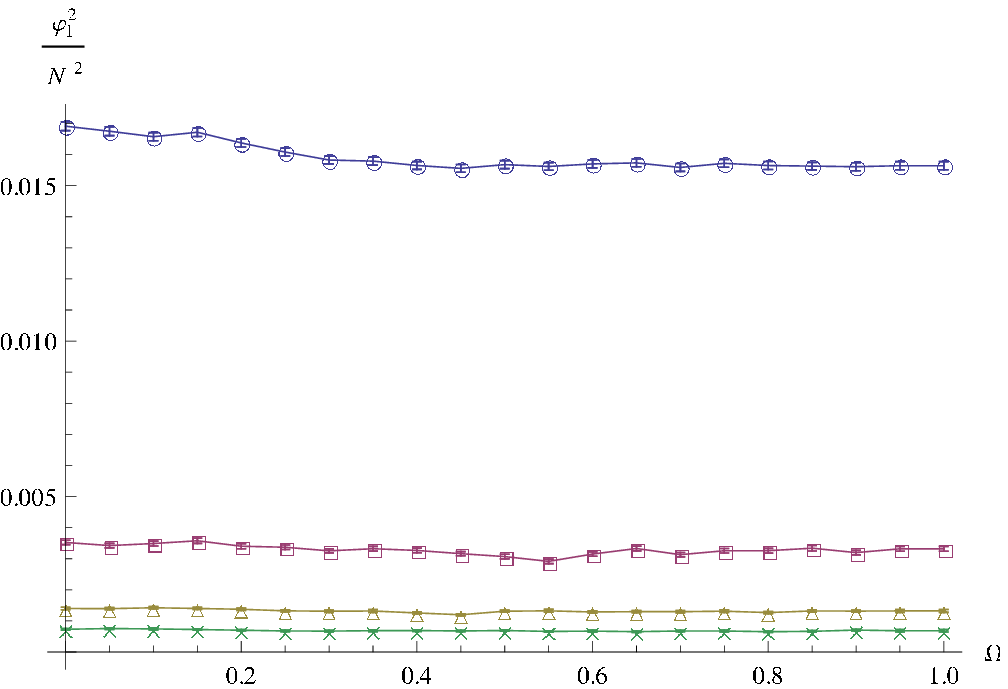}
\includegraphics[scale=0.4]{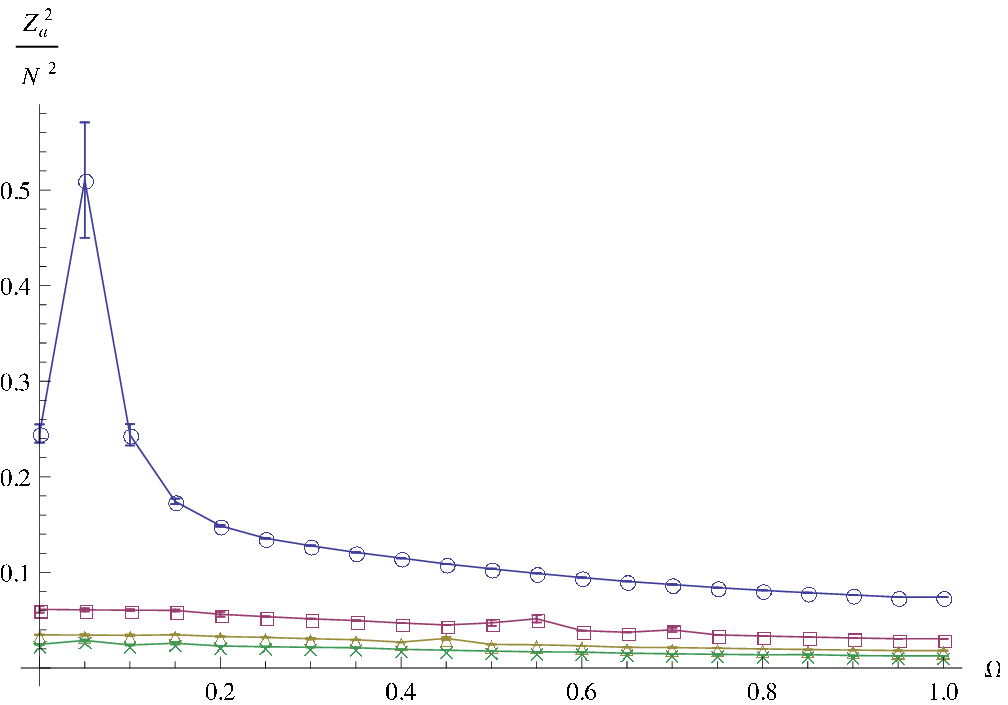}
\includegraphics[scale=0.4]{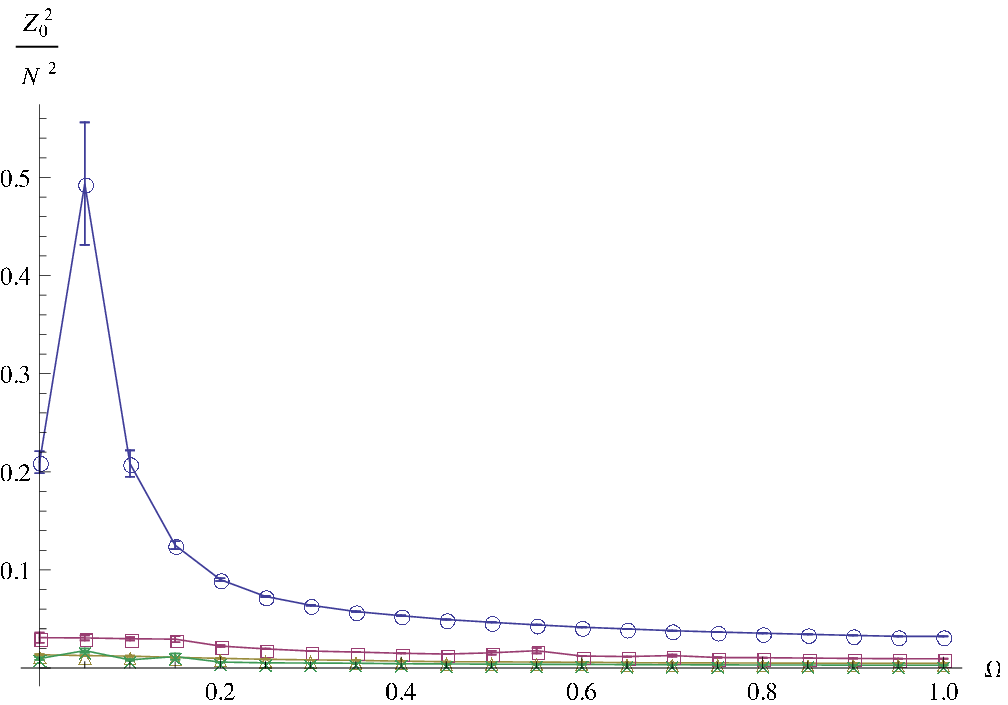}
\includegraphics[scale=0.4]{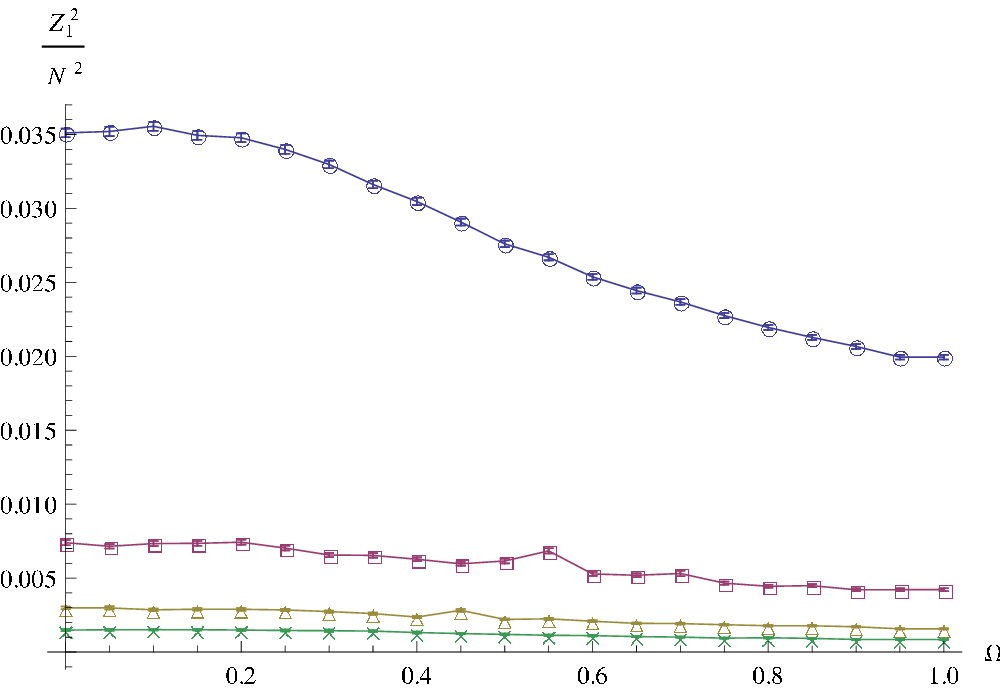}
\end{center}
\caption{\footnotesize Starting from the up left corner and from the left to the right the 	densities for $\varphi_a^2 $, $\varphi^2_0 $, $\varphi^2_1 $, $Z_{0a}^2$, $Z_{00}^2 $ and $Z_{01}^2$ for $\mu=1$ varying $\Omega$ and $N$. \normalsize}\label{Figure 9}
\end{figure}The values of the quantity of all the previous parameters decreases with $N$  but the dominance of the $\varphi_0$ on the total power of the field is independent by $N$. The peak  related  to $Z_0$ decrease, but if look at the single  graph for $N=20$ the spherical contribution approaching the point $\Omega=0$ it features a peak.
\newpage
Now we will analyze the  model for $\mu=0$; fig.\ref{Figure 10} shows the  graphs for total energy density and the contributions $V$, $D$, $F$. The slope of the total energy density seems to be constant. The $D$ contribution and the $F$ do not balance each other like in the previous case, but all the three contributions balance among them self to produce a constant sum. 
\begin{figure}[htb]
\begin{center}
\includegraphics[scale=0.4]{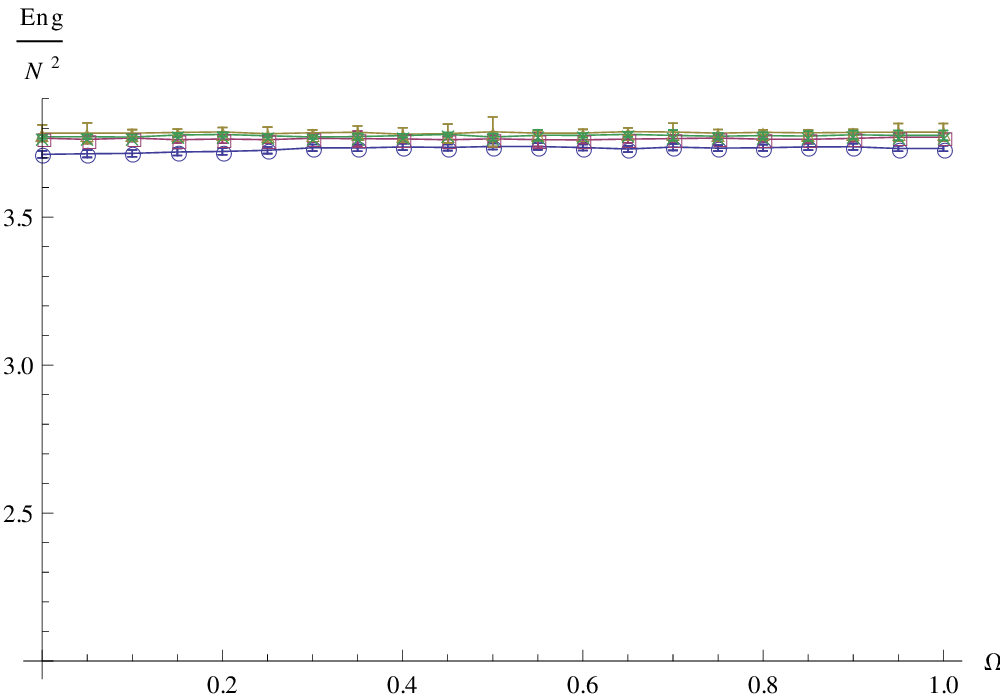}
\includegraphics[scale=0.4]{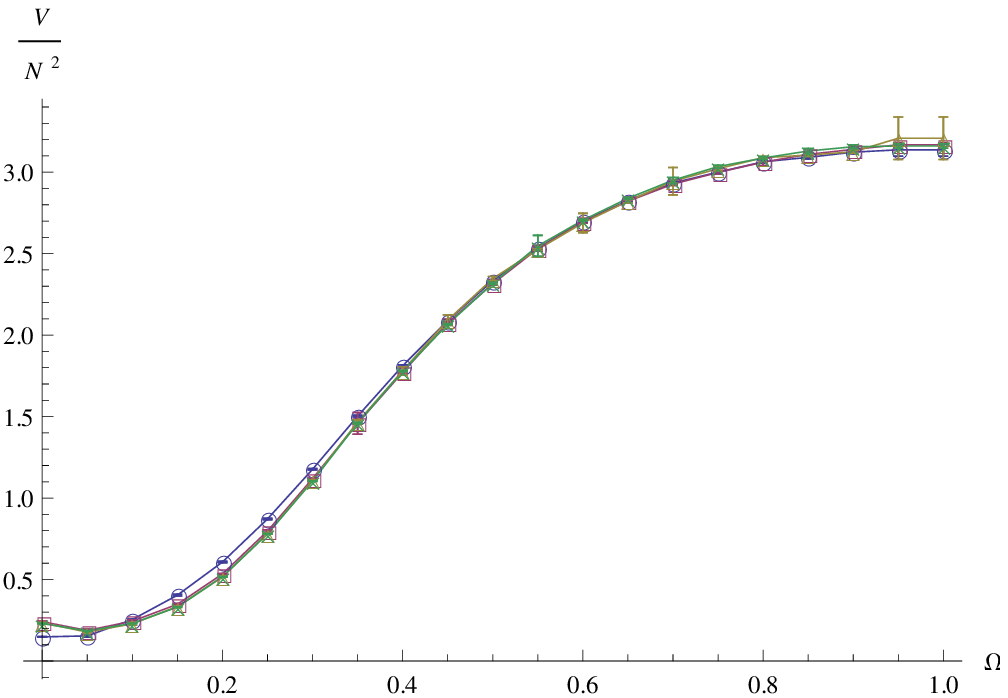}
\includegraphics[scale=0.4]{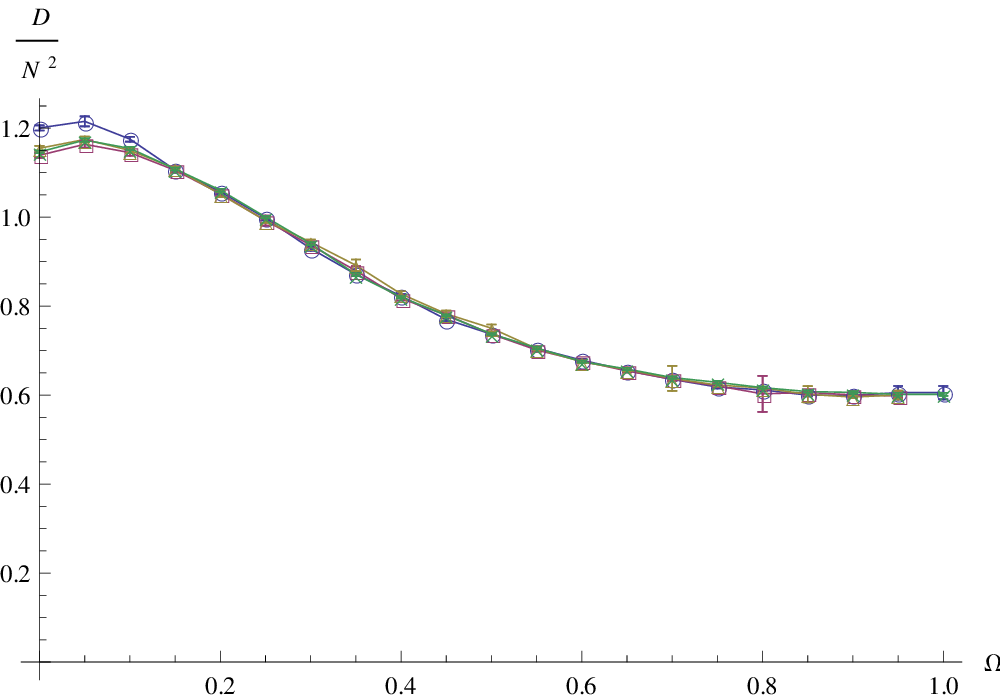}
\includegraphics[scale=0.4]{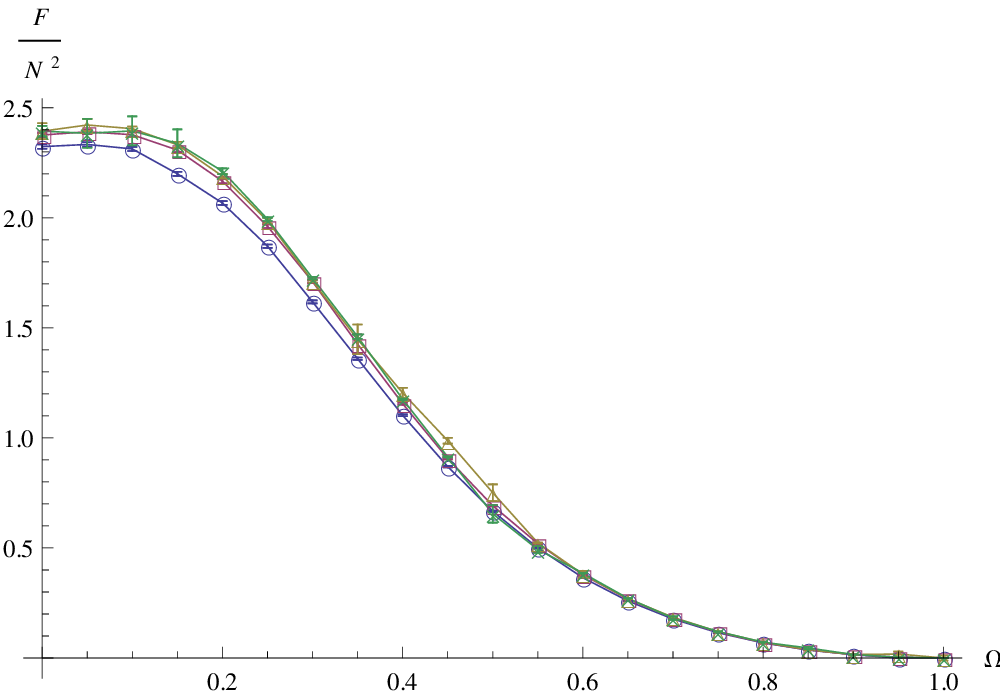}
\includegraphics[scale=0.45]{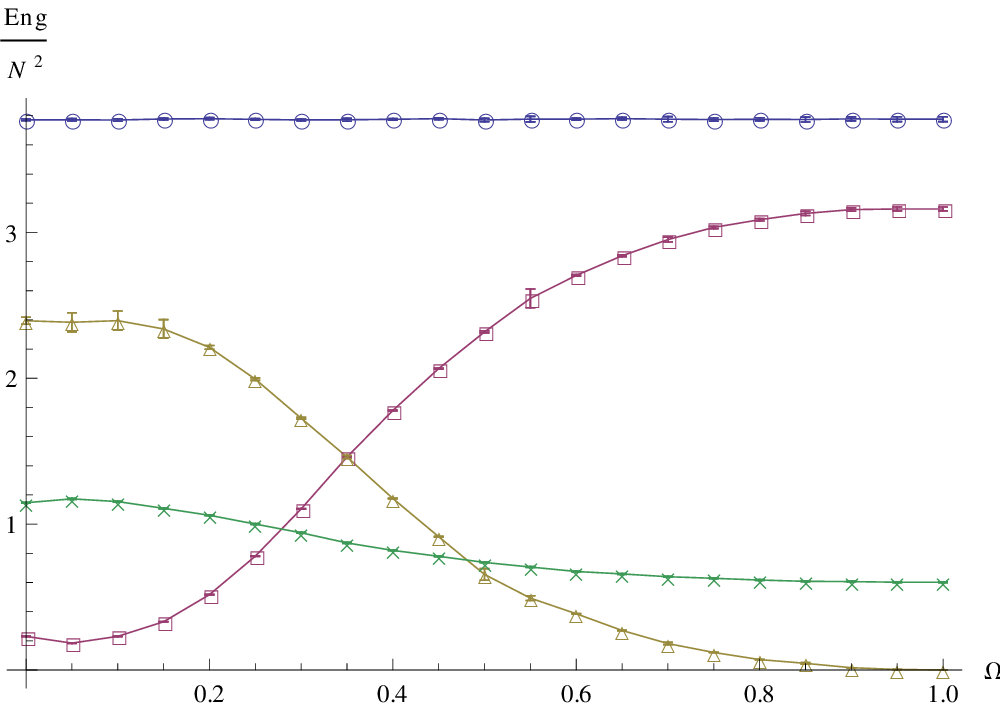}
\end{center}
\caption{\footnotesize Total energy density, various contributions and the comparison among them for $\mu=0$ varying $\Omega$ and $N$. From the left to the right  $E$, $V$, $D$, $F$ and comparison.\normalsize}\label{Figure 10}\end{figure}
\begin{figure}[htb]
\begin{center}
\includegraphics[scale=0.6]{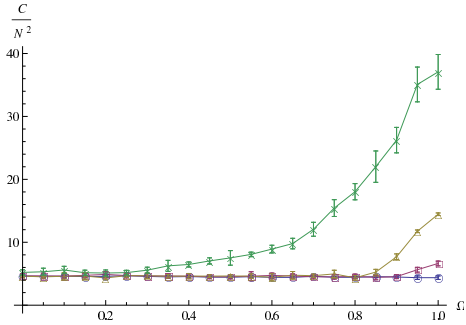}
\end{center}
\caption{\footnotesize Specific heat density for $\mu=0$ varying $\Omega$ and $N$.\normalsize}\label{Figure 11}
\end{figure}\newpage
The specific heat density shows fig.\ref{Figure 11} again the peak in $\Omega=1$ as $N$ increase.
For the other 	quantities  $\varphi_a^2 $, $\varphi^2_0 $, $\varphi^2_1 $ and  $Z_{0a}^2 $, $Z_{00}^2$,  $Z_{01}^2$ we have the same  behavior fig.\ref{Figure 14} of $\mu=1$ case, except for the oscillation appearing in the $Z_{0a}^2 $, $Z_{00}^2$ graphs close to zero, anyway it appears only for $N=5$.
\begin{figure}[htb]
\begin{center}
\includegraphics[scale=0.40]{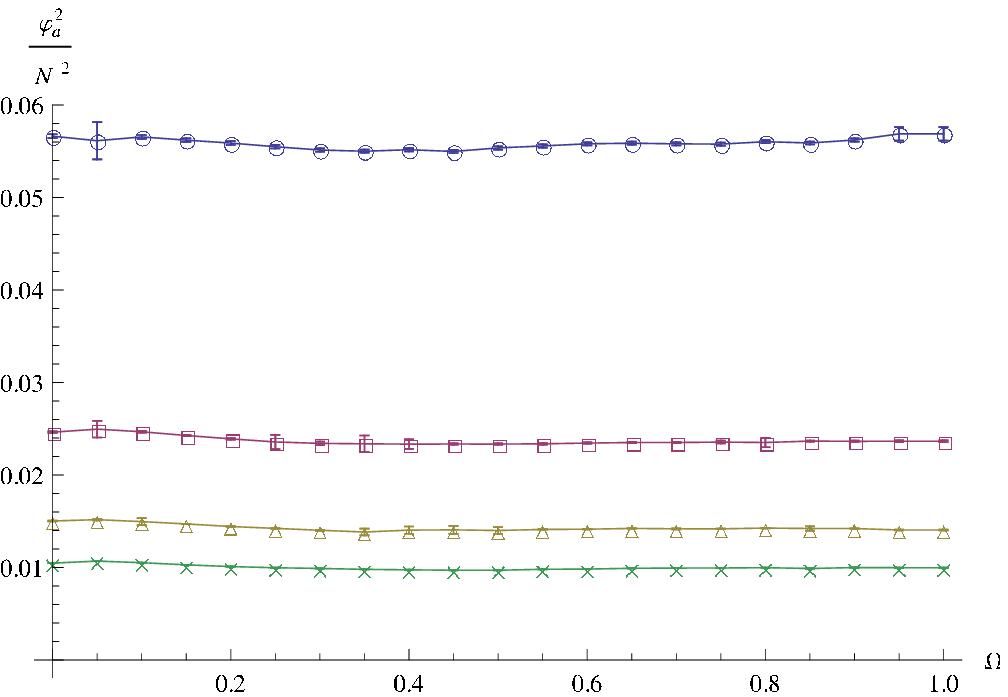}
\includegraphics[scale=0.40]{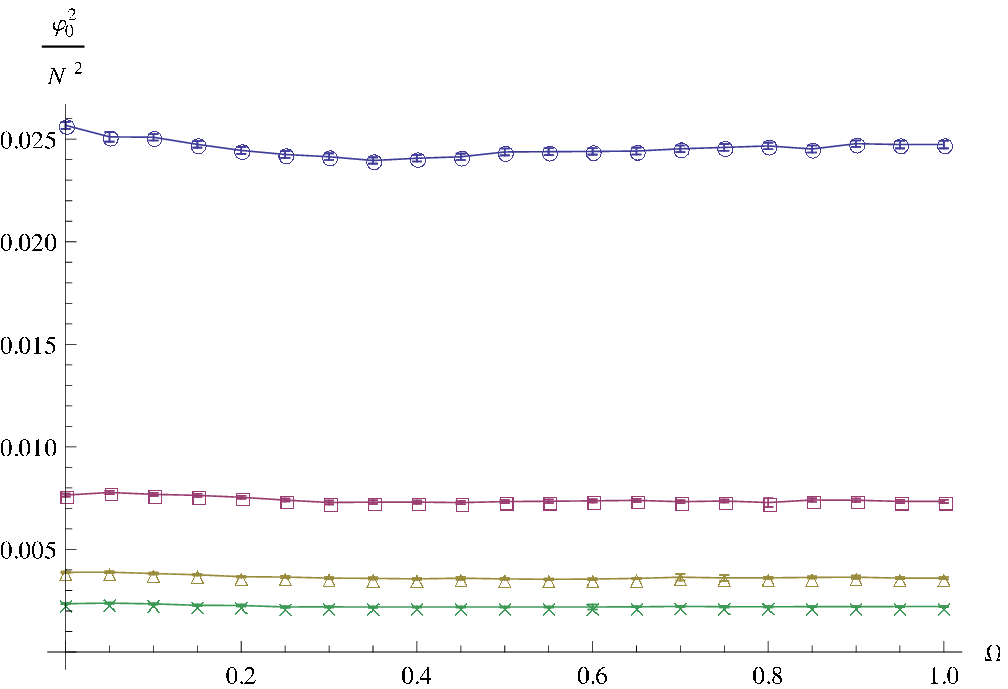}
\includegraphics[scale=0.40]{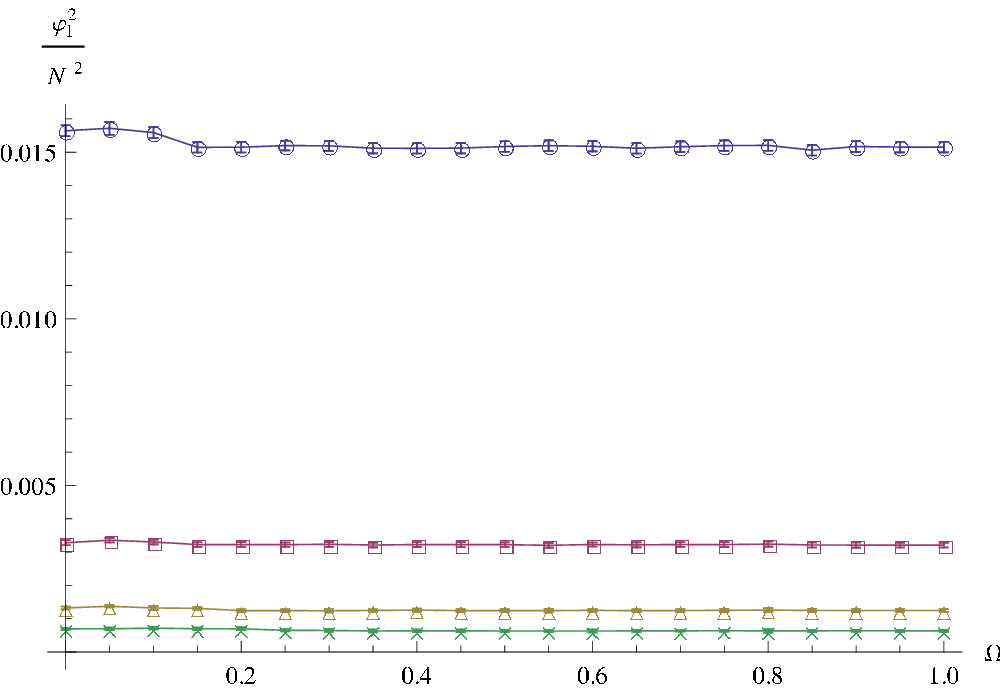}
\includegraphics[scale=0.40]{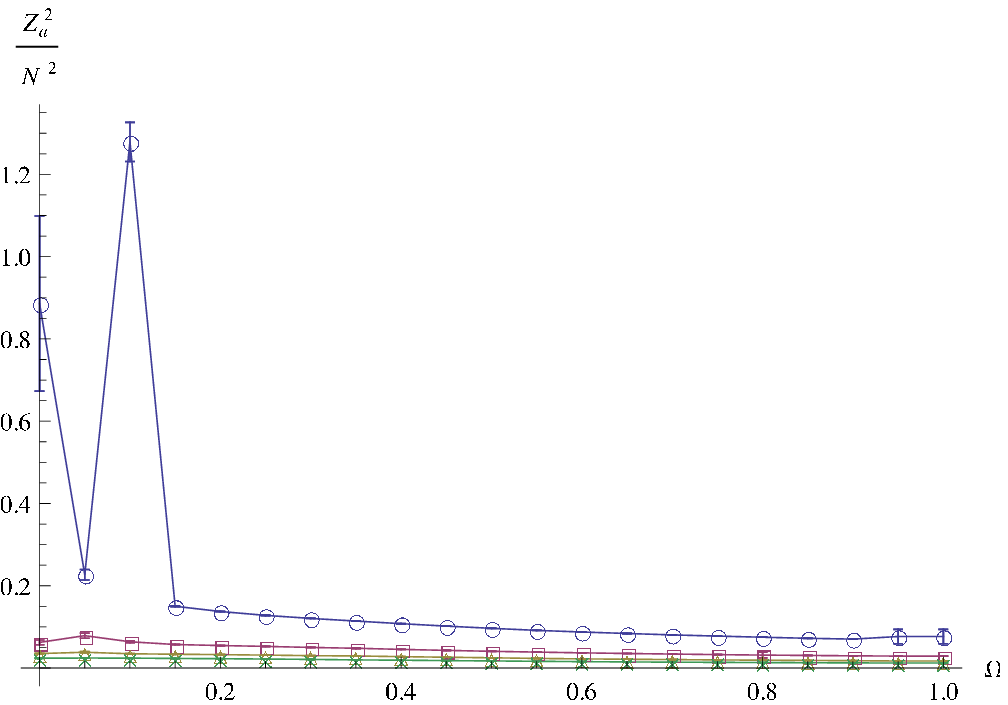}
\includegraphics[scale=0.40]{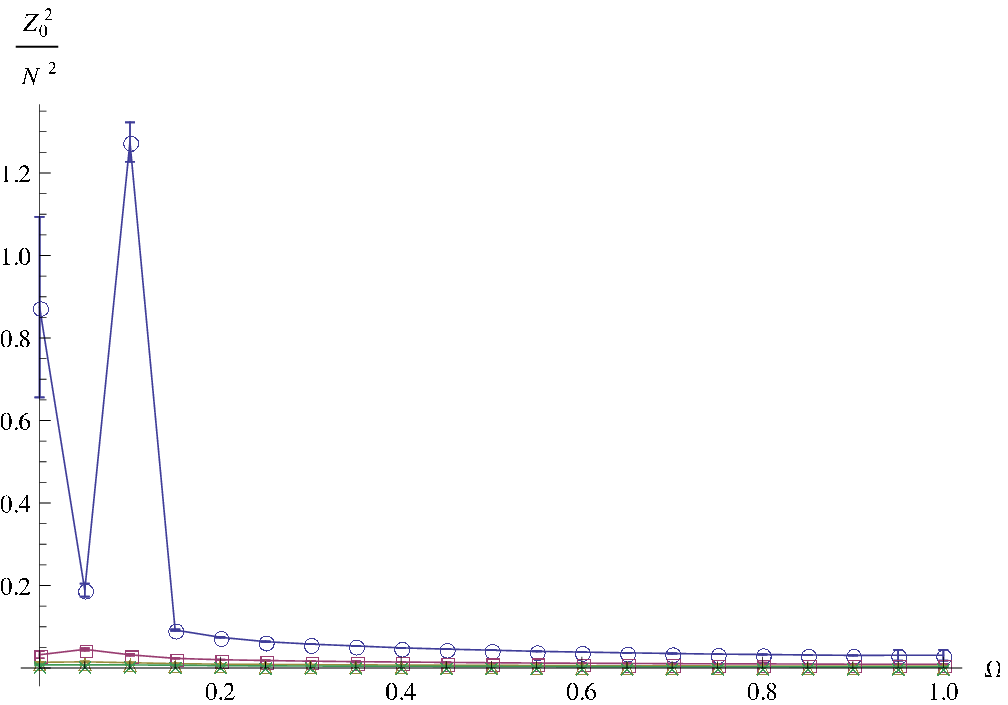}
\includegraphics[scale=0.40]{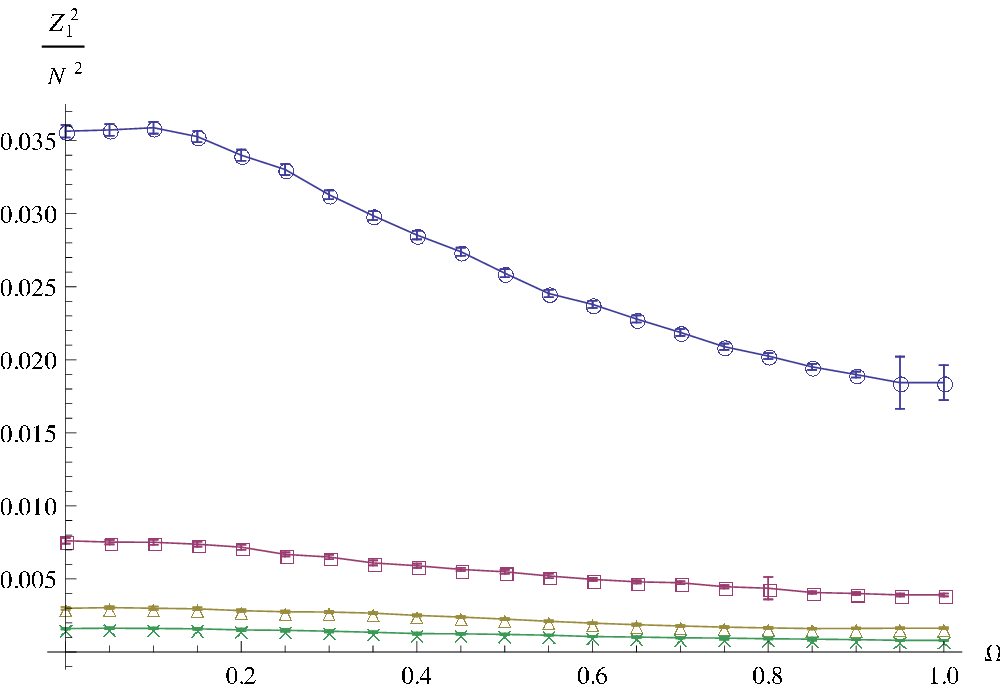}
\end{center}
\caption{\footnotesize Starting from the up left corner and from the left to the right the 	densities for $\varphi_a^2 $, $\varphi^2_0 $, $\varphi^2_1 $, $Z_{0a}^2$, $Z_{00}^2 $ and $Z_{01}^2$ for $\mu=0$ varying $\Omega$ and $N$. }\label{Figure 12}
\end{figure}
A complete different response of the system is described in the graphs for $\mu=3$, as we can see from fig.\ref{Figure 13}. The slope of total energy density is very similar to the $F$ component instead $D$. Beside, appears a maximum around  $\Omega=0.4$ for $N\to\infty$. This dramatic change of the graphs might be interpreted as consequence of a phase transition ruled  by the parameter $\mu$, actually in the next section we will find a peak in the specific heat density for some fixed $\Omega$ and varying $\mu\in[0,3]$. 
\begin{figure}[htb]
\begin{center}
\includegraphics[scale=0.45]{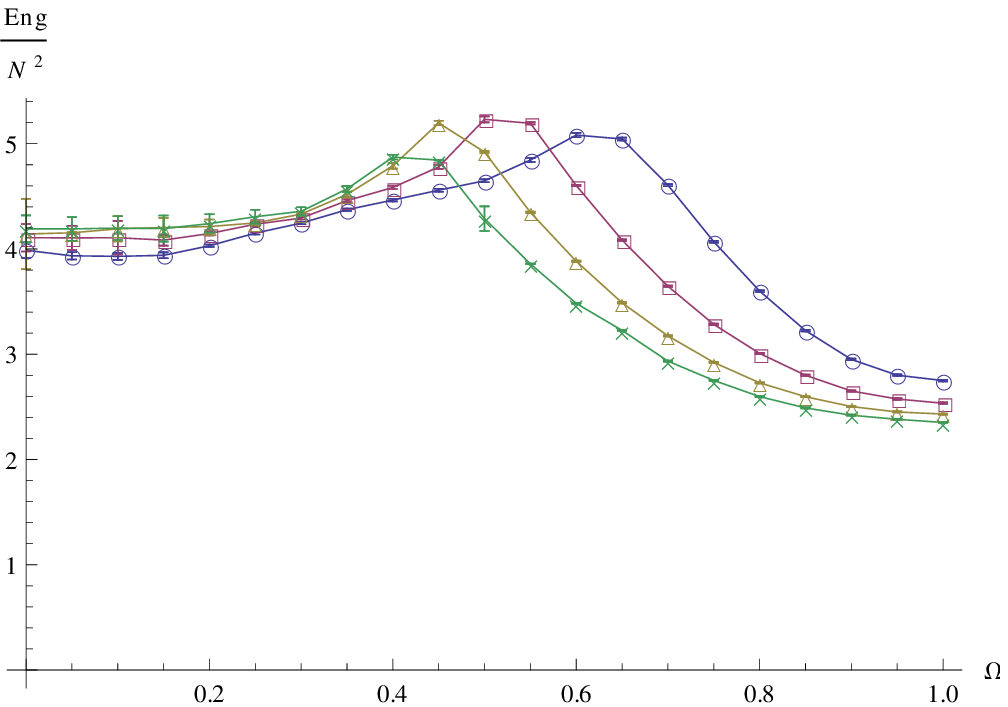}
\includegraphics[scale=0.45]{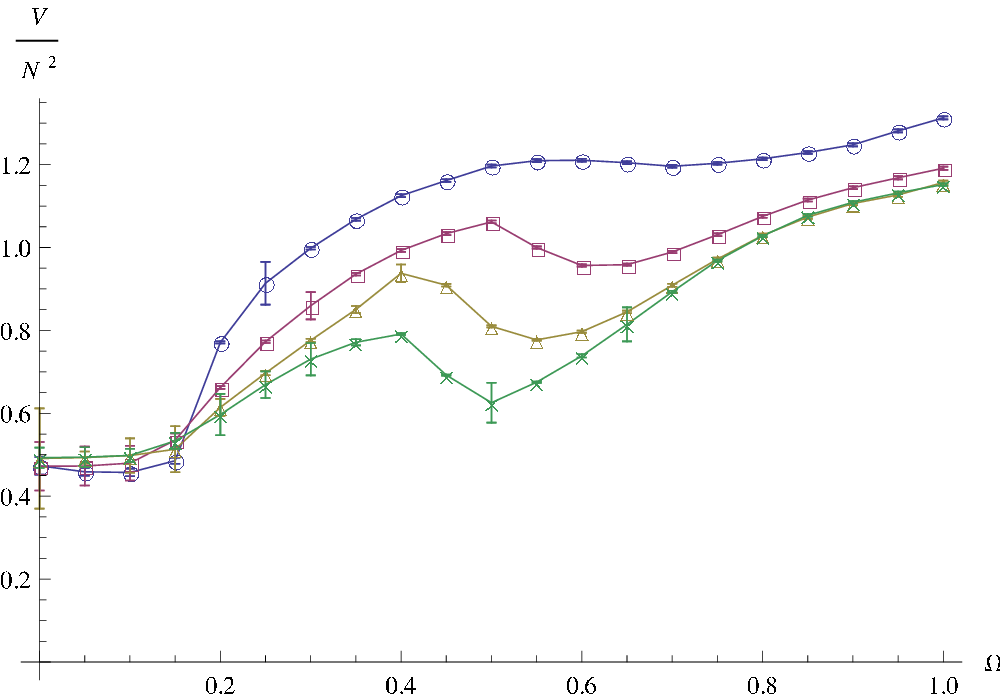}
\includegraphics[scale=0.45]{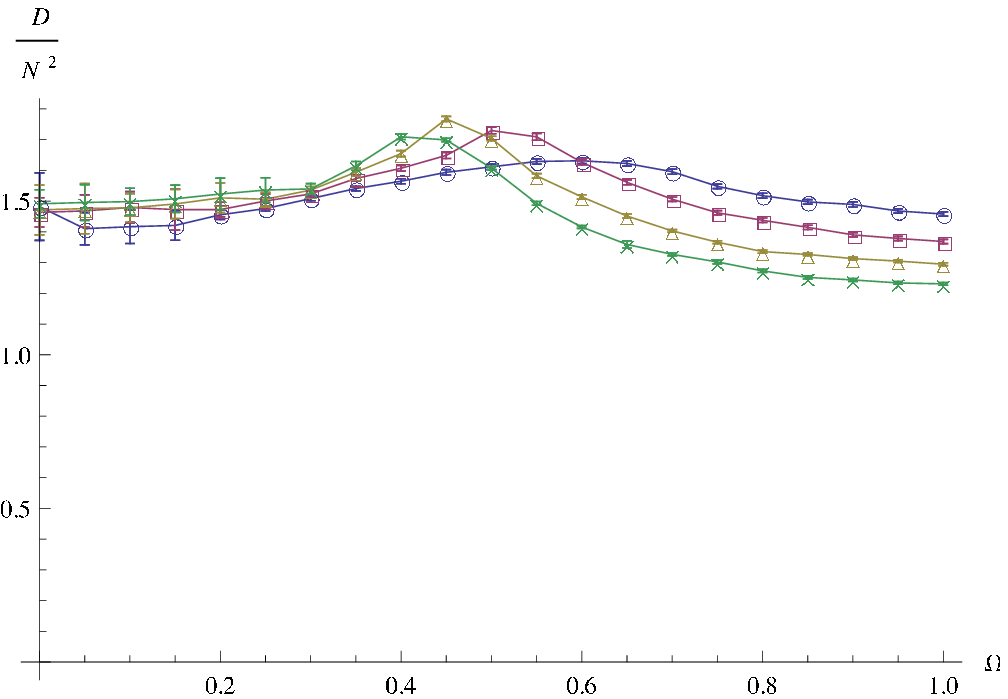}
\includegraphics[scale=0.45]{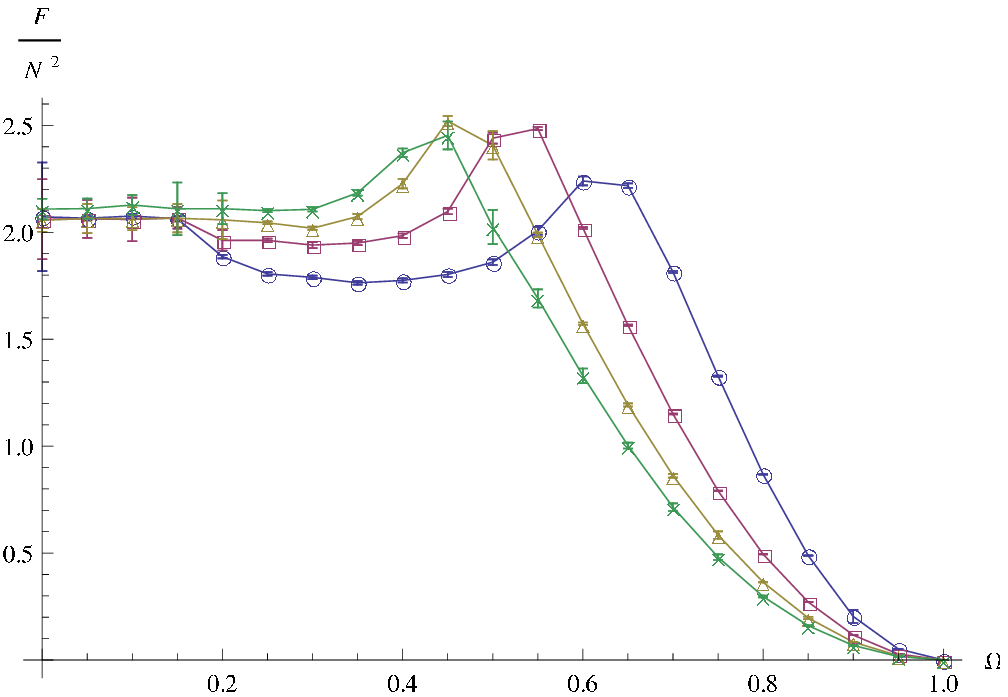}
\includegraphics[scale=0.45]{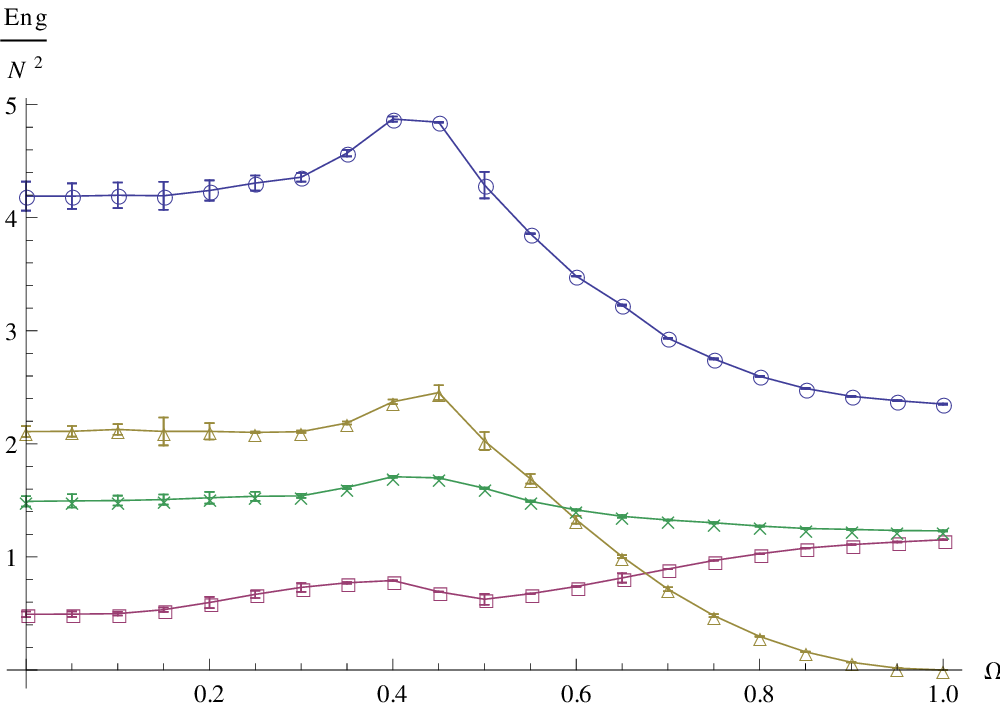}
\end{center}
\caption{\footnotesize Total energy density, various contributions and the comparison among them for $\mu=3$ varying $\Omega$ and $N$. From the left to the right  $E$, $V$, $D$, $F$ and comparison. \normalsize}\label{Figure 13}\end{figure}\newpage
Specific heat density displays fig.\ref{Figure 14} a strong change too, in fact instead the peak in $\Omega=1$, it  appears in the opposite side of the studied interval  in $\Omega=0$. This peak too, due to its grows  increasing $N$ could indicate a phase transition. 
\begin{figure}[htb]
\begin{center}
\includegraphics[scale=.6]{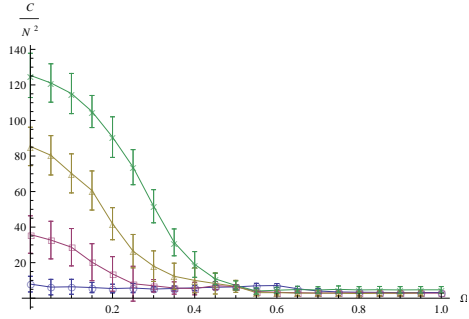}
\end{center}
\caption{\footnotesize Specific heat density for $\mu=3$ varying $\Omega$ and $N$.\normalsize}\label{Figure 14}
\end{figure}
\begin{figure}[htb]
\begin{center}
\includegraphics[scale=0.40]{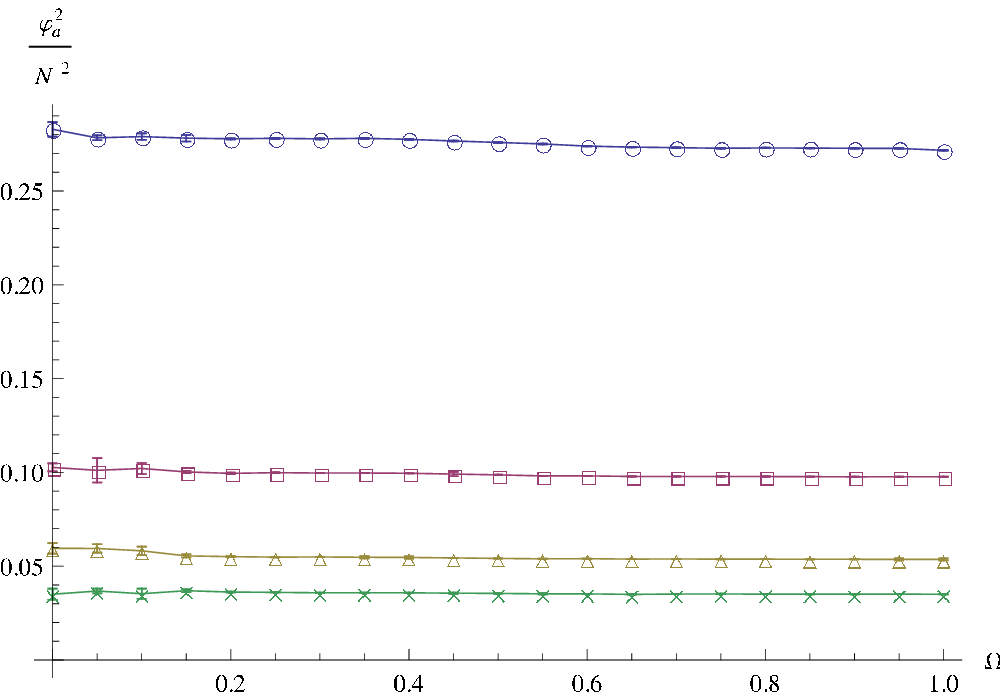}
\includegraphics[scale=0.40]{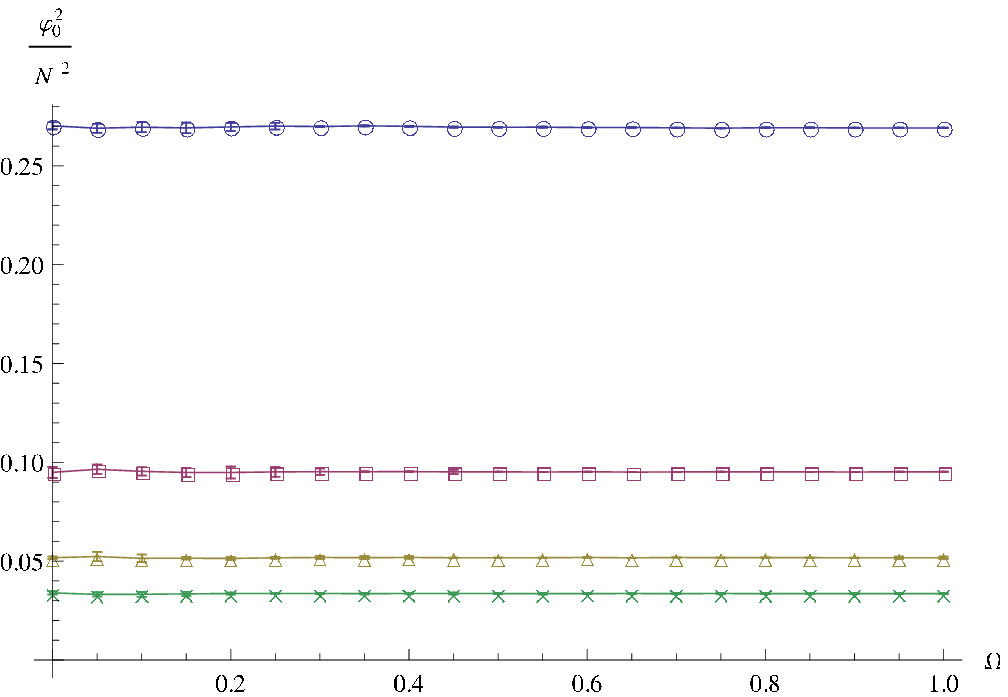}
\includegraphics[scale=0.40]{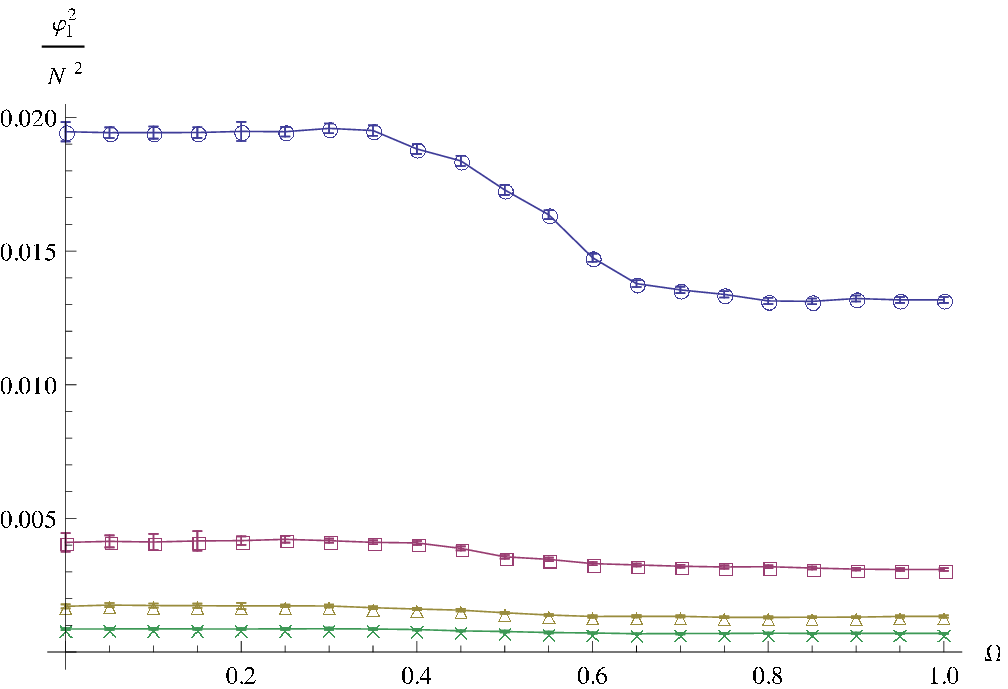}
\includegraphics[scale=0.40]{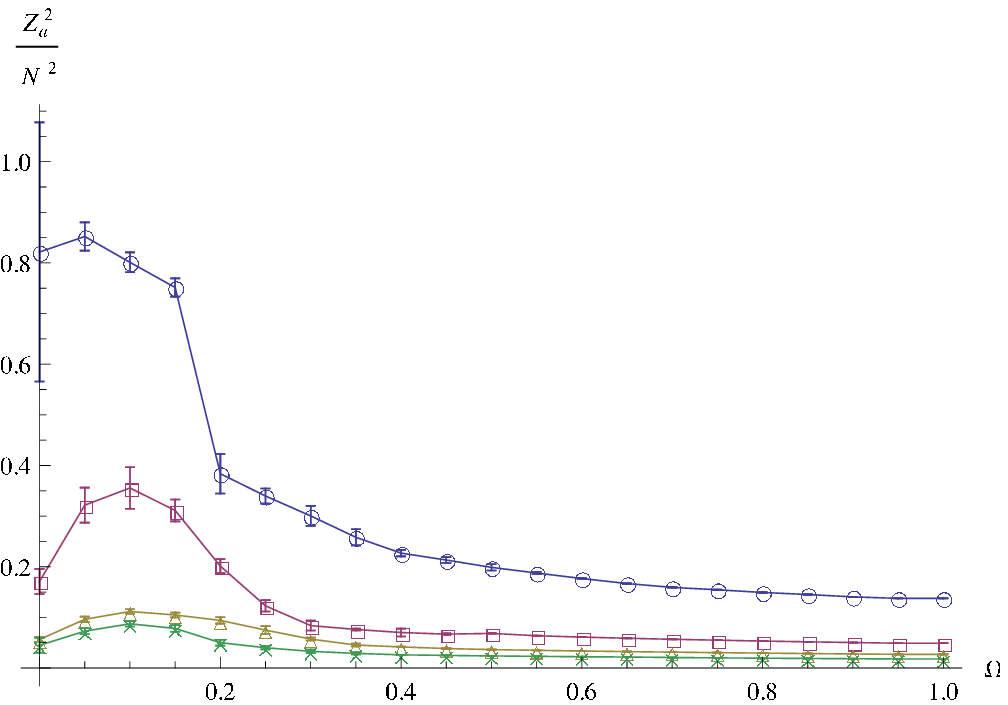}
\includegraphics[scale=0.40]{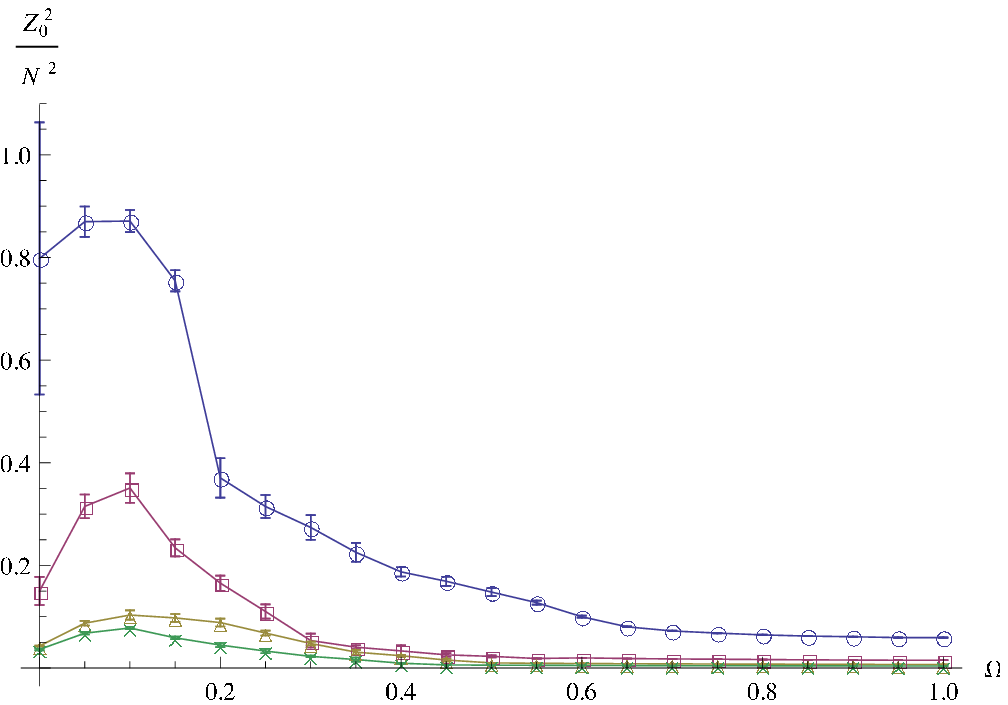}
\includegraphics[scale=0.40]{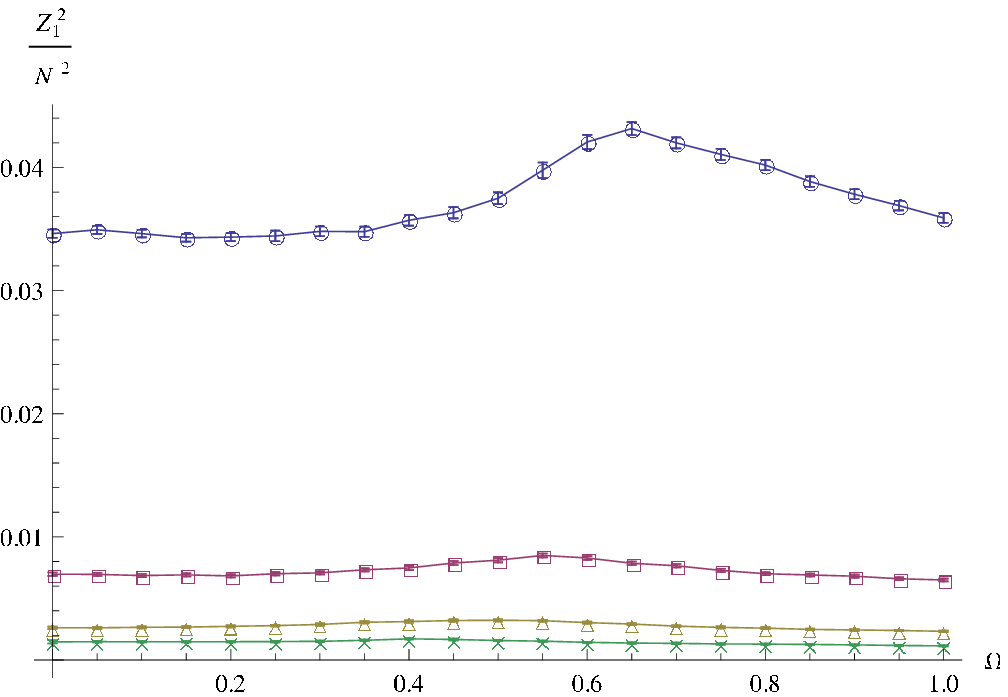}
\end{center}
\caption{\footnotesize Starting from the up left corner and from the left to the right the 	densities for $\varphi_a^2 $, $\varphi^2_0 $, $\varphi^2_1 $, $Z_{0a}^2$, $Z_{00}^2 $ and $Z_{01}^2$ for $\mu=3$ varying $\Omega$ and $N$.\normalsize}\label{Figure 15}
\end{figure}
The fig.\ref{Figure 15} describes the behavior of the order parameters 	densities $\varphi_a^2 $, $\varphi^2_0 $, $\varphi^2_1 $ and $Z_{0a}^2$, $Z_{00}^2 $, $Z_{01}^2$, they have a similar aspect to the previous relative graphs. For the $\psi$ field the spherical contribution remains dominant, beside  in the $\varphi^2_1 $ graph appears a deviation from the constant slope this deviation is evident for $N=5$ but still there for higher $N$. The order parameters for $Z_0$ display a peak close to the origin without oscillations even for $N=5$. This maximum for higher $N$ does not move closer to the origin, in other words, this shift in not due to the finite volume effect. Even for  $Z^2_{01} $ graph appears a deviation from the constant slope, a small peak which becomes shifted and smoother for higher $N$ 

\subsection{Varying $\mu$ }
\begin{figure}[htb]
\begin{center}
\includegraphics[scale=0.45]{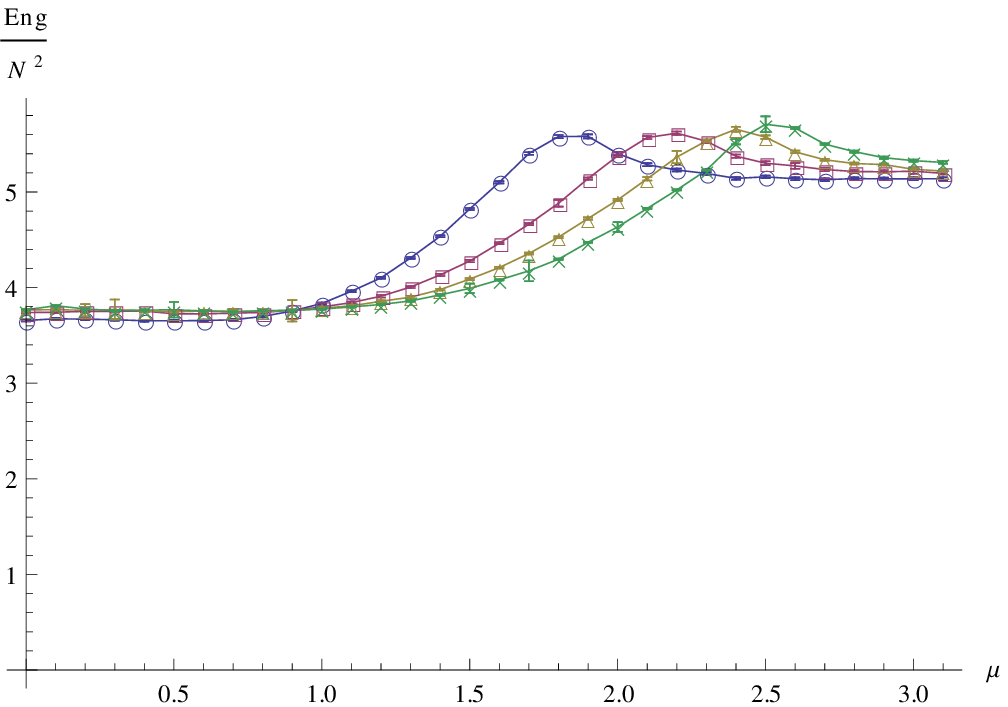}
\includegraphics[scale=0.45]{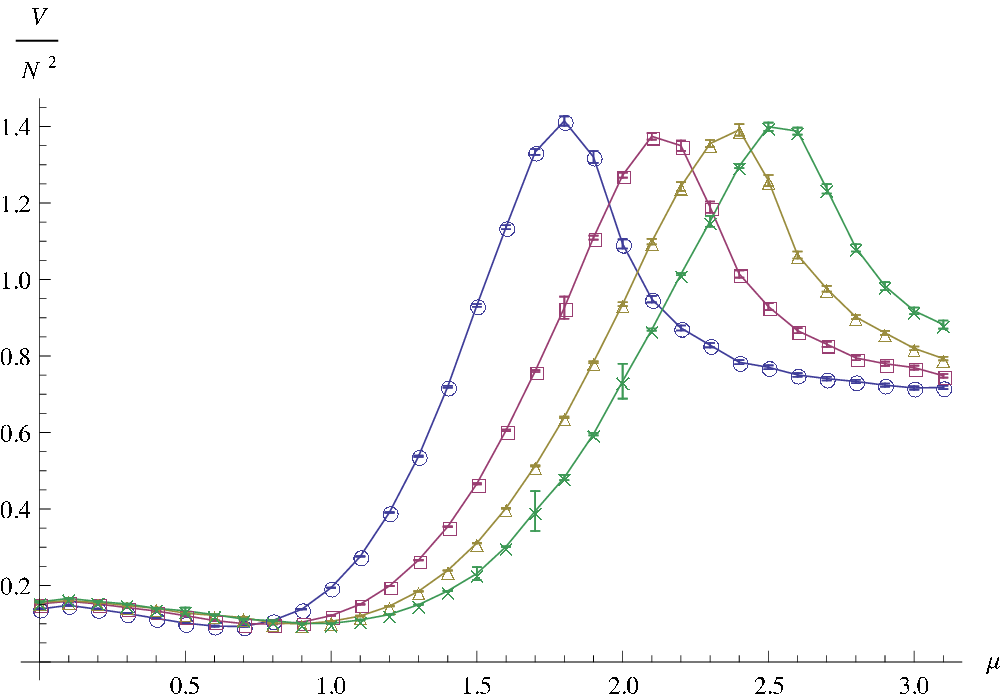}
\includegraphics[scale=0.45]{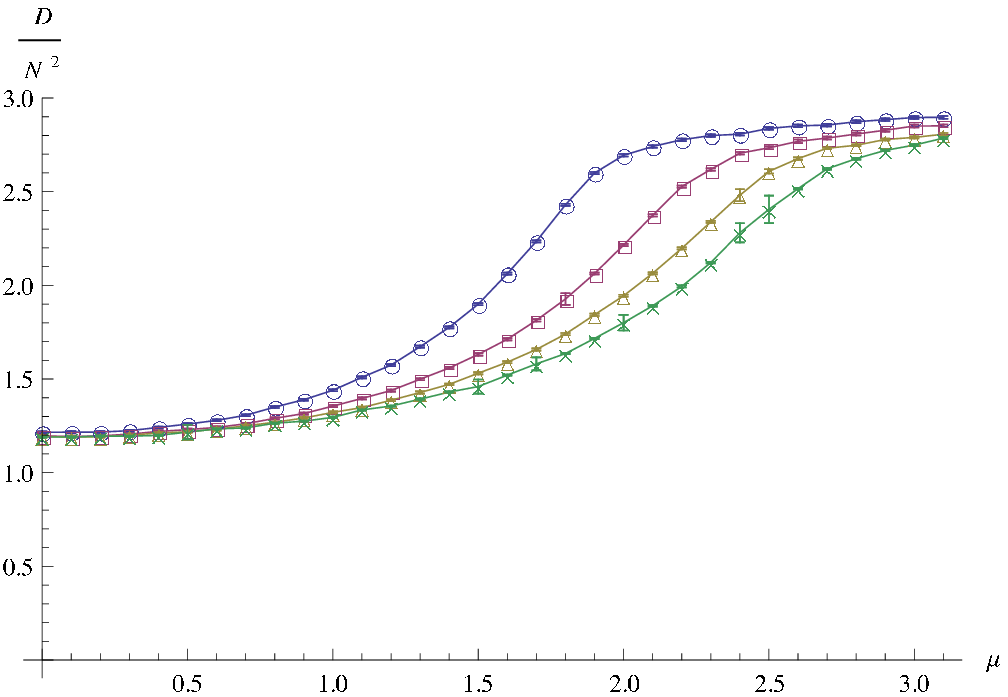}
\includegraphics[scale=0.45]{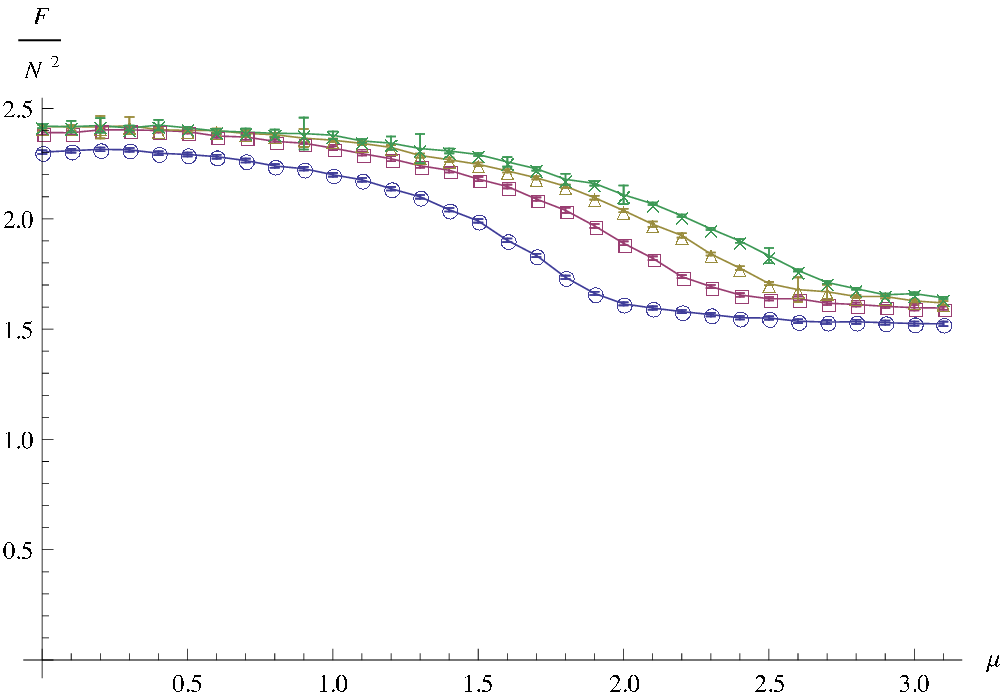}
\includegraphics[scale=0.45]{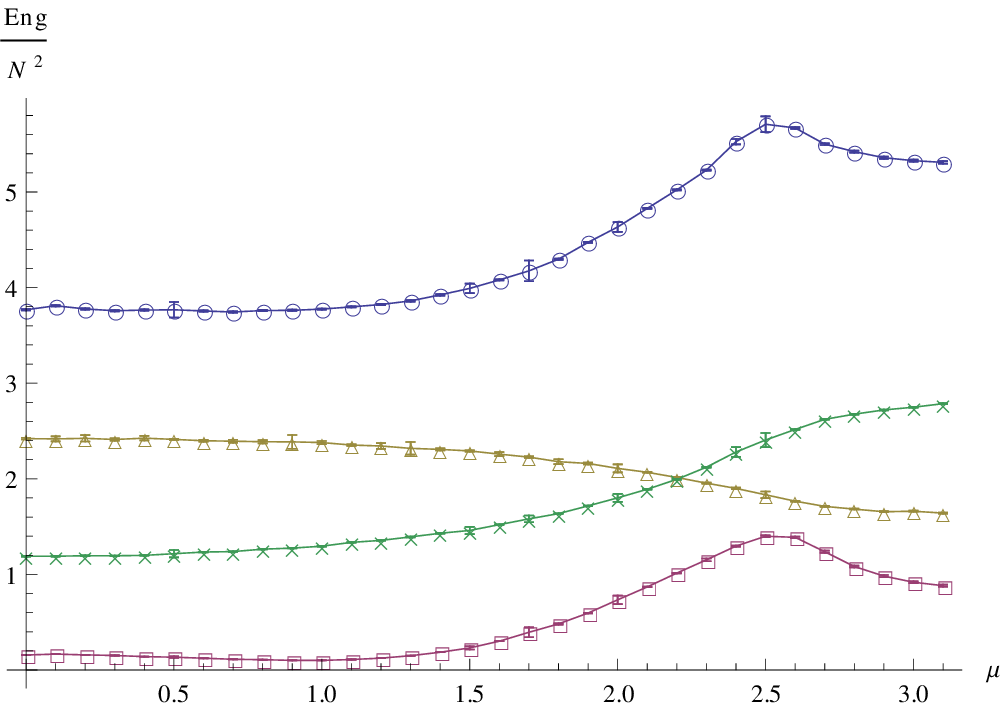}
\end{center}
\caption{\footnotesize The total energy density and the various contributions for $\Omega=0$ varying $\mu$ and $N$. From the left to the right  $E$, $V$, $D$, $F$ an comparison with $N=5$ (circle), $N=10$ (square), $N=15$ (triangle), $N=20$ (cross). For the 	comparison: $E$ (circle), $V$ (square), $D$ (triangle), $F$ (cross).\normalsize}\label{Figure 16}\end{figure}

In this section is analyzed  the response of the system varying $\mu\in[0,3]$ where $\Omega$ is fixed to $0,0.5,1$ and $\alpha$ is always zero. We start displaying the graphs fig.\ref{Figure 16} of the total energy density and of various contributions for $\Omega=0$. There is no evident discontinuity but appears a peak in the total energy density around $\mu\approx 2.5$ for $N=20$. Comparing all the contributions is easy to notice that the slope of the total energy is dictated by the curve $V$ of the potential part.    
\begin{figure}[htb]
\begin{center}
\includegraphics[scale=0.6]{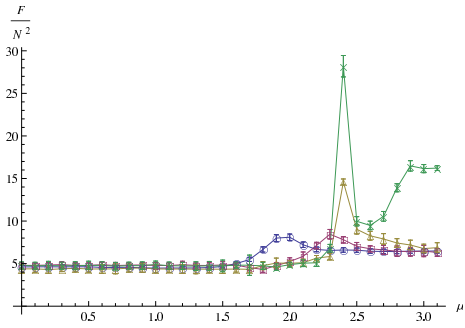}
\end{center}
\caption{\footnotesize Specific heat density for $\Omega=0$ varying $\mu$ and $N$.\normalsize}\label{Figure 17}
\end{figure}

\begin{figure}[htb]
\begin{center}
\includegraphics[scale=0.4]{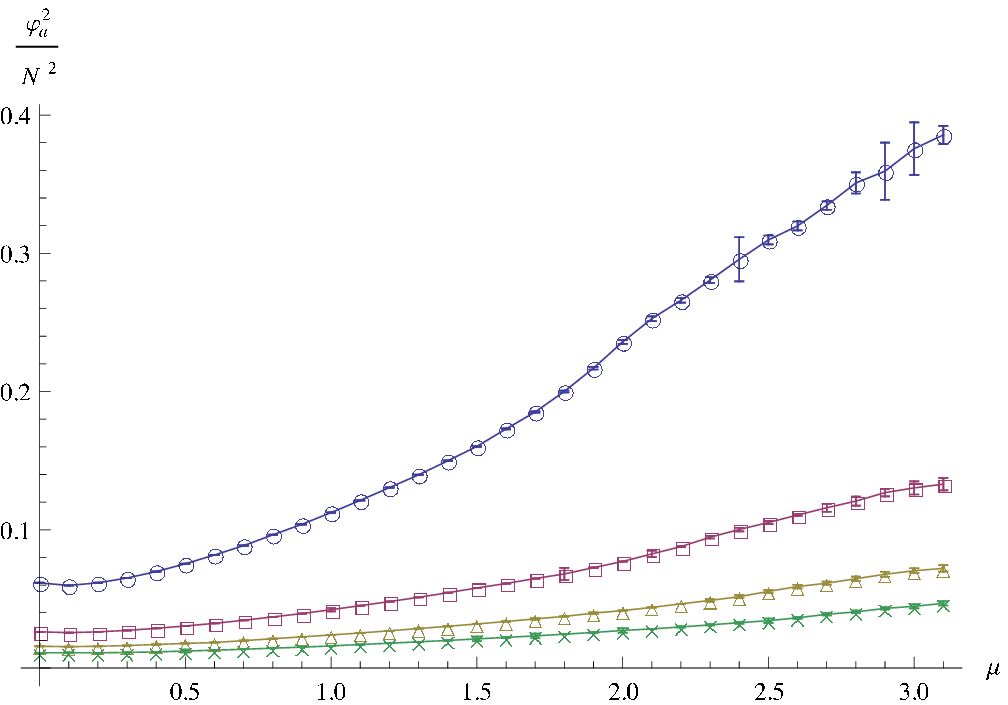}
\includegraphics[scale=0.4]{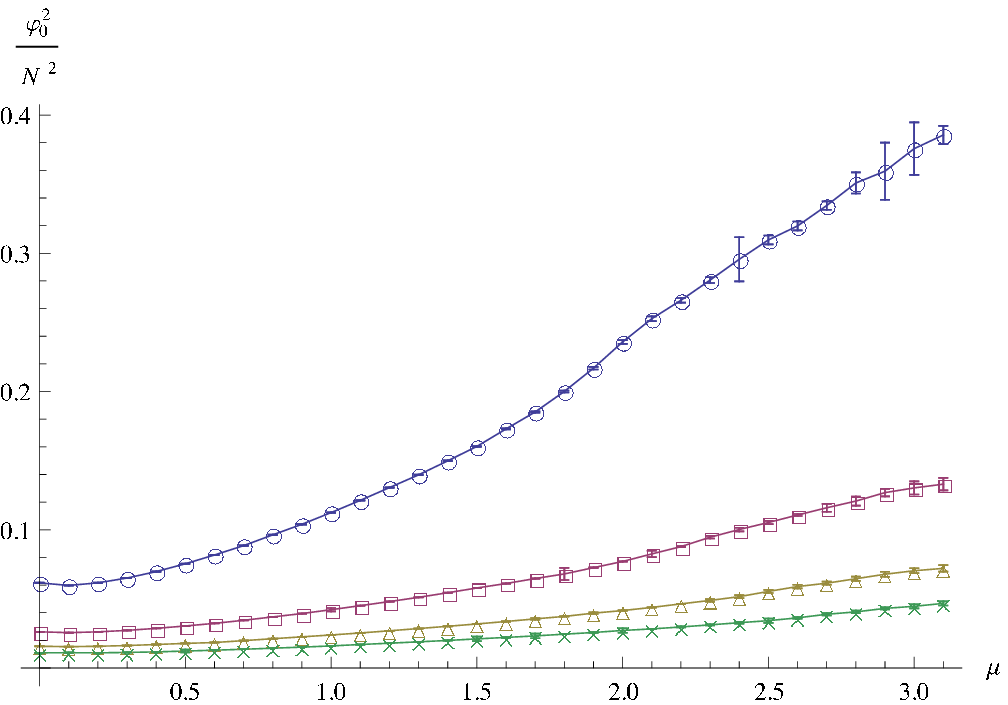}
\includegraphics[scale=0.4]{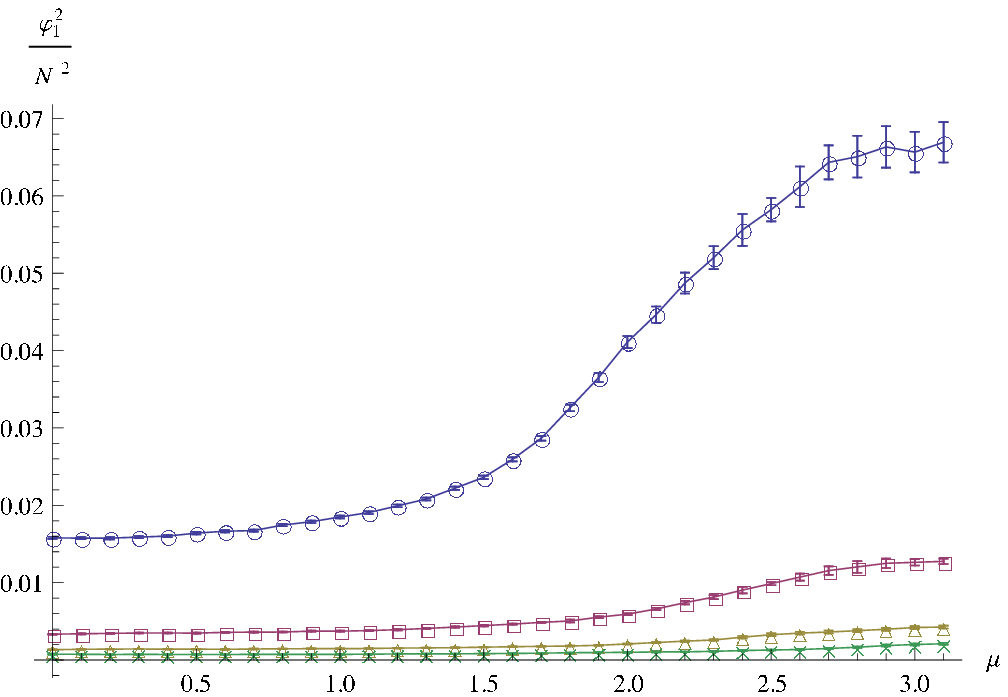}
\includegraphics[scale=0.4]{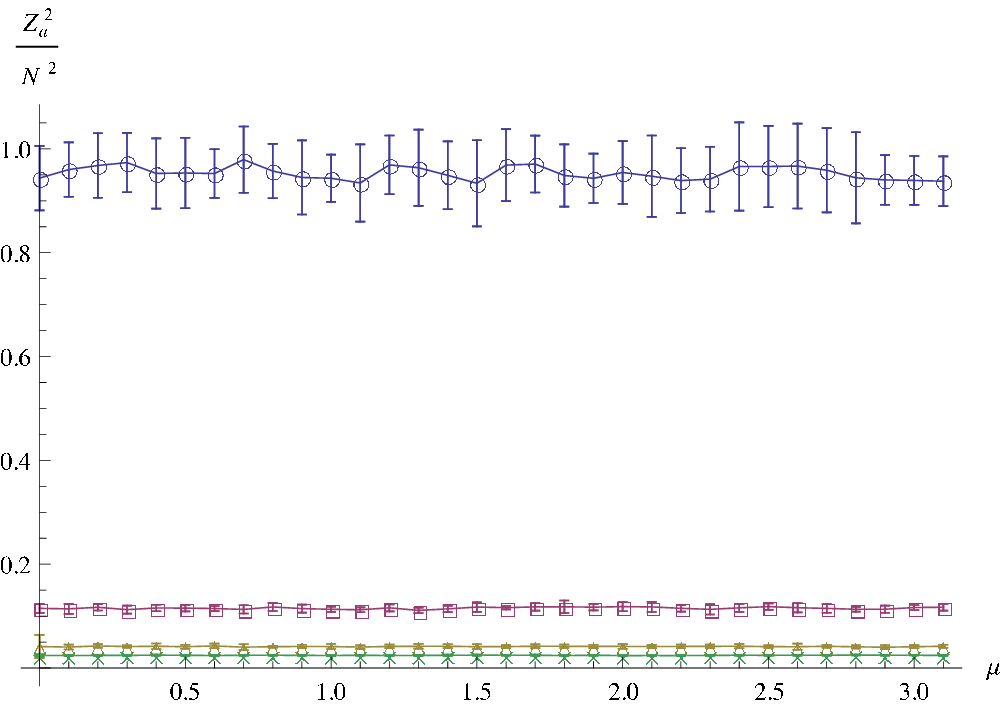}
\includegraphics[scale=0.4]{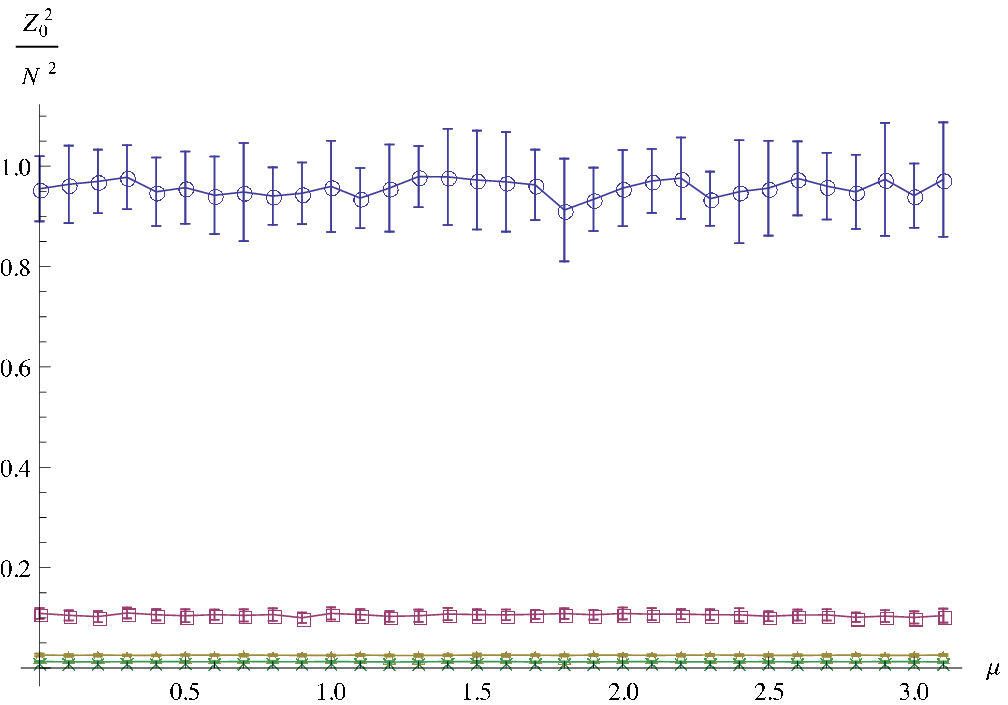}
\includegraphics[scale=0.4]{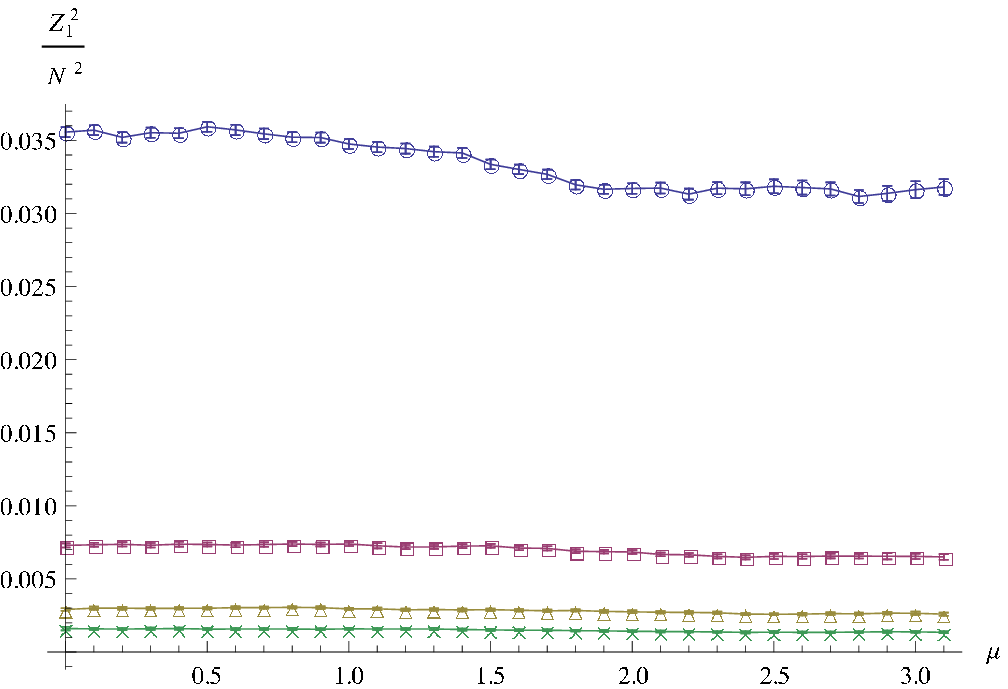}
\end{center}
\caption{\footnotesize Starting from the up left corner and from the left to the right the 	densities for $\varphi_a^2 $, $\varphi^2_0 $, $\varphi^2_1 $, $Z_{0a}^2$, $Z_{00}^2 $ and $Z_{01}^2$ for $\Omega=0$ varying $\mu$ and $N$. \normalsize}\label{Figure 18}
\end{figure}

As mentioned before, the specific heat density fig.\ref{Figure 17} features a peak  around $\mu\approx2.5$ for $N=20$ and again, due to this behavior as $N$ increase, we could relate this peak to a phase transition.
The plots for the quantities $\varphi_a^2 $ and $\varphi^2_0 $ denote a strong dependence on $\mu$, in particular the slope of $\varphi^2_0 $ seems mostly linear,  $\varphi^2_1 $ related graphs feature a similar behavior but the slope is no longer linear. Comparing the three graphs fig.\ref{Figure 18} we deduce that close to the origin
 the non spherical contribution is bigger the spherical  one, increasing $\mu$ this situation capsizes and $\varphi^2_0 $ becomes dominant respect 
$\varphi^2_1 $. 
The behavior of the $Z_0$ fields is quite different, referring to figure \ref{Figure 18}, the spherical contribution is always dominant for the all interval $\mu\in[0,3]$. The curves for $Z_{0a}^2$, $Z_{00}^2 $ are compatible to the constant slope, for $Z_{01}^2 $ we have  the same dependence on $\mu$ in particular there is a smooth descending step, however this step becomes smoother for bigger $N$. It is behooves to say that due to some cancellations effects the statistical errors are quite big and they can hide some dependence, anyway this results tell us about the dependence of the order parameter for $Z_i$ and in general of the system, on the two  choice $\Omega=0$ or $\Omega\neq0$.

\begin{figure}[htb]
\begin{center}
\includegraphics[scale=0.4]{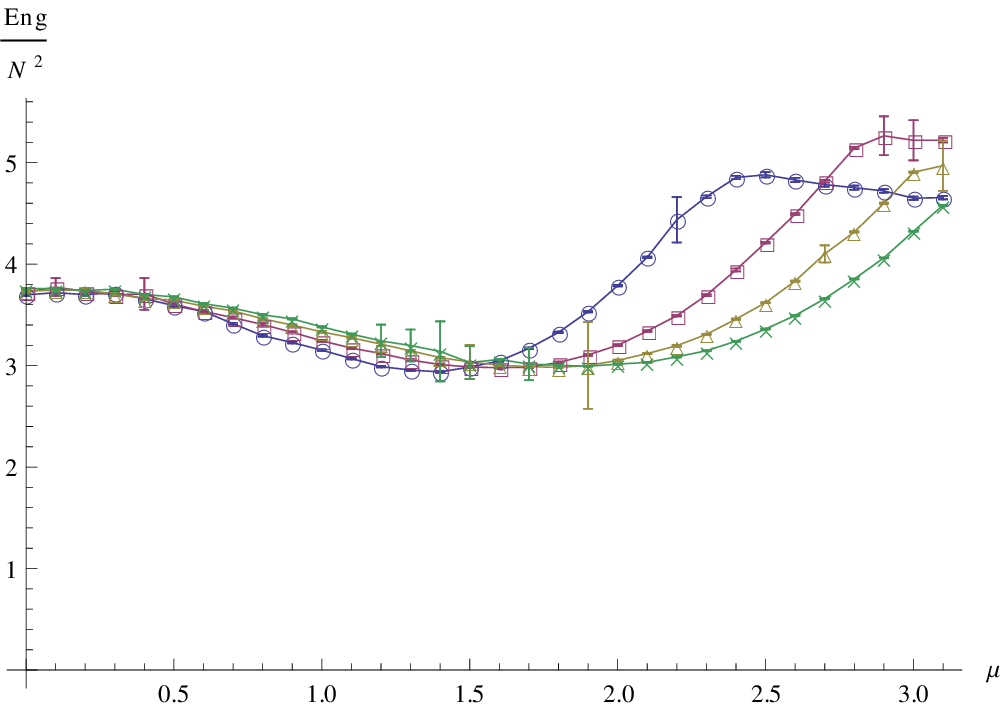}
\includegraphics[scale=0.4]{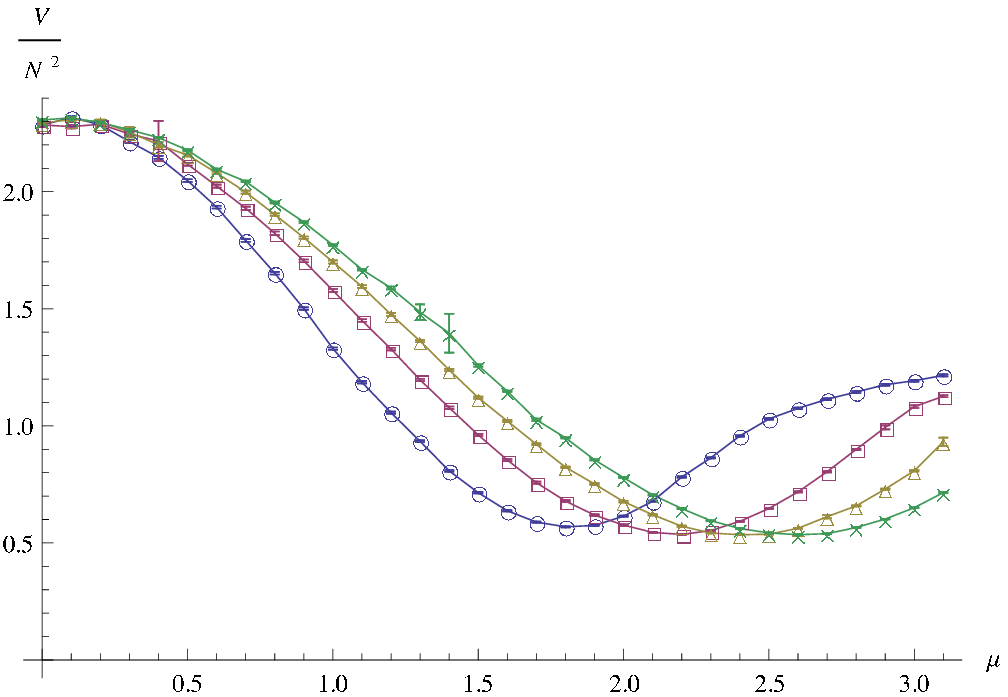}
\includegraphics[scale=0.4]{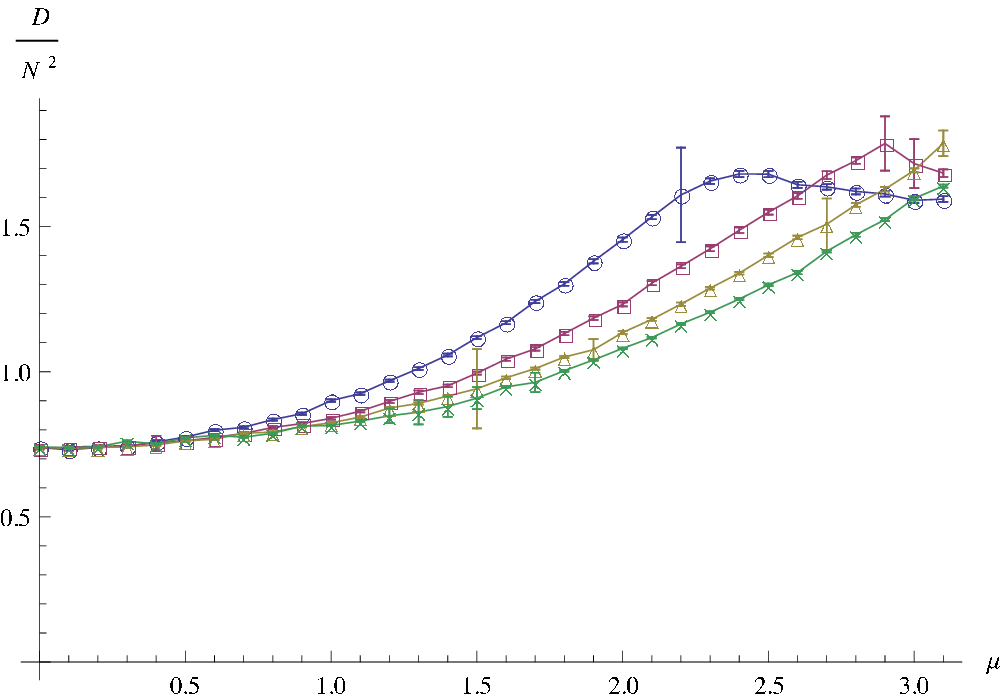}
\includegraphics[scale=0.4]{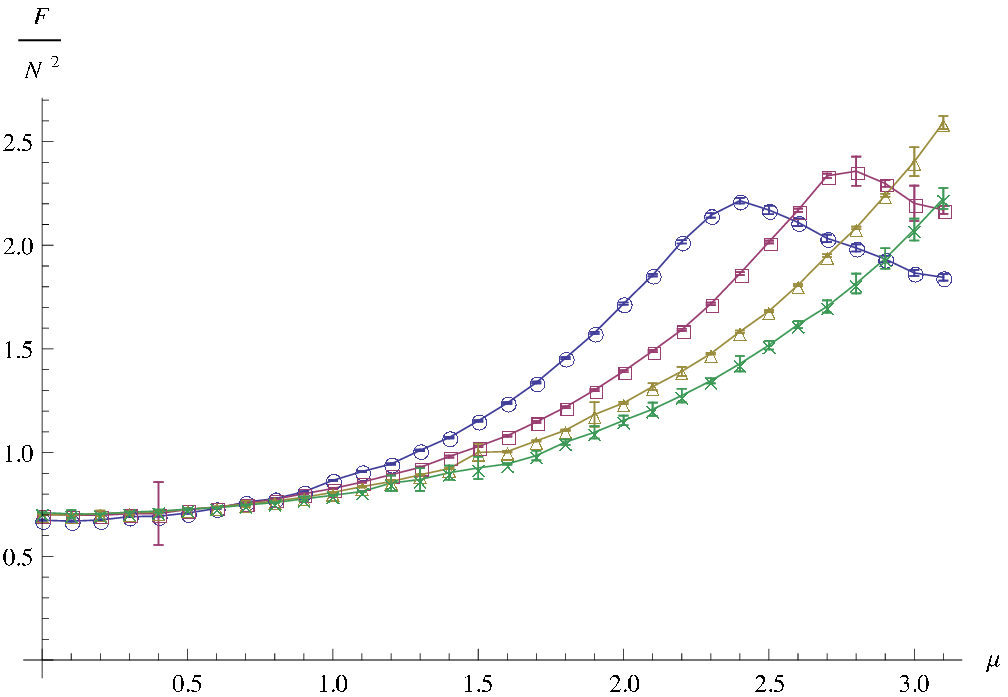}
\end{center}
\caption{\footnotesize Total energy density and contributions for $\Omega=0.5$ varying $\mu$ and $N$. From the left to the right  $E$, $V$, $D$, $F$.\normalsize}\label{Figure 19}\end{figure}

\begin{figure}[htb]
\begin{center}
\includegraphics[scale=0.7]{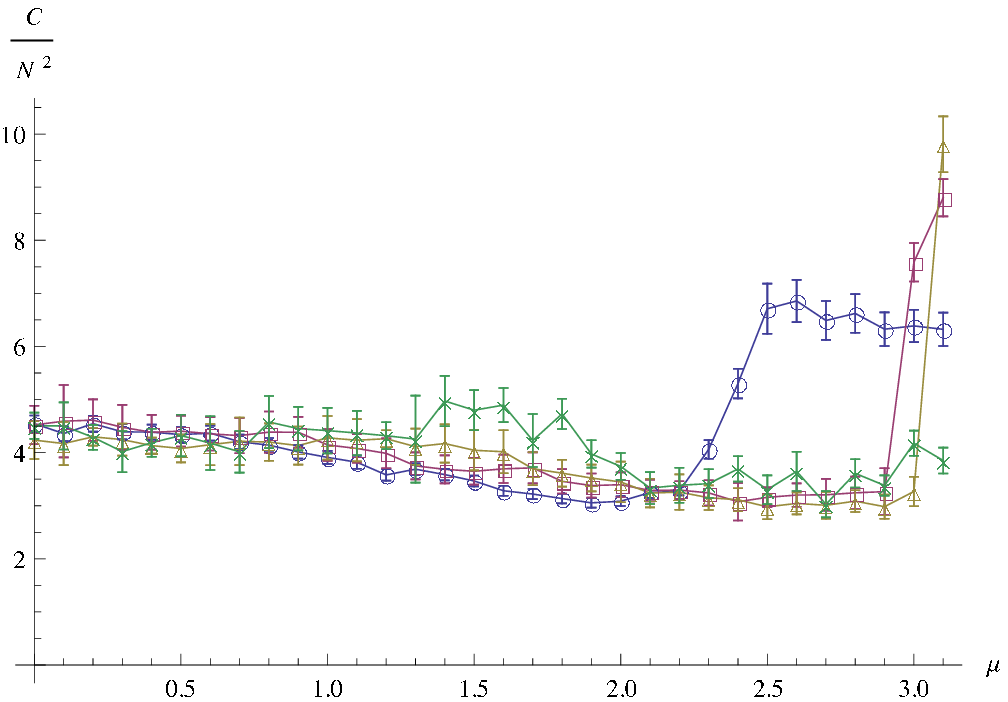}
\end{center}
\caption{\footnotesize Specific heat density for $\Omega=0.5$ varying $\mu$ and $N$.\normalsize}\label{Figure 20}
\end{figure}

Now we will analyze the  model for $\Omega=0.5$; as fig.\ref{Figure 19} shows the graphs have a different slope comparing to the previous case, the maximum of total energy density follow the  one of the $V$ component. If we focus ourself only on the total energy graph and we compare it with the one for $\Omega=0$, we notice a shift of the maximum for each $N$. In particular in fig.\ref{Figure 19} some maximum are moved outside the considered interval. 
We can find this shift very clearly looking at specific heat density graph fig.\ref{Figure 20}, we find again the peak as $N$ increase but it is shifted around $\mu\approx 3.3$. 
The  graphs fig.\ref{Figure 23} for $\varphi_a^2 $, $\varphi^2_0 $ have the same 
behavior of $\Omega=0.5$ case, excluding some fluctuations close to the origin for $N=5$ due to the finite volume effect, $\varphi^2_1 $ graph displays an almost constant curve. However, close to the origin, the spherical contribution and the first non spherical one are comparable.
\begin{figure}[htb]
\begin{center}
\includegraphics[scale=0.40]{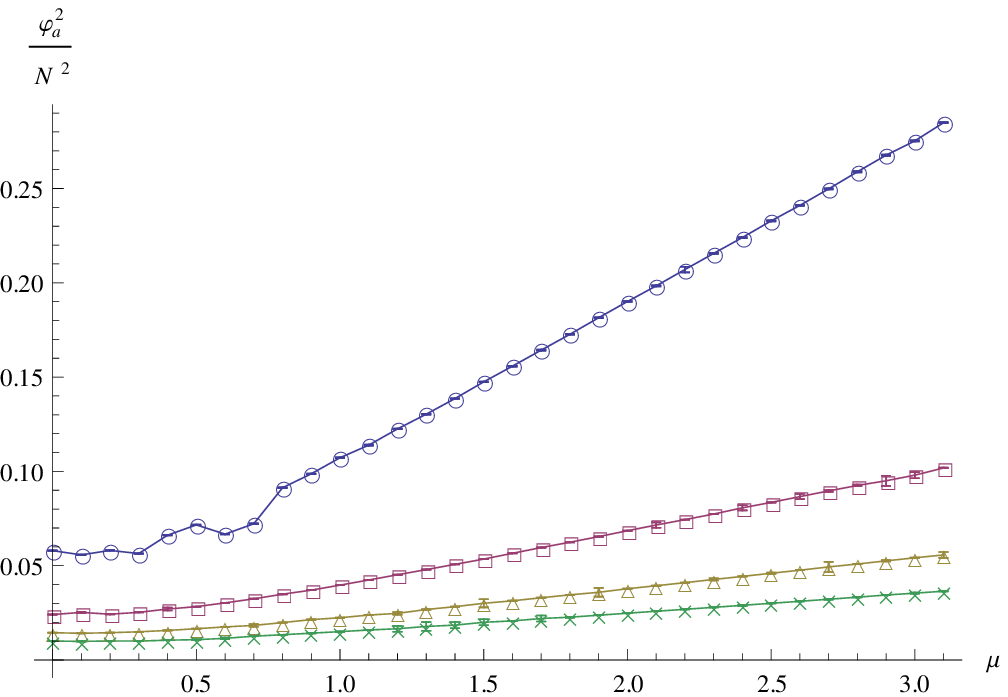}
\includegraphics[scale=0.40]{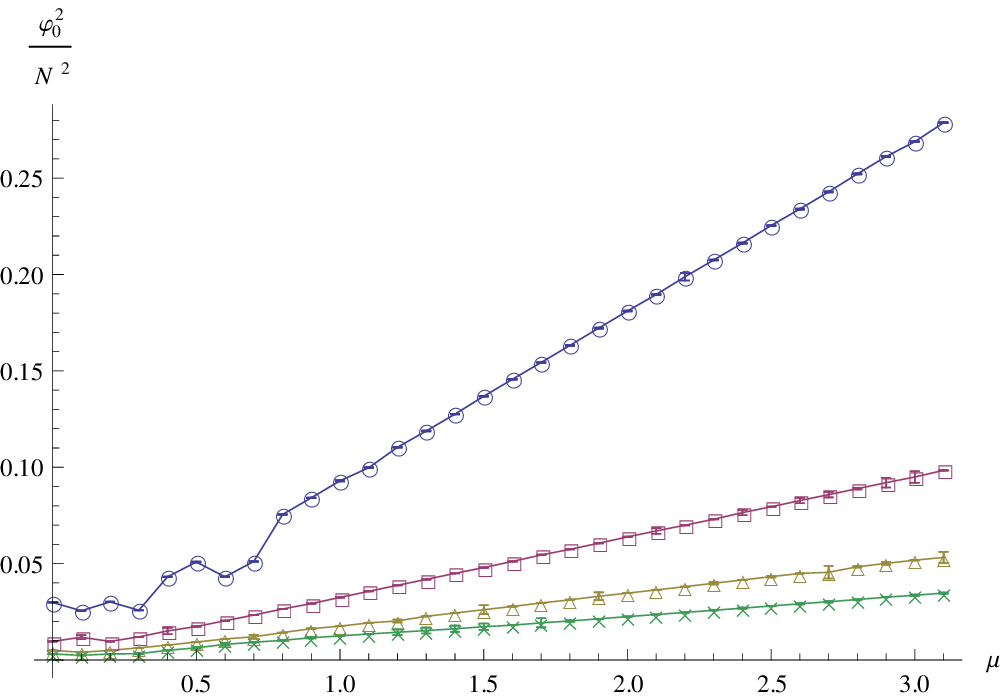}
\includegraphics[scale=0.40]{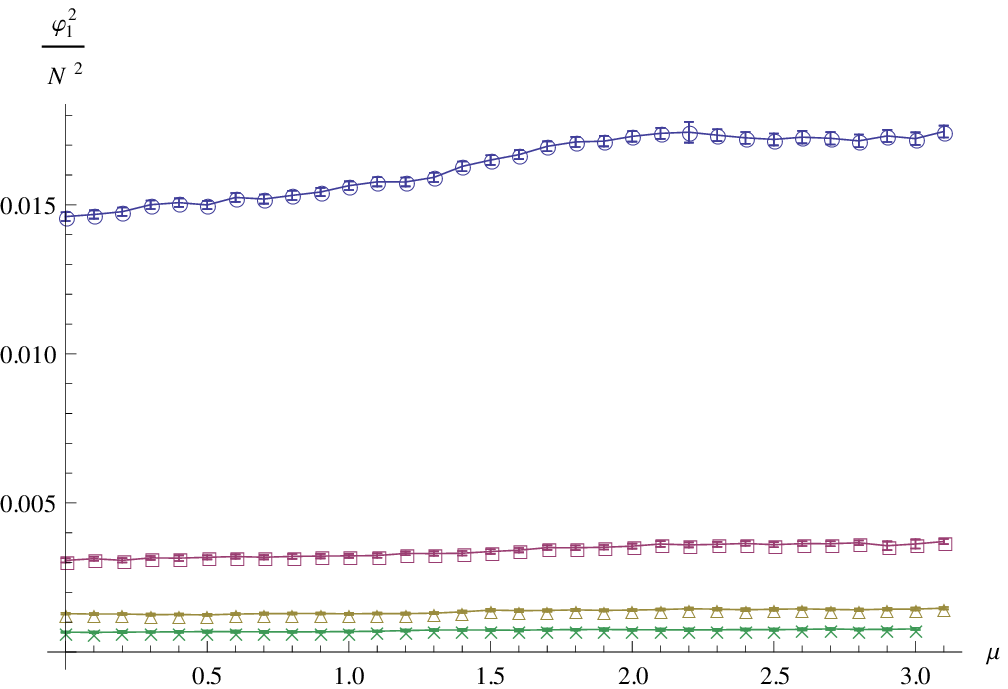}
\includegraphics[scale=0.40]{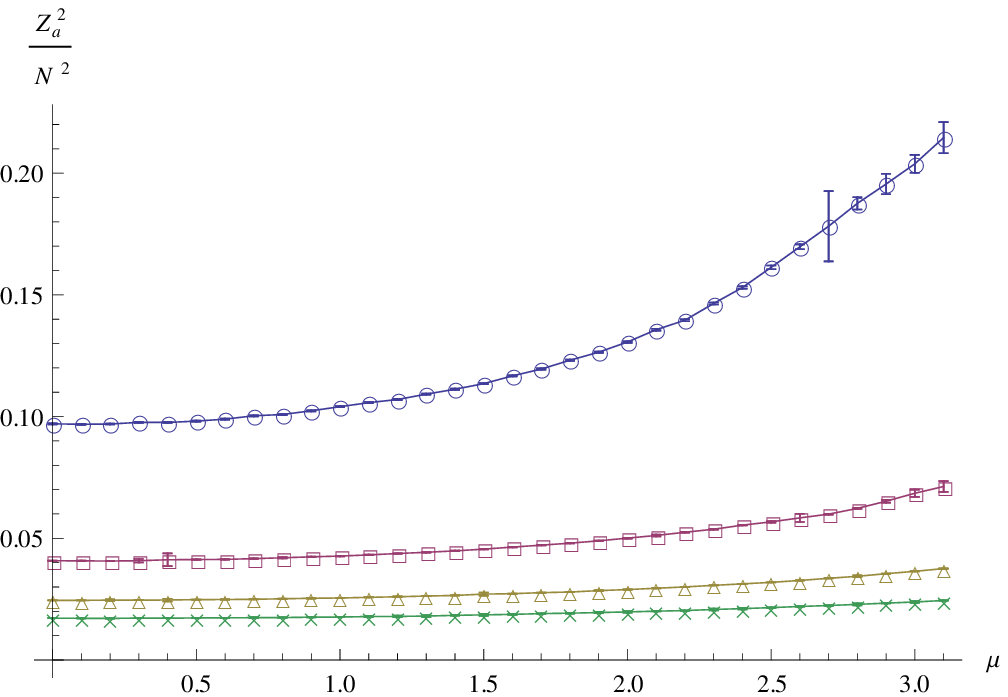}
\includegraphics[scale=0.40]{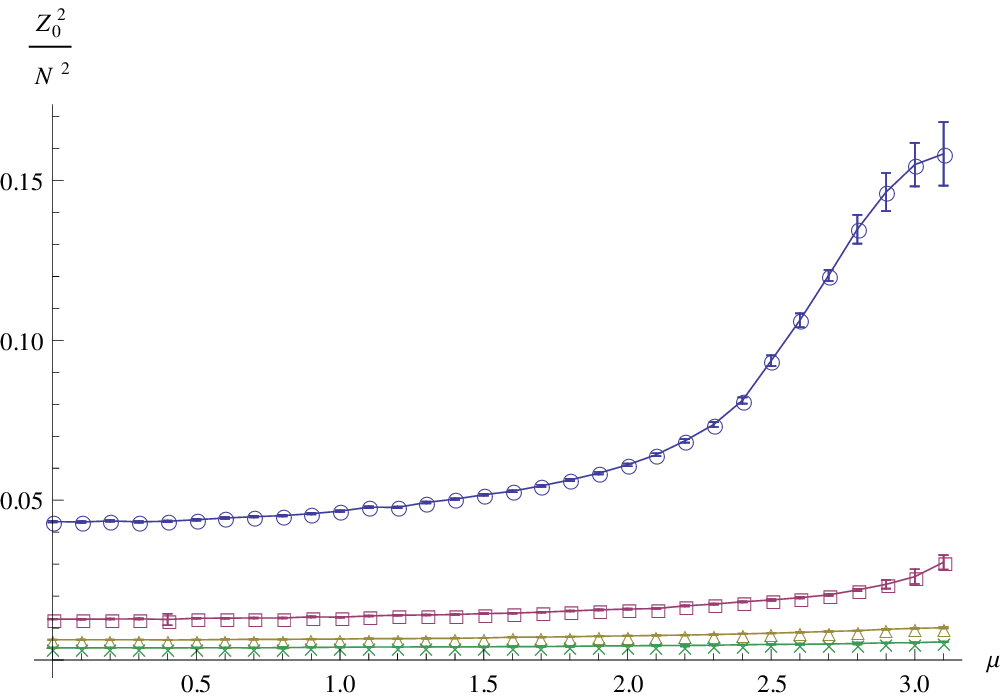}
\includegraphics[scale=0.40]{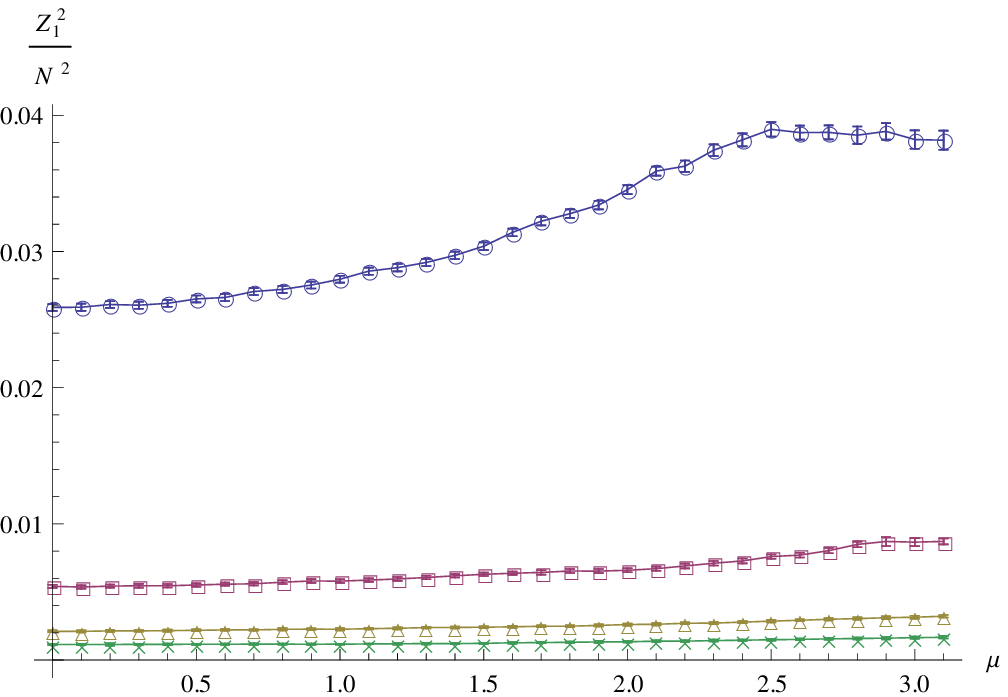}
\end{center}
\caption{\footnotesize Starting from the up left corner and from the left to the right the 	densities for $\varphi_a^2 $, $\varphi^2_0 $, $\varphi^2_1 $, $Z_{0a}^2$, $Z_{00}^2 $ and $Z_{01}^2$ for $\Omega=0.5$ varying $\mu$ and $N$.\normalsize}\label{Figure 21}
\end{figure}
The introduction of $\Omega\neq 0$ creates, in the $Z_0$ fields order parameters fig.\ref{Figure 21}, a dependence similar to the graphs for the $\psi$; the full power of the field density and the spherical contribution are no more constant and they grow increasing $\mu$. Even in this case the spherical contribution is always dominant excluding the region around $\mu=0$. 

\newpage
The last set of graphs for the 4-dimensional model are obtained fixing   $\Omega=1$, due to the vanishing of prefactor in front of the Yang-Mills part of the action  the $F$ contribution is always zero. The following diagrams for the energy and contributions show the absence of the previous peak and comparing again them with former graphs they seem a sort dilatation. 
\begin{figure}[htb]
\begin{center}
\includegraphics[scale=0.4]{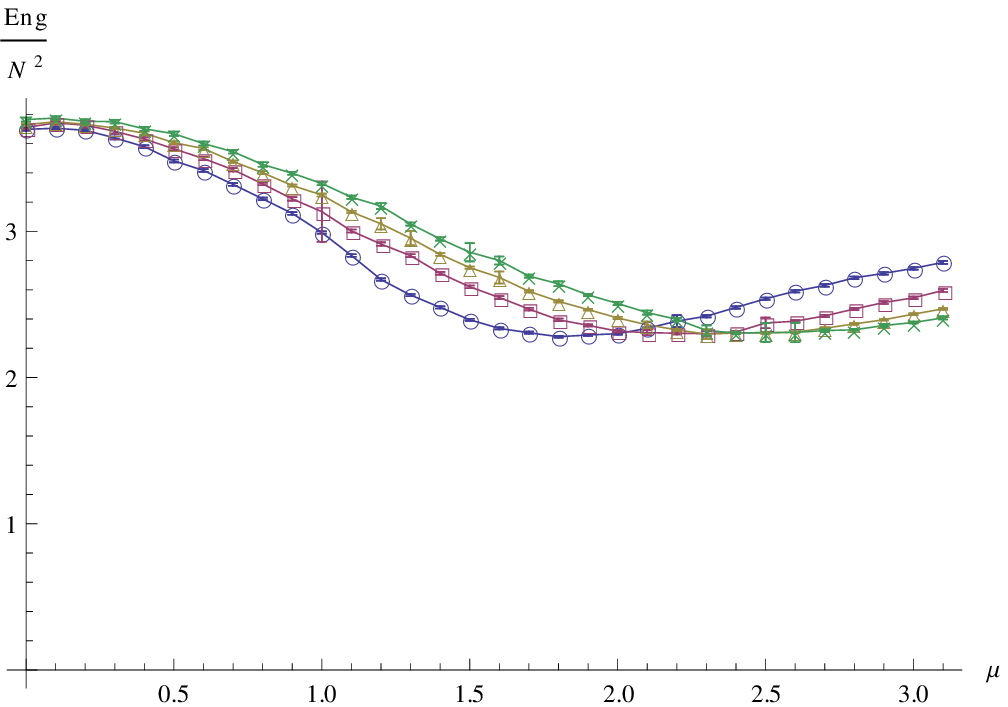}
\includegraphics[scale=0.4]{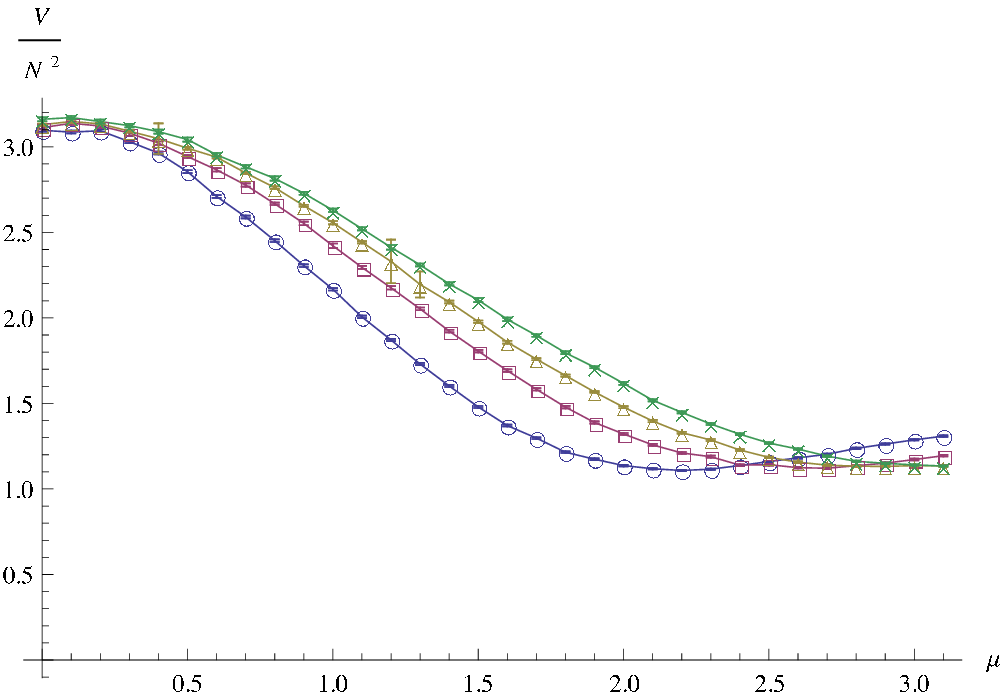}
\includegraphics[scale=0.4]{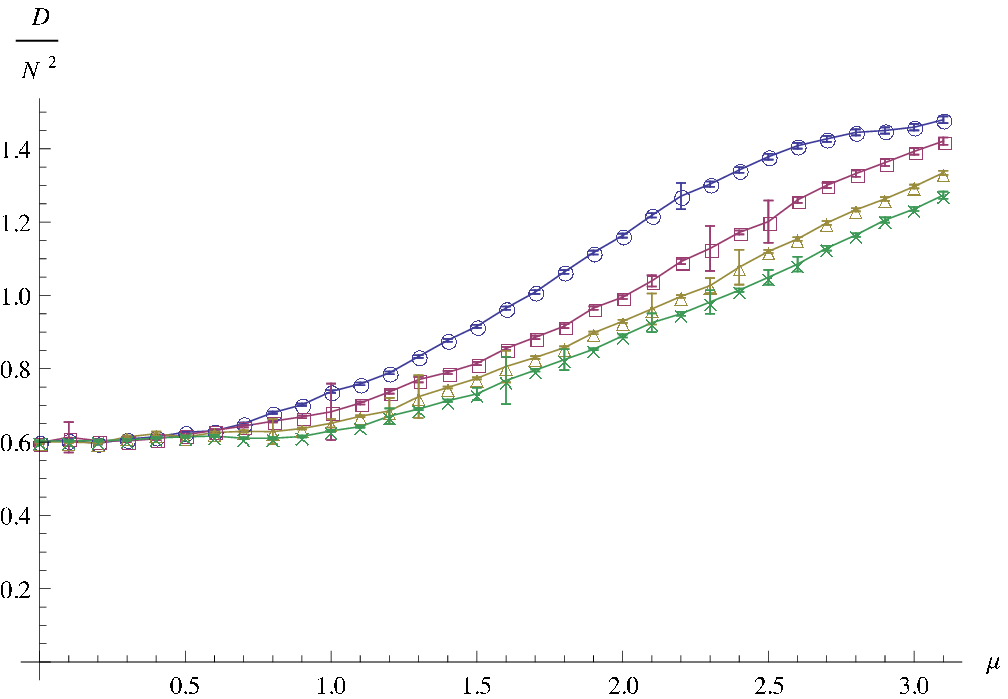}
\end{center}
\caption{\footnotesize Total energy density and the various contributions  for $\Omega=1$ varying $\mu$ and $N$. From the left to the right  $E$, $V$, $D$. \normalsize}\label{Figure 22}\end{figure}
The specific heat density does not show the peak in zero any more fig.\ref{Figure 23} and the curves does not show any particular point as $N$ increase, actually the peak can be found for higher $\mu$.  
\begin{figure}[htb]
\begin{center}
\includegraphics[scale=.7]{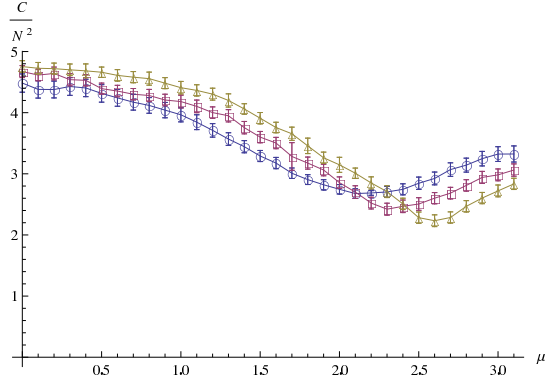}
\end{center}
\caption{\footnotesize Specific heat density for $\Omega=1$ varying $\mu$ and $N$.\normalsize}\label{Figure 23}
\end{figure}
\begin{figure}[htb]
\begin{center}
\includegraphics[scale=0.40]{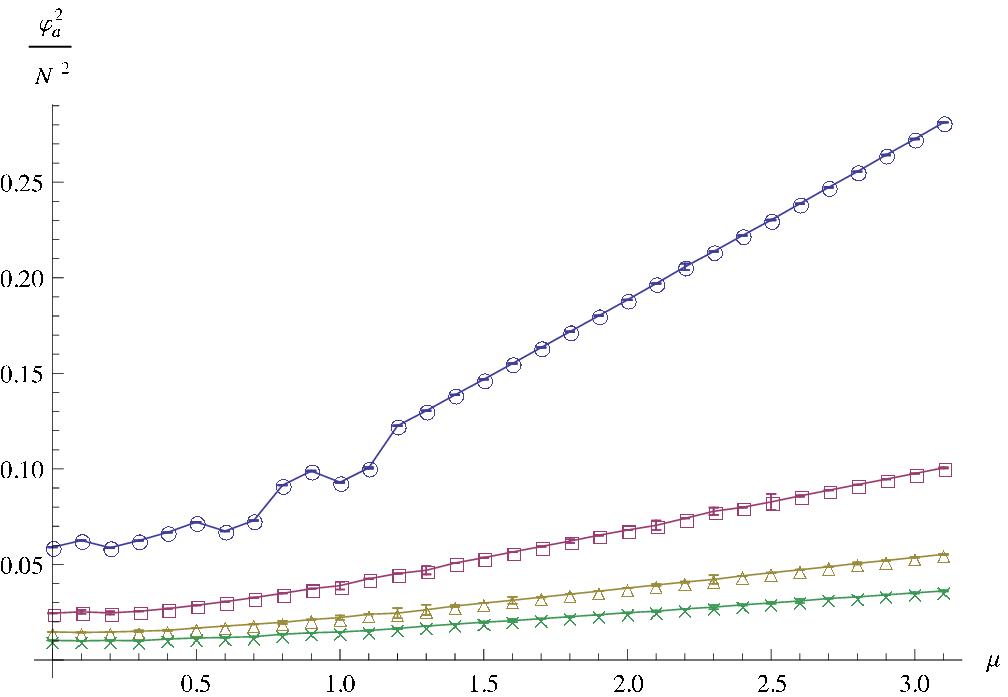}
\includegraphics[scale=0.40]{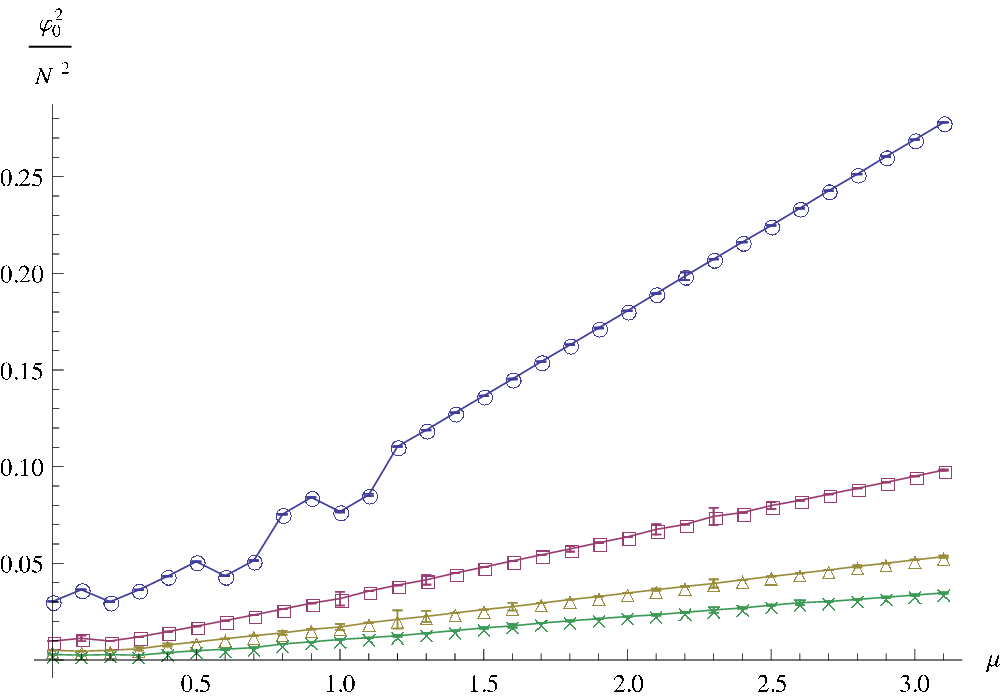}
\includegraphics[scale=0.40]{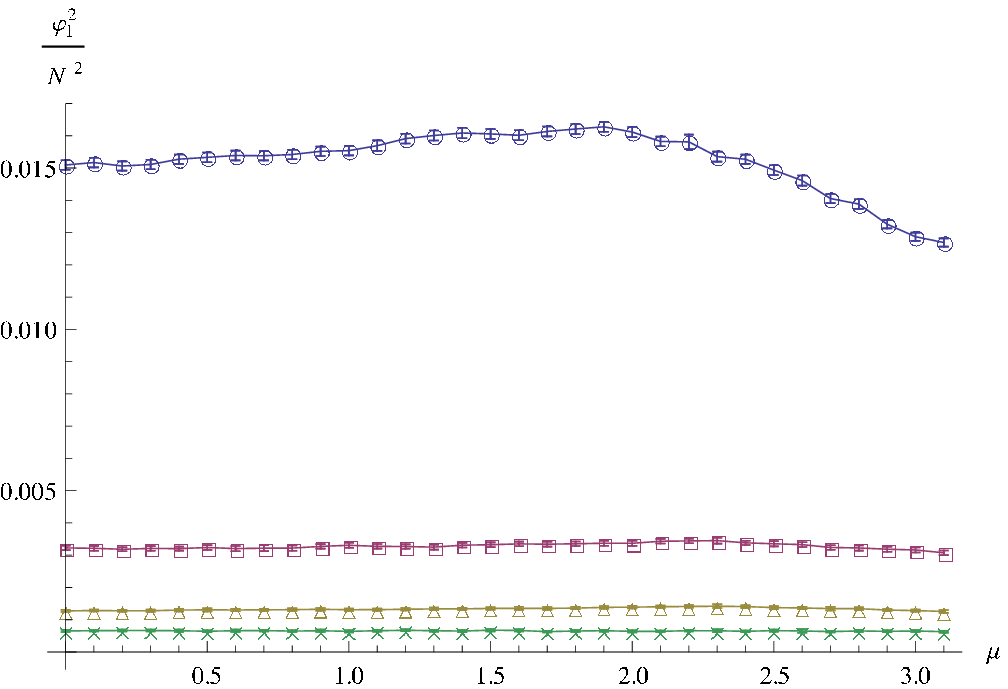}
\includegraphics[scale=0.40]{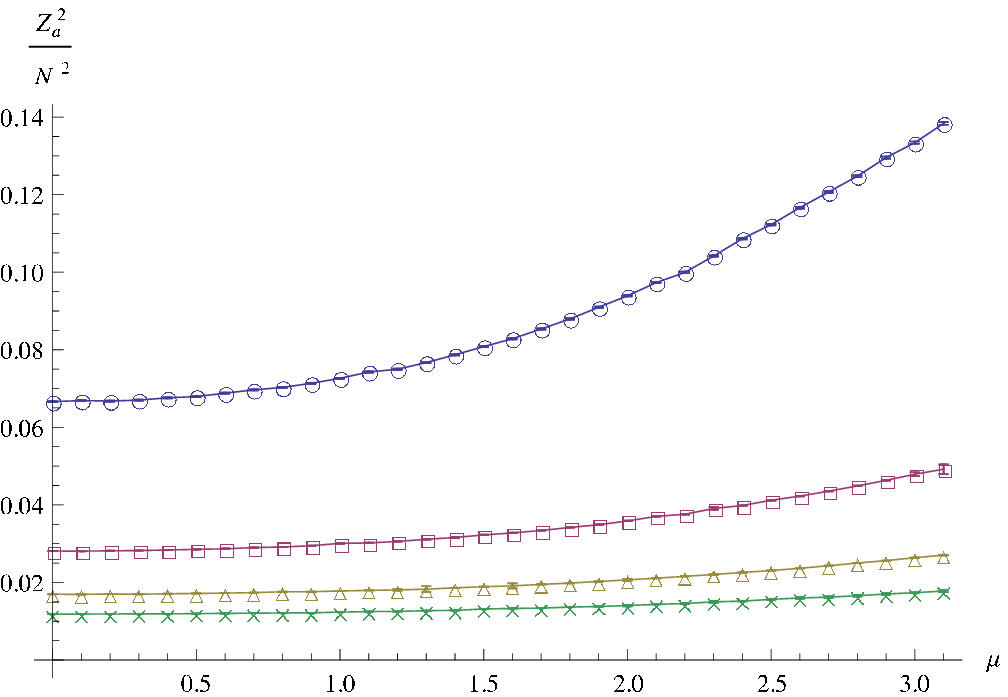}
\includegraphics[scale=0.40]{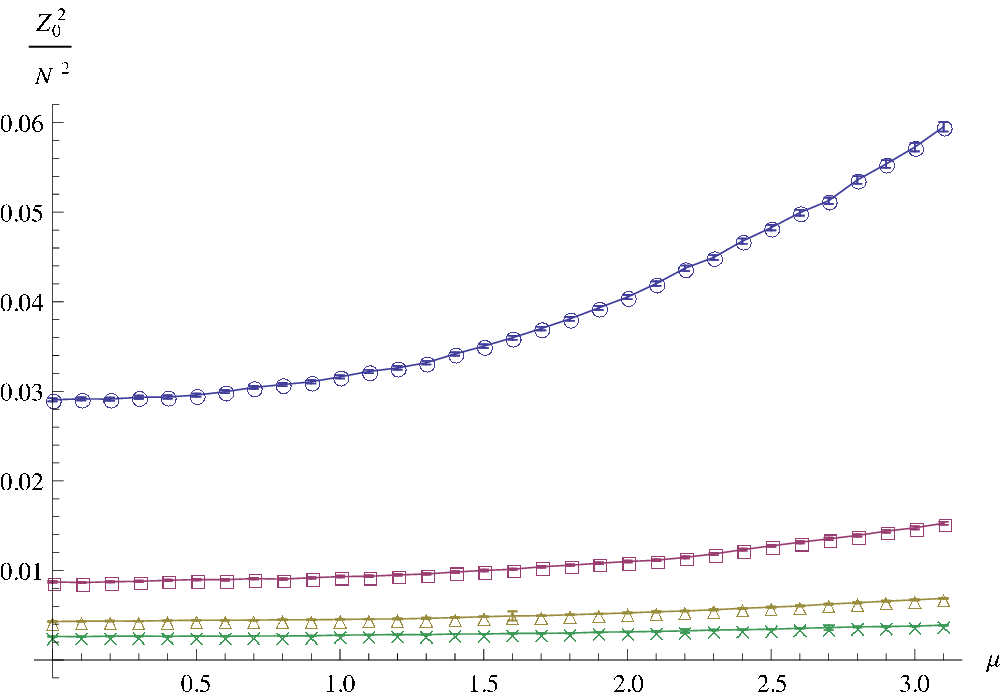}
\includegraphics[scale=0.40]{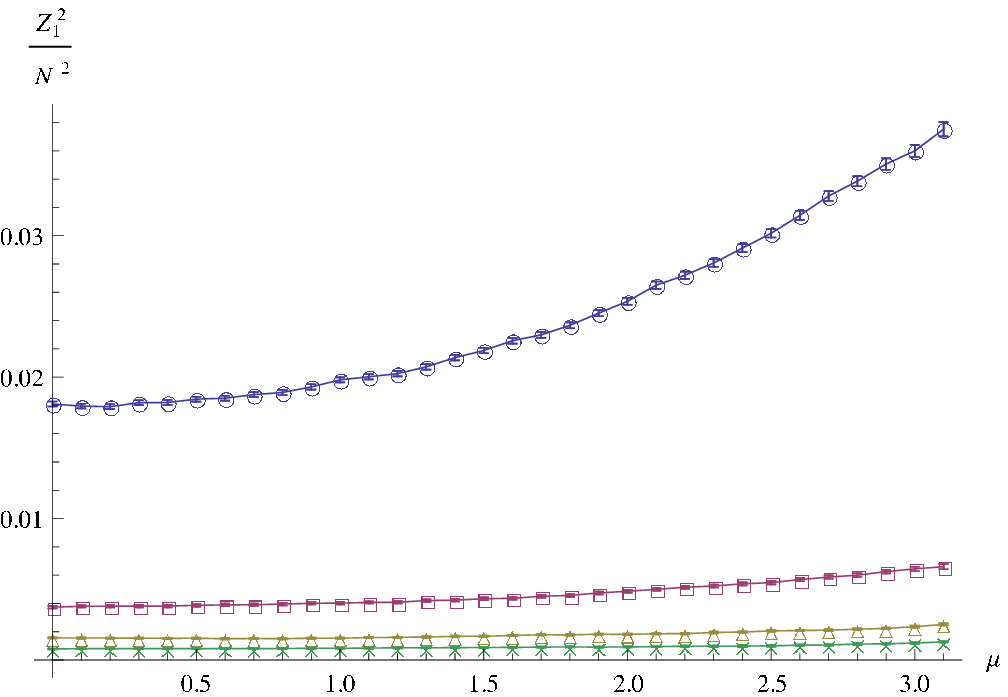}
\end{center}
\caption{\footnotesize Starting from the up left corner and from the left to the right the 	densities for $\varphi_a^2 $, $\varphi^2_0 $, $\varphi^2_1 $, $Z_{0a}^2$, $Z_{00}^2 $ and $Z_{01}^2$ for $\Omega=1$ varying $\mu$ and $N$.\normalsize}\label{Figure 24}
\end{figure}
At last in fig.\ref{Figure 24} we found a behavior of the order parameters density for $Z_0$ and $\psi$ fields similar to the former graphs for $\Omega=0.5$ and they are compatible with a dilatation of the previous diagrams.

\newpage
\section*{Conclusions and prospectives}
We have presented a study of a spectral action model constructed in order to extend to the Yang-Mills theories the non-commutative  Wulkenhaar-Grosse model.
The main aim of this work was to test a first Monte Carlo approach based on a non-perturbative regularization method. We have performed Monte Carlo simulations and obtained the values of the defined observables varying the parameters of the system. Despite the complexity of the approximated spectral action considered here we were able to obtain some reliable numerical results, we can conclude that a numerical approach of this kind of model using the matrix Moyal base seems feasible. The specific heat density shows various peaks indicating  phase transitions, in particular studying the behaviors for some fixed $\mu$  we found a peak around $\Omega=0$ for $\mu=\{0,1\}$ and peak in $\Omega=1$ for $ \mu=3$, beside we notice a huge change in the energy density and in its contributions between the cases $\mu=\{0,1\}$ and $\mu=3$. Other peaks in the specific heat density can be found varying $\mu$ and fixing $\Omega$, the graphs show that increasing $\Omega$  the peak in the specific heat, starting from  $\mu\approx 2.4$ for $\Omega=0$,  is moved towards higher $\mu$. The order parameters introduced show a strong dependence on the occurrence of $\Omega=0$ or $\Omega\neq0$. Referring to the fixed $\mu$ graphs we found a peak in the spherical contribution for the gauge fields $Z_i$, we can interpreted this slope as a sort of symmetry breaking introduced by $\Omega\neq0$. Additionally, varying $\mu$ and fixing $\Omega$ the other parameters display an increasing slope with  $\mu$ for all fields and all situation but one;  the graphs of  the order parameters concerning  $Z_{0a}$, $Z_{00}$  for $\Omega=0$ show a constant behavior. The  natural next steps in the numerical study of this model, could be the computation of the transition curves in order to separate the phase regions and classify them using eventually some additional  order parameters. Our treatment, forced by limited resource, was conducted conjecturing  that the system can fully described varying $\Omega$ in the range $[0,1]$ but since the L-S duality does not hold any more in our case, will be very useful to extend this range. Actually, the computed graphs does not show any periodicity in  $ \Omega\in[0,1]$ so there are some hints to infer that the range $[0,1]$ is not enough, however only a direct computation will  clarify this point.
Will be very interesting to do not require any more the condition $\mu^2>0$, in order to implement this change we have to conduct the calculation no more around the minimum of the actions. 
The expansion of the parameters space, together with the classification of the different phase regions allows us to compare our model with the results of the simulation conducted on the fuzzy spaces, looking in particular about the occurrence of the so called non-uniformly ordered phase which is connected with the UV/IR mixing. Since we have constructed our model starting from a renormalizable one this study is very desirable. 

\section*{Acknowledgements}
This work has been supported by the Marie Curie Research Training Network MRTN-CT-2006-031962 in Noncommutative Geometry, EU-NCG.
I wish here to acknowledge all those who contributed and help me making this work possible. In particular
to the Dublin Institute for Advanced Studies To  Prof. Raimar Wulkenhaar for all the guidance, and support, I  am very grateful for his
careful review of my  work and his very useful comments.

\end{document}